\documentclass{article}

 \usepackage[main, final]{neurips_2026}

\usepackage[utf8]{inputenc} 
\usepackage[T1]{fontenc}    
\usepackage{hyperref}       
\usepackage{url}            
\usepackage{booktabs}       
\usepackage{amsfonts}       
\usepackage{nicefrac}       
\usepackage{microtype}      
\usepackage{xcolor}         

\usepackage{amsmath}
\usepackage{amssymb}
\usepackage{mathtools}
\usepackage{amsthm}

\usepackage{bm}
\usepackage{multirow}

\usepackage{wrapfig}
\usepackage{color}
\usepackage{colortbl}
\usepackage{booktabs}
\usepackage[dvipsnames,table]{xcolor}

\newcommand{\thickhline}{\specialrule{\heavyrulewidth}{0pt}{0pt}}

\title{Hybrid Lagrangian-Eulerian Model for Lagrangian Fluid Simulation}

\author{%
  Ruoyan Li\\
  University of California, Los Angeles \\
  \AND
  Wei Wang \\
  University of California, Los Angeles \\
  \And
  Yizhou Sun \\
  University of California, Los Angeles \\
}

\begin{document}

\maketitle

\begin{abstract}
  Pure Lagrangian neural simulators offer geometric flexibility and exact advection, making them well-suited for modeling moving domains and free surfaces. However, the absence of a fixed global reference frame introduces two severe limitations: a spatial bottleneck, in which model capacity is wasted on uniform regions because the dense particle neighborhoods required for stable gradients are applied indiscriminately, and rapid temporal drift, caused by purely local message passing that lacks a global anchor. Inspired by classical hybrid numerical solvers, we propose a \textbf{Hybrid Lagrangian–Eulerian} neural simulator that augments Lagrangian dynamics with an Eulerian representation. To address the spatial bottleneck, we introduce adaptive downsampling that eliminates kinematic redundancy, preserving micro-scale details on particles while aggregating compressed features onto Eulerian nodes to resolve large-scale dynamics. To counter temporal drift, we employ a cross-attention mechanism that queries these Eulerian features, using the fixed grid as a stable spatial anchor to correct trajectory deviations at every timestep. Comprehensive experiments show that this hierarchical, cross-attended design substantially suppresses error accumulation, establishing a new state-of-the-art for accuracy and rollout stability in Lagrangian fluid simulation.
\end{abstract}

\section{Introduction}
Neural network–based physical simulators have recently achieved remarkable progress in fluid dynamics \citep{wang2024recentadvancesmachinelearning, wang2025fdbenchmodularfairbenchmark}. These models are typically built on either an Eulerian formulation, where dynamics are resolved on fixed spatial domains, or a Lagrangian formulation, which models the fluid as a set of discrete particles that carry and update physical quantities along their moving trajectories. Lagrangian representations are highly favored for their physical intuition and geometric flexibility. By tracking discrete particles, these methods naturally handle complex phenomena like moving domains and free surfaces, which are notoriously difficult to capture on fixed grids~\citep{monaghan1992smoothed, Gingold:1977sh}. Furthermore, since particles carry their properties as they move, advection is treated exactly and computational effort is concentrated where mass exists, avoiding the numerical diffusion and wasted resolution that plague grid-based schemes.  For example, the dam break dataset in Section~\ref{sec:main_results} and the stretching droplet dataset in Section~\ref{sec:elliptical} are particularly well-suited for a Lagrangian representation. An Eulerian approach would be highly inefficient here, as tracking the water splashes would require a prohibitively large and dense grid, and the stretching droplet would demand costly re-meshing at every time step.

However, the pure Lagrangian neural simulator introduces fundamental modeling challenges, primarily stemming from the lack of a fixed global reference frame. This absence cascades into two intrinsic challenges. First, without a background grid, Lagrangian methods suffer from a severe spatial and modeling bottleneck. While an Eulerian grid can calculate stable finite differences with as few as two neighbors, particle-based methods rely on kernel averaging that often requires a dense neighborhood of 30–50 particles to yield stable gradients \citep{Gingold:1977sh}. Consequently, maintaining grid-equivalent fidelity requires a drastically higher particle count \citep{Agertz_2007, Price_2012, Dehnen_2012}. Because of this dense packing, local particle motions often exhibit indistinguishable kinematics, forcing the neural solver to expend significant representational capacity on parsing highly correlated, nearly uniform local dynamics rather than capturing the more complex interactions. Second, the lack of a global anchor causes rapid temporal error accumulation. Because particles update their states based solely on local message passing, any slight deviation in a trajectory causes incorrect neighborhood connections in the next timestep. Without a fixed Eulerian grid to anchor and correct these local mistakes, topological errors compound geometrically, causing the simulation to drift significantly.

In classical computational physics, this precise dilemma led to the development of hybrid methods, most notably the Particle-in-Cell (PIC)~\citep{harlow1955machine, dawson1983particle, tskhakaya2008pic} method. PIC fundamentally improves Lagrangian simulation by leveraging an Eulerian grid as a global computational anchor. Instead of resolving complex PDE dynamics strictly between unstructured particle neighbors, PIC transfers particle properties onto a fixed background grid. The governing equations are then solved stably and efficiently on this Eulerian grid, and the resulting physical fields are interpolated back to update the particles. This approach anchors the drifting particle trajectories to a robust global structure while preserving the Lagrangian advantages. Inspired by this classical synergy, we investigate \textbf{how Eulerian representations can be similarly leveraged to enhance the accuracy and robustness of Lagrangian neural simulators.}

We propose a \textbf{Hybrid Lagrangian–Eulerian} neural model that downsamples indistinguishable particles and aggregates their information onto a set of Eulerian nodes, where the governing equations are solved efficiently. We further employ a cross-attention mechanism that facilitates information exchange between the Lagrangian and Eulerian features. The downsampling operation mitigates the computational burden of high-density representations by eliminating redundancy among particles with indistinguishable kinematics. This architecture also eases modeling difficulty through a hierarchical decomposition, as the downsampler handles micro-scale details while Eulerian aggregation resolves the underlying PDE at a coarse level. Additionally, because Eulerian nodes never move and their features are tied to fixed spatial locations, the Eulerian branch acts as a spatial anchor. Cross-attention re-grounds the Lagrangian update at every step against this stable reference, suppressing the error accumulation that causes pure-Lagrangian surrogates to drift into unphysical particle distributions over long rollouts. Ultimately, our design exploits the Eulerian representation to enhance the accuracy and stability of Lagrangian neural simulators.

Our contributions are as follows: \textbf{(i) Problem Identification:} We identify the fundamental limitations of pure Lagrangian neural simulators, specifically, spatial bottlenecks and temporal drift, and explain the benefits of a global Eulerian reference frame; \textbf{(ii) Practical Solution:} Inspired by classical hybrid methods like PIC, we present a \textbf{Hybrid Lagrangian–Eulerian} model that unifies the strengths of both representations within a single neural simulator; \textbf{(iii) Experimental Validation:} We conduct comprehensive experiments to validate that our model significantly enhances accuracy and stability over baseline methods.

\section{Related Work} 

Several prior works have explored Lagrangian fluid simulations. A prominent line of research focuses on purely graph-based methods to model particle interactions. GNS~\citep{sanchezgonzalez2020learningsimulatecomplexphysics} applies message-passing graph neural networks to learn the dynamics of fluids, granular media, and deformable bodies. Building on this foundation, several works have introduced physical constraints and solver mechanics to improve these networks. \citet{toshev2024neuralsphimprovedneural} improves upon GNS by incorporating various components from standard SPH solvers. Furthermore, \citet{toshev2023learning} introduces an architecture that enforces Euclidean symmetries, including translation, rotation, and reflection invariance, resulting in improved generalization, while \citet{prantl2022guaranteed} presents a novel method for guaranteeing linear momentum in the neural simulator. Subsequent works have sought to enhance efficiency and scalability by incorporating particle-grid transfers. \citet{rochman-sharabi2025a} integrates the material point method into a neural framework, enabling learned particle–grid transfers and updates for efficient emulation of high-fidelity simulations. \citet{xu2025hybridneuralmpminteractivefluid} integrates a neural simulator with the classical material-point method~\citep{johnson1996legacy, brackbill1986flip, sulsky1994particle, nairn2003material} specifically to achieve real-time simulation. To standardize the evaluation of this growing body of literature, LagrangeBench~\citep{toshev2024lagrangebench} provides a comprehensive benchmark for particle-based fluid simulation learning, systematically comparing neural models across multiple physical configurations. Finally, \citet{alkin2024upt} proposes a framework that generalizes across various discretizations. However, this approach falls under the domain of field learning as it requires ground truth particle coordinates to query velocity and cannot directly predict future particle positions. Due to this fundamental difference in problem formulation, it is not included in our comparative analysis.

In the domain of Eulerian simulation, significant progress has been made using deep learning to accelerate solvers on fixed spatial discretizations. This includes (1) grid-based methods~\citep{li2021fourier, li2023physicsinformedneuraloperatorlearning, brandstetter2022message, williams2023a, TACCARI2022104169, tran2023factorized, huang2024diffusionpdegenerativepdesolvingpartial, cao2021choosetransformerfouriergalerkin, lu2021learning}, and (2) mesh-based techniques designed to generalize across irregular geometries~\citep{pfaff2021learning, wu2024Transolver, luo2025transolver, cao2023efficient, li2023geometryinformed, wu2023LSM, wang2024latentneuraloperatorsolving, hao2023gnot}. While these approaches offer efficiency for problems with static topology, they do not address the specific challenges of Lagrangian tracking and dynamic neighborhood configurations central to the particle-based methods discussed in this work.

Beyond forward simulation, neural models are widely utilized for foundational models~\citep{hao2024dpot, mccabe2024multiple, ye2024pdeformer, herde2024poseidon, shen2024ups, zhouunisolver}, fluid field reconstruction~\citep{li2025flow, zhong2023sparse, mo2024reconstructingunsteadyflowssparse, yadav2025rfpinns, jing2024airflow, He_2022_flow}, ML-assisted classical solver~\citep{list2022learned, sun2023neural, greenfeld2019learning, sappl2019deep}, aerodynamic shape optimization~\citep{elrefaie2025drivaernet, NEURIPS2024_013cf29a}, and inverse design~\citep{behrmann2019invertible, teng2019invertible, kruse2021benchmarking}. While these works fall within the broader area of data-driven neural PDE modeling, they address computational tasks distinct from the sequential forward time-stepping of Lagrangian particles and therefore lie outside the primary scope of our work.

\section{Problem Statement}
We follow the setup in LagrangeBench~\citep{toshev2024lagrangebench} and FD-Bench~\citep{wang2025fdbenchmodularfairbenchmark} to focus on the weakly-compressible Navier–Stokes equation.
\begin{align}
\frac{\mathrm{d} \rho}{\mathrm{d} t} = - \rho \left( \nabla \cdot \dot{\bm{x}} \right), \quad
    \frac{\mathrm{d} \dot{\bm{x}}}{\mathrm{d} t} = -\frac{1}{\rho} \nabla p
    + \frac{1}{\mathrm{Re}} \nabla^2 \dot{\bm{x}}
    + \frac{1}{\rho} \bm{F},
\end{align}
where $\rho$ is density, $\dot{\bm{x}}$ is velocity, $p$ is pressure, $\mathrm{Re}$ is the Reynolds number, and $\bm{F}$ is an external force field. This choice is made solely due to the availability of extensive benchmark datasets. Our design is not tied to this particular equation, and we expect the model to generalize to other fluid scenarios. We adopt a particle-based representation, where the solution is discretized into a set of particles carrying physical properties, denoted by $\bm{x} \in \mathbb{R}^{N_\mathrm{lag} \times d}$, where $N_\mathrm{lag}$ represents the number of Lagrangian particles. 
Here, $d$ denotes the spatial dimension, e.g., 2D or 3D. Our objective is to learn a neural surrogate model $\Psi$ that, given $\tau$ consecutive frames of particle positions, predicts the positions in the next frame: $\bm{x}_{t+1} = \Psi \left( \bm{x}_{(t-\tau+1):t} \right)$. 

\begin{figure*}[t]
    \centering
    \includegraphics[width=\textwidth]{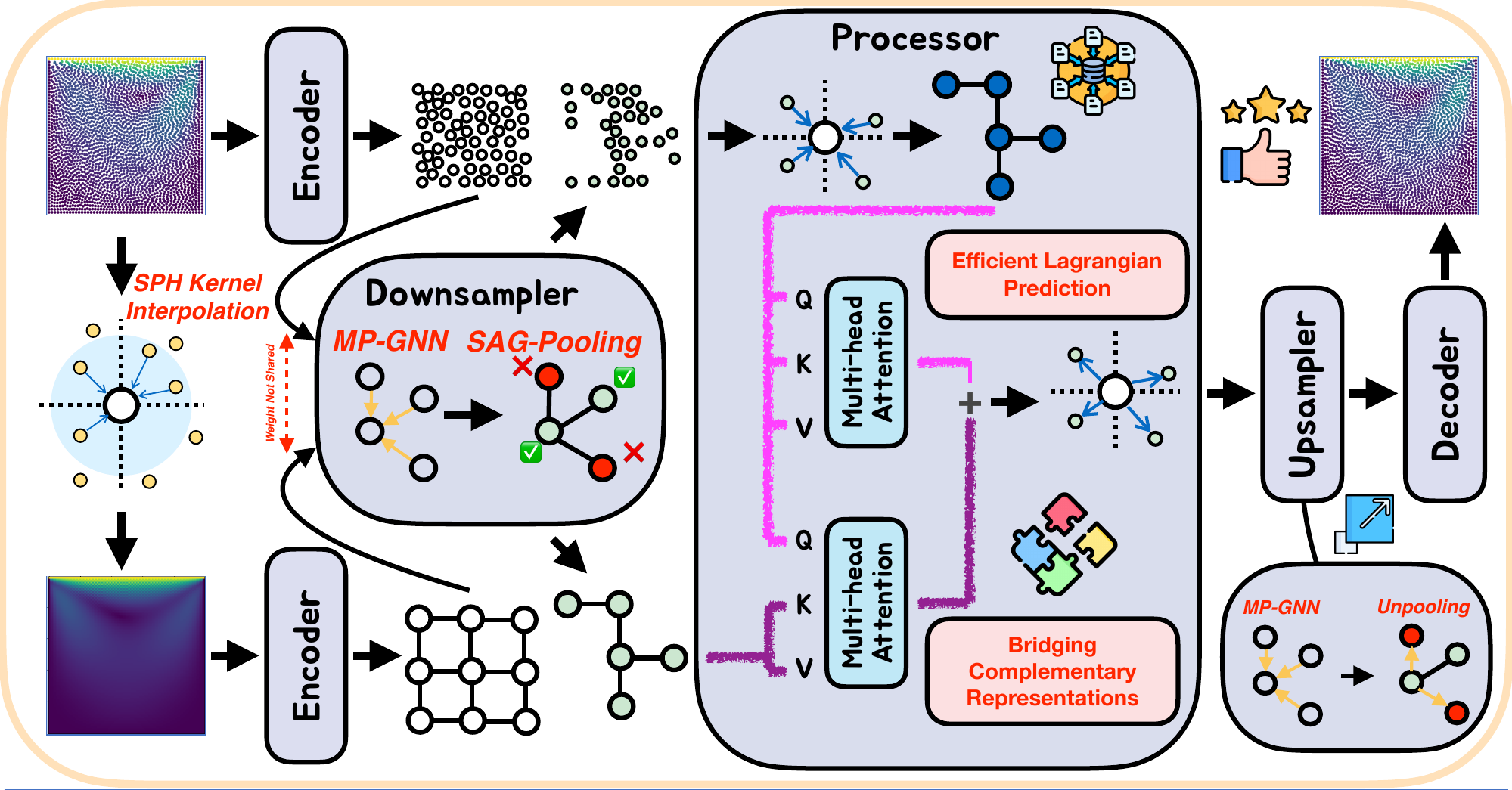}
    \caption{Overall framework of the proposed model. We employ an Encoder – Downsampler – Processor – Upsampler – Decoder architecture. Lagrangian inputs are converted to Eulerian graphs, encoded via MLPs, and downsampled into sparse representations. The processor solves coarse PDE dynamics by particle aggregation and cross-attention, after which the upsampler and decoder predict final kinematics.}
    \label{fig:framework}
\end{figure*}

\section{Methodology}
In this section, we present our \textbf{Hybrid Lagrangian–Eulerian} model, which adopts an Encoder – Downsampler – Processor – Upsampler – Decoder architecture. We describe the role of each component in the following subsections and Figure~\ref{fig:framework}.

\textbf{Preprocess} Given the initial particle positions $\bm{x}_{(t-\tau+1):t}$, we compute the velocities $\bm{v}_{(t-\tau+2):t}$ using the backward finite difference method. We then construct the Lagrangian graph $\mathcal{G}_\mathrm{lag} = (\mathcal{V}_\mathrm{lag}, Ec\mathrm{lag})$, whose feature vector is defined as $\mathcal{V}_\mathrm{lag}^i = \left[ \bm{x}_{t}^i \,\|\, \bm{v}_{(t-\tau+2):t}^i \,\|\, \mathrm{type}^i \,\|\, \Theta \right]$, where $\mathrm{type}^i$ denotes the type of particle $i$, and $\Theta$ represents global variables (both are optional). 
The edge set $E_\mathrm{lag}$ contains directed edge indices and edge attributes, where the attributes include the relative position vector and its norm. The edge indices are determined via a radius graph.

We also define a corresponding Eulerian graph. Let $\bm{\tilde{x}} \in \mathbb{R}^{N_\mathrm{eu} \times d}$ denote the positions of grid or mesh points in the spatial domain $\Omega$, where $N_\mathrm{eu}$ represents the number of Eulerian nodes. 
Unlike the Lagrangian particle positions, which evolve over time, the Eulerian nodes remain fixed. We choose $N_\mathrm{eu}$ such that $N_\mathrm{eu} \ll N_\mathrm{lag}$. To obtain Eulerian velocities, we follow \citet{ramachandran2021a} and aggregate the particle velocities using a quintic kernel  
$W: \mathbb{R}_{\ge 0} \times \mathbb{R}_{>0} \to \mathbb{R}_{\ge 0}$, defined by
\begin{align}
W(r,h) = \frac{\sigma_2}{h^{2}} \left[ (3-q)^5 - 6(2-q)^5 + 15(1-q)^5 \right]_+,
\end{align}
where $q = \frac{r}{h}$, $\sigma_2 = \frac{7}{478\pi}$, and $[\,\cdot\,]_+ := \max(\cdot,0)$. The Eulerian velocity aggregation is given by
\begin{align}
    \bm{\tilde{v}}_j
    = \frac{\sum_{i=1}^{N} \bm{v}_i\, W(\|\bm{\tilde{x}}_j - \bm{x}_i\|, h)}{\sum_{i=1}^{N} W(\|\bm{\tilde{x}}_j - \bm{x}_i\|, h) + \varepsilon},
\end{align}
where $\bm{\tilde{v}}_j$ denotes the velocity at Eulerian node $j$. The Eulerian interaction graph is defined as $\mathcal{G}_\mathrm{eu} = (\mathcal{V}_\mathrm{eu}, E_\mathrm{eu})$, where $\mathcal{V}_{\mathrm{eu}}^i = \left[ \bm{\tilde{x}}_{t}^i \,\|\, \bm{\tilde{v}}_{(t-\tau+2):t}^i \right]$, and $E_{\mathrm{eu}}$ contains directed edge indices that connect each node to its immediate horizontal and vertical neighbors, with edge features encoding the relative position vector and its norm. For regularly spaced Eulerian grids, omitting edge attributes and node positions does not negatively impact the model.

\textbf{Encoder: } In the encoding stage, we employ four distinct MLPs denoted by $\mathcal{E}_{\mathrm{lag}, \mathcal{V}}$, $\mathcal{E}_{\mathrm{lag}, E}$, $\mathcal{E}_{\mathrm{eu}, \mathcal{V}}$, and $\mathcal{E}_{\mathrm{eu}, E}$ to embed node and edge features into latent space. Formally,
\begin{align}
    \bm{h}_{\mathrm{lag}} &= \mathcal{E}_{\mathrm{lag}, \mathcal{V}}(\mathcal{V}_{\mathrm{lag}}), 
    &\quad 
    \bm{e}_{\mathrm{lag}} &= \mathcal{E}_{\mathrm{lag}, E}(E_{\mathrm{lag}}), \\ 
    \bm{h}_{\mathrm{eu}}  &= \mathcal{E}_{\mathrm{eu}, \mathcal{V}}(\mathcal{V}_{\mathrm{eu}}), 
    &\quad 
    \bm{e}_{\mathrm{eu}}  &= \mathcal{E}_{\mathrm{eu}, E}(E_{\mathrm{eu}}).
\end{align}
Here, $\bm{h}_{\mathrm{lag}} \in \mathbb{R}^{N_{\mathrm{lag}} \times d_{\mathrm{latent}}}$ and $\bm{h}_{\mathrm{eu}} \in \mathbb{R}^{N_{\mathrm{eu}} \times d_{\mathrm{latent}}}$.

\textbf{Downsampler: } Our downsampler consists of $M$ successive blocks, each comprising a SAGPooling~\citep{pmlr-v97-lee19c} operation followed by message passing. Given a node feature $\bm{h} \in \mathbb{R}^{N \times d_{\mathrm{latent}}}$ as well as edge index $E$ with latent features $\bm{e}$. SAGPooling first computes a scalar importance score for each node via a self-attention GNN: $\bm{s} = \mathrm{GNN}(\bm{h}, \bm{e})$. Then, it selects the top-$k$ nodes with the highest importance scores: $\operatorname{TopK}(\mathbf{s}, k)$, where $k = \lceil \lambda N \rceil$ and $\lambda$ is the downsmpling rate, and retain only the corresponding node features and the induced subgraph.
Next, we apply a message passing GNN with the update rule $\mathbf{m}_i = \sum_{(j \to i) \in E} \phi_m(\mathbf{h}_i, \mathbf{h}_j, \mathbf{e}_{i,j})$ and $\mathbf{h}_i \leftarrow \phi_h(\mathbf{h}_i, \mathbf{m}_i)$, where $\phi_m$ and $\phi_h$ are parametrized by MLPs.

We apply this downsampling procedure independently to both the Lagrangian and Eulerian graphs, yielding latent node embeddings $\mathbf{q}_{\mathrm{lag}} \in \mathbb{R}^{\tilde{N}_{\mathrm{lag}} \times d_{\mathrm{latent}}}$ and $\mathbf{q}_{\mathrm{eu}} \in \mathbb{R}^{\tilde{N}_{\mathrm{eu}} \times d_{\mathrm{latent}}}$, with associated particle or node coordinates $\mathbf{y}$ and $\tilde{\mathbf{y}}$, respectively. By construction, $\tilde{N}_{\mathrm{eu}} \ll \tilde{N}_{\mathrm{lag}}$. To reduce the number of hyperparameters, we use a shared pooling rate $\lambda$ for both the Lagrangian and Eulerian graphs. Empirically, this choice is sufficient to achieve strong performance. Nevertheless, if computational resources permit, we recommend assigning separate rates to the two graphs and performing hyperparameter tuning.

The downsampler plays a critical role by reducing the redundancy inherent in particle-based fluid representations. In regions where the flow is smooth or nearly uniform, such as laminar zones, the dynamics of individual particles become indistinguishable. They move together or remain nearly static, offering little new information if treated separately. The downsampler condenses these redundant particles into a smaller set of representative super-nodes, capturing their shared properties without expending computation on unnecessary features. This reduction not only improves efficiency but also ensures that the model allocates its capacity to regions with complex, highly variable dynamics, where detailed interactions matter. The processor, introduced in the next section, then operates on this compressed representation, solving the governing equation at a global level. Without this downsampler, the processor would be forced to carry the full burden of redundant particle interactions, wasting computation while diluting attention away from the regions that matter most.

\textbf{Processor: } We define a learnable aggregation kernel $\mathcal{K}: \mathbb{R}^d \times \mathbb{R}^d \to \mathbb{R}_{>0}$, parametrized by MLPS, that maps a Lagrangian particle position and an Eulerian node position to a positive weight. For each particle $i$, $\mathcal{K}$ assigns a weight to each Eulerian node $j$, satisfying the normalization condition $\sum_{j=1}^{\tilde{N}_{\mathrm{eu}}} \mathcal{K}(\bm{y}_i, \tilde{\bm{y}}_j) = 1$. Let $\bm{W} \in \mathbb{R}^{\tilde{N}_{\mathrm{lag}} \times \tilde{N}_{\mathrm{eu}}}$ denote the weight matrix produced by this kernel. Using $\bm{W}$, we aggregate the downsampled Lagrangian particles onto the downsampled Eulerian nodes:
\begin{align*}
    \mathbf{Z}_{\mathrm{lag}} = \bm{W}^\top \bm{q}_{\mathrm{lag}} \in \mathbb{R}^{\tilde{N}_{\mathrm{eu}} \times d_{\mathrm{latent}}}.
\end{align*}
We then apply multi-head self-attention to the aggregated Lagrangian feature,
\begin{align}
\bm{A} = 
    \mathrm{MHA}\!\left(\mathrm{Query} = \mathbf{Z}_{\mathrm{lag}},\ \mathrm{Key} = \mathbf{Z}_{\mathrm{lag}},\ \mathrm{Value} = \mathbf{Z}_{\mathrm{lag}}\right),
\end{align}
producing coarse predictions based solely on the aggregated Lagrangian representation. Next, we perform cross-attention between the aggregated Lagrangian feature and the downsampled Eulerian feature,
\begin{align}
\bm{C} = \mathrm{MHA}\!\left(\mathrm{Query} = \bm{Z}_{\mathrm{lag}},\ \mathrm{Key} = \bm{q}_{\mathrm{eu}},\ \mathrm{Value} = \bm{q}_{\mathrm{eu}}\right),
\end{align}
allowing the model to jointly consider Lagrangian and Eulerian features. The outputs are then combined and scattered back to the downsampled Lagrangian nodes:
\begin{align}
    \bm{F} = \bm{A} + \alpha\,\bm{C}, \quad \hat{\bm{q}}_{\mathrm{lag}} = \bm{W}\,\bm{F} \in \mathbb{R}^{\tilde{N}_{\mathrm{lag}} \times d_{\mathrm{latent}}},
    \label{eq:eu_plus}
\end{align}
where $\alpha$ is a learnable parameter.

The processor is inspired by PIC~\citep{harlow1955machine, dawson1983particle, tskhakaya2008pic}, adapted to the hierarchical latent space. In classical PIC, particles are projected onto a grid using voxelization to solve field equations. The learnable aggregation kernel $\mathcal{K}$ plays the role of this projection operator, while the self-attention on aggregated Lagrangian feature functions as a latent field solver on a coarse level.



The Lagrangian formulation naturally tracks the material derivative along particle trajectories, providing an accurate representation of advection-dominated dynamics, while the Eulerian formulation encodes the spatial structure of the flow field on a fixed stencil, facilitating the computation of differential operators such as divergence and vorticity that govern incompressibility. Self-attention on aggregated Lagrangian representation alone can only mix Lagrangian-derived information, whereas cross-attention is the channel by which each Lagrangian query token retrieves the field-derived quantities it cannot reconstruct internally. This complementarity is also what makes the design stable over long autoregressive rollouts. Pure-Lagrangian neural surrogates are known to drift, since small per-step errors compound into unphysical particle distributions over hundreds of steps. Because Eulerian nodes never move and their features are tied to fixed spatial locations, the Eulerian branch acts as a spatial anchor. Cross-attention re-grounds the Lagrangian update at every step against this stable, field-based reference, suppressing error accumulation and improving long-horizon fidelity.

We repeat this layer for $L$ steps. Note that the downsampled Lagrangian particles and Eulerian nodes are fairly small. The calculation of the weight matrix and aggregation does not incur significant computational overhead.

\textbf{Upsampler: }
With $\hat{\bm{q}}_{\mathrm{lag}}$, we reconstruct the full particle latent representation $\hat{\bm{h}}_{\mathrm{lag}} \in \mathbb{R}^{N_{\mathrm{lag}} \times d_{\mathrm{latent}}}$ by executing $M$ successive upsampling operations, each coupled with message-passing steps, while integrating a residual link from the downsampling stage.

\textbf{Decoder: }
The latent state $\hat{\bm{h}}_{\mathrm{lag}}$ is passed through a learnable MLP decoder to produce the predicted position, velocity, or acceleration. If the network output is velocity or acceleration, position is computed with Euler integration.

\textbf{Remark: }
While NeuralMPM~\citep{rochman-sharabi2025a} shares a conceptual link with our work in that it also aggregates particles to Eulerian nodes, our work is not simply an extension of this method with additional components. The voxelization approach used in this prior work is highly restrictive, limiting it exclusively to regular grids, and they only perform a single aggregation step at the beginning of the process before scattering back at the end. In contrast, our method is highly adaptive. We utilize a completely different aggregation strategy that performs distinct particle aggregations at each layer, which we believe significantly enhances the model's expressivity. Furthermore, because our Eulerian representation is defined on a general mesh graph rather than a fixed grid, it naturally extends to arbitrary domains and adapts to complex geometries and deformable domains, a capability we demonstrate in subsequent experiments. By combining the hierarchical, layer-wise aggregation with a cross-attention mechanism, our framework achieves geometric generality, efficiency, and fidelity in a manner entirely separate from existing methods.

\begin{table*}[!t]
\centering
\scriptsize{%
\resizebox{\linewidth}{!}{%
    \setlength\tabcolsep{3pt}%
    \renewcommand\arraystretch{1.4}%
\begin{tabular}{l | l || c c c c c c c || c}
\hline\thickhline
\rowcolor{CadetBlue!20}
& Metric & GNS & SEGNN & GraphUnet & GraphTransformer & AdvDIFFormer & NeuralMPM & PhysicsNFP & \cellcolor[HTML]{FFFFE0}\textbf{Ours} \\
\hline

\multirow{3}{*}{\rotatebox{90}{\parbox{1cm}{\centering \textit{2D TGV}}}}%
& \cellcolor{green!3} $\mathrm{MSE}$ & \cellcolor{green!3} $7.920 \times 10^{-3}$ & \cellcolor{green!3} $1.260 \times 10^{-2}$ & \cellcolor{green!3} $5.356 \times 10^{-3}$ & \cellcolor{green!3} $2.963 \times 10^{-2}$ & \cellcolor{green!3} $1.586 \times 10^{-2}$ & \cellcolor{green!3} $3.051 \times 10^{-2}$  & \cellcolor{green!3} $7.140 \times 10^{-3}$ & \cellcolor[HTML]{FFFFE0}\bm{$4.551 \times 10^{-3}$}$_{\textcolor{ForestGreen}{\uparrow 15.02\%}}$ \\
& \cellcolor{pink!10} $\mathrm{KE}$ & \cellcolor{pink!10} $1.795 \times 10^{-3}$ & \cellcolor{pink!10} $2.854 \times 10^{-3}$ & \cellcolor{pink!10} $1.246 \times 10^{-3}$  & \cellcolor{pink!10} $4.729 \times 10^{-3}$ & \cellcolor{pink!10} $2.339 \times 10^{-3}$  & \cellcolor{pink!10} $6.402 \times 10^{-3}$   & \cellcolor{pink!10} $1.652 \times 10^{-3}$ & \cellcolor[HTML]{FFFFE0}\bm{$1.082 \times 10^{-3}$}$_{\textcolor{ForestGreen}{\uparrow 13.20\%}}$ \\
& \cellcolor{blue!3} $\mathrm{Sinkhorn}$ & \cellcolor{blue!3} $4.150 \times 10^{-5}$ & \cellcolor{blue!3} $3.722 \times 10^{-4}$ & \cellcolor{blue!3} $3.767 \times 10^{-5}$  & \cellcolor{blue!3} $3.099 \times 10^{-3}$ & \cellcolor{blue!3} $1.744 \times 10^{-4}$ & \cellcolor{blue!3} $9.607 \times 10^{-4}$  & \cellcolor{blue!3} $3.323 \times 10^{-5}$  & \cellcolor[HTML]{FFFFE0}\bm{$1.422 \times 10^{-5}$}$_{\textcolor{ForestGreen}{\uparrow 57.22\%}}$ \\

\hline

\multirow{3}{*}{\rotatebox{90}{\parbox{1cm}{\centering \textit{2D LDC}}}}%
& \cellcolor{green!3} $\mathrm{MSE}$ & \cellcolor{green!3} $3.542 \times 10^{-5}$ & \cellcolor{green!3} $6.293 \times 10^{-6}$ & \cellcolor{green!3} $4.413 \times 10^{-6}$  & \cellcolor{green!3} $6.490 \times 10^{-4}$ & \cellcolor{green!3} $3.955 \times 10^{-5}$    & \cellcolor{green!3} $2.308 \times 10^{-3}$ & \cellcolor{green!3} $2.629 \times 10^{-6}$ & \cellcolor[HTML]{FFFFE0}\bm{$1.868 \times 10^{-6}$}$_{\textcolor{ForestGreen}{\uparrow 28.94\%}}$ \\
& \cellcolor{pink!10} $\mathrm{KE}$ & \cellcolor{pink!10} $7.880 \times 10^{-6}$ & \cellcolor{pink!10} $1.043 \times 10^{-5}$ & \cellcolor{pink!10} $1.163 \times 10^{-7}$  & \cellcolor{pink!10} $1.635 \times 10^{-4}$ & \cellcolor{pink!10} $1.316 \times 10^{-5}$     & \cellcolor{pink!10} $5.724 \times 10^{-2}$ & \cellcolor{pink!10} $1.068 \times 10^{-7}$ & \cellcolor[HTML]{FFFFE0}\bm{$7.240 \times 10^{-8}$}$_{\textcolor{ForestGreen}{\uparrow 32.23\%}}$ \\
& \cellcolor{blue!3} $\mathrm{Sinkhorn}$ & \cellcolor{blue!3} $8.239 \times 10^{-6}$ & \cellcolor{blue!3} $1.388 \times 10^{-5}$ & \cellcolor{blue!3} $2.227 \times 10^{-6}$  & \cellcolor{blue!3} $1.803 \times 10^{-5}$ & \cellcolor{blue!3} $2.214 \times 10^{-5}$   & \cellcolor{blue!3} $3.172 \times 10^{-4}$ & \cellcolor{blue!3} $8.768 \times 10^{-7}$ & \cellcolor[HTML]{FFFFE0}\bm{$4.675 \times 10^{-7}$}$_{\textcolor{ForestGreen}{\uparrow 46.68\%}}$ \\

\hline

\multirow{3}{*}{\rotatebox{90}{\parbox{1cm}{\centering \textit{2D RPF}}}}%
& \cellcolor{green!3} $\mathrm{MSE}$ & \cellcolor{green!3} $2.629 \times 10^{-1}$ & \cellcolor{green!3} $5.779 \times 10^{-2}$  & \cellcolor{green!3} $4.067 \times 10^{-2}$ & \cellcolor{green!3} $8.835 \times 10^{-1}$   & \cellcolor{green!3} $2.842 \times 10^{-1}$  & \cellcolor{green!3} $1.044 \times 10^{-1}$ & \cellcolor{green!3} $3.862 \times 10^{-2}$ & \cellcolor[HTML]{FFFFE0}\bm{$2.426 \times 10^{-2}$}$_{\textcolor{ForestGreen}{\uparrow 37.19\%}}$ \\
& \cellcolor{pink!10} $\mathrm{KE}$ & \cellcolor{pink!10} $7.961 \times 10^{-2}$ & \cellcolor{pink!10} $6.374 \times 10^{-2}$ & \cellcolor{pink!10} $2.471 \times 10^{-2}$  & \cellcolor{pink!10}  $4.104 \times 10^{0}$ & \cellcolor{pink!10} $5.848 \times 10^{-2}$ & \cellcolor{pink!10} $2.694 \times 10^{-2}$ & \cellcolor{pink!10} $3.181 \times 10^{-2}$ & \cellcolor[HTML]{FFFFE0}\bm{$1.595 \times 10^{-2}$}$_{\textcolor{ForestGreen}{\uparrow 35.42\%}}$ \\
& \cellcolor{blue!3} $\mathrm{Sinkhorn}$ & \cellcolor{blue!3} $3.627 \times 10^{-2}$ & \cellcolor{blue!3} $2.134 \times 10^{-2}$ & \cellcolor{blue!3} $7.909 \times 10^{-3}$  & \cellcolor{blue!3} $4.120 \times 10^{-1}$ & \cellcolor{blue!3} $4.812 \times 10^{-2}$ & \cellcolor{blue!3} $2.163 \times 10^{-2}$ & \cellcolor{blue!3} $1.684 \times 10^{-2}$ & \cellcolor[HTML]{FFFFE0}\bm{$6.140 \times 10^{-3}$}$_{\textcolor{ForestGreen}{\uparrow 22.37\%}}$ \\

\hline

\multirow{3}{*}{\rotatebox{90}{\parbox{1cm}{\centering \textit{2D DAM}}}}%
& \cellcolor{green!3} $\mathrm{MSE}$ & \cellcolor{green!3} $2.363 \times 10^{-4}$ & \cellcolor{green!3} $9.245 \times 10^{-4}$ & \cellcolor{green!3} $1.505 \times 10^{-4}$  & \cellcolor{green!3} $4.552 \times 10^{-3}$ & \cellcolor{green!3} $6.360 \times 10^{-3}$  & \cellcolor{green!3} $2.010 \times 10^{-4}$ & \cellcolor{green!3} $1.458 \times 10^{-4}$ & \cellcolor[HTML]{FFFFE0}\bm{$9.171 \times 10^{-5}$}$_{\textcolor{ForestGreen}{\uparrow 37.10\%}}$ \\
& \cellcolor{pink!10} $\mathrm{KE}$ & \cellcolor{pink!10} $9.461 \times 10^{-7}$ & \cellcolor{pink!10} $1.817 \times 10^{-6}$ & \cellcolor{pink!10} $9.235 \times 10^{-7}$  & \cellcolor{pink!10} $3.783 \times 10^{-4}$  & \cellcolor{pink!10} $6.238 \times 10^{-4}$  & \cellcolor{pink!10} $6.080 \times 10^{-6}$ & \cellcolor{pink!10} $7.856 \times 10^{-7}$ & \cellcolor[HTML]{FFFFE0}\bm{$4.709 \times 10^{-7}$}$_{\textcolor{ForestGreen}{\uparrow 40.06\%}}$ \\
& \cellcolor{blue!3} $\mathrm{Sinkhorn}$ & \cellcolor{blue!3} $2.131 \times 10^{-4}$ & \cellcolor{blue!3} $1.935 \times 10^{-4}$ & \cellcolor{blue!3} $3.710 \times 10^{-4}$  & \cellcolor{blue!3} $7.751 \times 10^{-3}$ & \cellcolor{blue!3} $8.843 \times 10^{-3}$ & \cellcolor{blue!3} $2.196 \times 10^{-4}$ & \cellcolor{blue!3} $1.886 \times 10^{-4}$ & \cellcolor[HTML]{FFFFE0}\bm{$1.216 \times 10^{-4}$}$_{\textcolor{ForestGreen}{\uparrow 35.55\%}}$ \\

\hline

\end{tabular}}}%
\caption{Comparison of MSE, Kinetic Energy (KE) error, and Sinkhorn divergence across five rollout steps on 2D datasets. The relative performance gain against the strongest baseline is computed as $\frac{\text{Baseline} - \text{Ours}}{\text{Baseline}} \times 100\%$.}
\label{tab:main_results}
\end{table*}

\section{Experiment}
\textbf{Dataset} We use Lagrangian particle-based datasets~\citep{toshev2024lagrangebench} solved based on the weakly-compressible Navier–Stokes equations using SPH~\citep{desbrun1996smoothed, hoover2006smooth, monaghan1992smoothed} with the quintic kernel. The \textit{Taylor–Green Vortex (TGV)} contains a periodic-domain simulation initialized from an analytical solution without any external forcing. The \textit{Lid-Driven Cavity Flow (LDC)} introduces no-slip boundaries, including a moving top lid that drives the internal circulation, creating sharp shear layers and corner vortices. \textit{Reverse Poiseuille Flow (RPF)} features opposing constant body forces in the upper and lower halves of the domain. \textit{Dam Break (DAM)} captures the transient free-surface flow following the sudden release of a water column into an open downstream region, resulting in splashing and wave breaking. The datasets are sampled every 100 steps of the ground truth solver. Thus, a five-step rollout corresponds to 500 simulation steps. We refer the readers to Appendix~\ref{appendix:datasets} for additional details on the datasets.

\textbf{Task Setup and Baselines}
To evaluate our model, we benchmark it against a range of strong baselines. We include GNS~\citep{sanchezgonzalez2020learningsimulatecomplexphysics} and SEGNN~\citep{brandstetter2021geometric}, the top-performing models reported in LagrangeBench \citep{toshev2024lagrangebench}. Graph-Unet~\citep{gao2019graph} and GraphTransformer~\citep{dwivedi2021generalizationtransformernetworksgraphs} are also considered due to their wide adoption and proven effectiveness in graph-based learning tasks. In addition, we evaluate against AdvDIFFormer~\citep{wu2025supercharging} and PhysicsNFP~\citep{jiang2025topologyaware}, which are recently proposed architectures that have achieved competitive results in diverse physical simulation settings. We also include NeuralMPM~\citep{rochman-sharabi2025a}, as it represents the work most closely related to our approach. We refer to Appendix~\ref{appendix:model_training_and_implementation_details} for detailed training and implementation of all models.

\begin{wraptable}{r}{0.5\textwidth}
\centering
\scriptsize{%
\resizebox{\linewidth}{!}{%
    \setlength\tabcolsep{3pt}%
    \renewcommand\arraystretch{1.4}%
\begin{tabular}{l | l || c c|| c}
\hline\thickhline
\rowcolor{CadetBlue!20}
& Metric & GraphUnet & PhysicsNFP & \cellcolor[HTML]{FFFFE0}\textbf{Ours} \\
\hline

\multirow{3}{*}{\rotatebox{90}{\parbox{1cm}{\centering \textit{3D TGV}}}}%
& \cellcolor{green!3} $\mathrm{MSE}$ & \cellcolor{green!3} $4.512 \times 10^{-1}$ & \cellcolor{green!3} $3.034 \times 10^{0}$ & \cellcolor[HTML]{FFFFE0}\bm{$3.649 \times 10^{-1}$}$_{\textcolor{ForestGreen}{\uparrow 19.14\%}}$ \\
& \cellcolor{pink!10} $\mathrm{KE}$ & \cellcolor{pink!3} $3.719 \times 10^{0}$ & \cellcolor{pink!3} $1.176 \times 10^{2}$ & \cellcolor[HTML]{FFFFE0}\bm{$3.215 \times 10^{0}$}$_{\textcolor{ForestGreen}{\uparrow 13.56\%}}$ \\
& \cellcolor{blue!3} $\mathrm{Sinkhorn}$ & \cellcolor{blue!3} $8.799 \times 10^{-2}$ & \cellcolor{blue!3} $7.560 \times 10^{-1}$ & \cellcolor[HTML]{FFFFE0}\bm{$1.094 \times 10^{-2}$}$_{\textcolor{ForestGreen}{\uparrow 87.56\%}}$ \\

\hline

\multirow{3}{*}{\rotatebox{90}{\parbox{1cm}{\centering \textit{3D LDC}}}}%
& \cellcolor{green!3} $\mathrm{MSE}$ & \cellcolor{green!3} $4.810 \times 10^{-3}$ & \cellcolor{green!3} $2.455 \times 10^{-2}$ & \cellcolor[HTML]{FFFFE0}\bm{$3.059 \times 10^{-3}$}$_{\textcolor{ForestGreen}{\uparrow 36.40\%}}$ \\
& \cellcolor{pink!10} $\mathrm{KE}$ & \cellcolor{pink!3} $1.464 \times 10^{-4}$ & \cellcolor{pink!3} $1.807 \times 10^{-2}$ & \cellcolor[HTML]{FFFFE0}\bm{$1.271 \times 10^{-4}$}$_{\textcolor{ForestGreen}{\uparrow 13.17\%}}$ \\
& \cellcolor{blue!3} $\mathrm{Sinkhorn}$ & \cellcolor{blue!3} $4.647 \times 10^{-4}$ & \cellcolor{blue!3} $3.452 \times 10^{-2}$ & \cellcolor[HTML]{FFFFE0}\bm{$2.436 \times 10^{-4}$}$_{\textcolor{ForestGreen}{\uparrow 47.59\%}}$ \\

\hline

\multirow{3}{*}{\rotatebox{90}{\parbox{1cm}{\centering \textit{3D RPF}}}}%
& \cellcolor{green!3} $\mathrm{MSE}$ & \cellcolor{green!3} $2.161 \times 10^{-2}$ & \cellcolor{green!3} $6.386 \times 10^{-2}$ & \cellcolor[HTML]{FFFFE0}\bm{$1.810 \times 10^{-2}$}$_{\textcolor{ForestGreen}{\uparrow 16.25\%}}$ \\
& \cellcolor{pink!10} $\mathrm{KE}$ & \cellcolor{pink!3} $1.167 \times 10^{-2}$ & \cellcolor{pink!3} $7.377 \times 10^{-2}$ & \cellcolor[HTML]{FFFFE0}\bm{$1.022 \times 10^{-2}$}$_{\textcolor{ForestGreen}{\uparrow 12.38\%}}$ \\
& \cellcolor{blue!3} $\mathrm{Sinkhorn}$ & \cellcolor{blue!3} $3.721 \times 10^{-4}$ & \cellcolor{blue!3} $8.551 \times 10^{-3}$ & \cellcolor[HTML]{FFFFE0}\bm{$3.018 \times 10^{-4}$}$_{\textcolor{ForestGreen}{\uparrow 18.89\%}}$ \\

\hline

\end{tabular}}}%
\caption{Comparison of MSE, Kinetic Energy (KE) error, and sinkhorn divergence across five rollout steps on \textbf{3D datasets}. The relative performance gain against the strongest baseline is computed as $\frac{\text{Baseline} - \text{Ours}}{\text{Baseline}} \times 100\%$.}
\label{tab:3d_results}
\end{wraptable}

Model performance is evaluated over five-step rollouts using three metrics. The mean squared error (MSE) measures the accuracy of trajectory predictions. The kinetic energy error (KE) captures global discrepancies in the fluid’s energy. Finally, the Sinkhorn divergence quantifies the distance between predicted and reference particle distributions using optimal transport.

All models are configured to directly predict particle positions from unperturbed inputs. We deliberately exclude auxiliary training strategies found in prior literature~\citep{pfaff2021learning, sanchezgonzalez2020learningsimulatecomplexphysics} to isolate the specific contributions of our proposed modules.

\begin{wraptable}{r}{0.5\textwidth}
\centering
\scriptsize{%
\resizebox{\linewidth}{!}{%
    \setlength\tabcolsep{3pt}%
    \renewcommand\arraystretch{1.4}%
\begin{tabular}{l | l || c c|| c}
\hline\thickhline
\rowcolor{CadetBlue!20}
& Metric & GraphUnet & PhysicsNFP & \cellcolor[HTML]{FFFFE0}\textbf{Ours} \\
\hline

\multirow{3}{*}{\rotatebox{90}{\parbox{1cm}{\centering \textit{3600}}}}%
& \cellcolor{green!3} $\mathrm{MSE}$ & \cellcolor{green!3} $5.356 \times 10^{-3}$ & \cellcolor{green!3} $7.140 \times 10^{-3}$ & \cellcolor[HTML]{FFFFE0}\bm{$4.551 \times 10^{-3}$}$_{\textcolor{ForestGreen}{\uparrow 15.02\%}}$ \\
& \cellcolor{pink!10} $\mathrm{KE}$ & \cellcolor{pink!3} $1.246 \times 10^{-3}$ & \cellcolor{pink!3} $1.652 \times 10^{-3}$ & \cellcolor[HTML]{FFFFE0}\bm{$1.082 \times 10^{-3}$}$_{\textcolor{ForestGreen}{\uparrow 13.20\%}}$ \\
& \cellcolor{blue!3} $\mathrm{Sinkhorn}$ & \cellcolor{blue!3} $3.767 \times 10^{-5}$ & \cellcolor{blue!3} $3.323 \times 10^{-5}$ & \cellcolor[HTML]{FFFFE0}\bm{$1.422 \times 10^{-5}$}$_{\textcolor{ForestGreen}{\uparrow 57.22\%}}$ \\

\hline

\multirow{3}{*}{\rotatebox{90}{\parbox{1cm}{\centering \textit{4489}}}}%
& \cellcolor{green!3} $\mathrm{MSE}$ & \cellcolor{green!3} $6.396 \times 10^{-3}$ & \cellcolor{green!3} $1.623 \times 10^{-2}$ & \cellcolor[HTML]{FFFFE0}\bm{$4.572 \times 10^{-3}$}$_{\textcolor{ForestGreen}{\uparrow 28.52\%}}$ \\
& \cellcolor{pink!10} $\mathrm{KE}$ & \cellcolor{pink!3} $1.530 \times 10^{-3}$ & \cellcolor{pink!3} $4.895 \times 10^{-3}$ & \cellcolor[HTML]{FFFFE0}\bm{$1.087 \times 10^{-3}$}$_{\textcolor{ForestGreen}{\uparrow 28.97\%}}$ \\
& \cellcolor{blue!3} $\mathrm{Sinkhorn}$ & \cellcolor{blue!3} $2.229 \times 10^{-5}$ & \cellcolor{blue!3} $1.397 \times 10^{-3}$ & \cellcolor[HTML]{FFFFE0}\bm{$1.171 \times 10^{-5}$}$_{\textcolor{ForestGreen}{\uparrow 47.47\%}}$ \\

\hline

\multirow{3}{*}{\rotatebox{90}{\parbox{1cm}{\centering \textit{10000}}}}%
& \cellcolor{green!3} $\mathrm{MSE}$ & \cellcolor{green!3} $6.798 \times 10^{-3}$ & \cellcolor{green!3} $1.796 \times 10^{-2}$ & \cellcolor[HTML]{FFFFE0}\bm{$5.999 \times 10^{-3}$}$_{\textcolor{ForestGreen}{\uparrow 11.75\%}}$ \\
& \cellcolor{pink!10} $\mathrm{KE}$ & \cellcolor{pink!3} $1.697 \times 10^{-3}$ & \cellcolor{pink!3} $6.252 \times 10^{-3}$ & \cellcolor[HTML]{FFFFE0}\bm{$1.468 \times 10^{-3}$}$_{\textcolor{ForestGreen}{\uparrow 13.47\%}}$ \\
& \cellcolor{blue!3} $\mathrm{Sinkhorn}$ & \cellcolor{blue!3} $2.444 \times 10^{-5}$ & \cellcolor{blue!3} $1.540 \times 10^{-3}$ & \cellcolor[HTML]{FFFFE0}\bm{$9.640 \times 10^{-6}$}$_{\textcolor{ForestGreen}{\uparrow 60.55\%}}$ \\

\hline

\multirow{3}{*}{\rotatebox{90}{\parbox{1cm}{\centering \textit{40000}}}}%
& \cellcolor{green!3} $\mathrm{MSE}$ & \cellcolor{green!3} $6.923 \times 10^{-3}$ & \cellcolor{green!3} $1.669 \times 10^{-2}$ & \cellcolor[HTML]{FFFFE0}\bm{$5.063 \times 10^{-3}$}$_{\textcolor{ForestGreen}{\uparrow 26.87\%}}$ \\
& \cellcolor{pink!10} $\mathrm{KE}$ & \cellcolor{pink!3} $1.403 \times 10^{-3}$ & \cellcolor{pink!3} $6.631 \times 10^{-3}$ & \cellcolor[HTML]{FFFFE0}\bm{$1.205 \times 10^{-3}$}$_{\textcolor{ForestGreen}{\uparrow 14.10\%}}$ \\
& \cellcolor{blue!3} $\mathrm{Sinkhorn}$ & \cellcolor{blue!3} $5.086 \times 10^{-6}$ & \cellcolor{blue!3} $2.183 \times 10^{-3}$ & \cellcolor[HTML]{FFFFE0}\bm{$4.295 \times 10^{-6}$}$_{\textcolor{ForestGreen}{\uparrow 15.55\%}}$ \\

\hline

\end{tabular}}}%
\caption{Comparison of MSE, Kinetic Energy (KE) error, and Sinkhorn divergence on \textbf{higher resolution simulations}. The relative performance gain against the strongest baseline is computed as $\frac{\text{Baseline} - \text{Ours}}{\text{Baseline}} \times 100\%$.}
\label{tab:scalability_results}
\end{wraptable}

\subsection{Main Results}
\label{sec:main_results}
Table~\ref{tab:main_results} summarizes the quantitative results for the 2D experiments, where our proposed model demonstrates consistently superior predictive accuracy. In particular, our method significantly outperforms NeuralMPM, validating the efficacy of our architectural design choices. The most substantial gains are observed in the Sinkhorn divergence metric, with improvements of $57.22\%$ and $46.68\%$ on the TGV and LDC datasets, respectively. This indicates that our model is far more effective at capturing the underlying distributional geometry and global structure of the data than the baseline methods, rather than merely fitting local features. This success is driven by the downsampler’s ability to prioritize high-importance areas, while the Eulerian cross-attention ensures the predictions remain grounded in a consistent global structure.

\textbf{3D Datasets}
To further assess the effectiveness of our approach, we compare the proposed model against the two strongest baselines, GraphUnet and PhysicsNFP, on challenging 3D datasets. The results, summarized in Table~\ref{tab:3d_results}, demonstrate that our model consistently surpasses both baselines across evaluation metrics. This indicates that the advantages of our design generalize beyond 2D settings and remain robust even in the more complex 3D scenarios.

\subsection{Higher Resolution and Scalability Simulation}
\label{sec:scalability}
High-resolution simulations are essential for providing a detailed and accurate description of fluid flow, enabling more precise queries of velocity and pressure at arbitrary locations. However, they introduce two compounding challenges: the inherent difficulty of modeling complex micro-scale details and the computational burden of processing massive particle counts. Our approach is specifically designed to address these distinct hurdles. To manage the large particle counts, our downsampler condenses indistinguishable particles, effectively reducing computational overhead. Simultaneously, our hierarchical framework allows for the separate modeling of micro-scale details and coarse-level large structures. We validate this capability using the \textit{Taylor Green Vortex}, varying particle spacing from 0.02 to 0.005, resulting in 3,600 to 40,000 particles. This experiment serves as a scalability test for both the model’s scalability and its ability to resolve intricate flow features. We refer the readers to Appendix~\ref{appendix:resolution_data} for dataset details.

The results are summarized in Table~\ref{tab:scalability_results}, where we compare our proposed model against the two strongest baselines, GraphUnet and PhysicsNFP, introduced in Section~\ref{sec:main_results}. Our model consistently outperforms both baselines. In particular, PhysicsNFP shows a noticeable drop in performance as the resolution increases, indicating that hierarchical architectures such as GraphUnet and our model are better suited for simulations with larger particle counts.

\begin{wraptable}{r}{0.5\textwidth}
\centering
\scriptsize{%
\resizebox{\linewidth}{!}{%
    \setlength\tabcolsep{3pt}%
    \renewcommand\arraystretch{1.4}%
\begin{tabular}{l || c c|| c}
\hline\thickhline
\rowcolor{CadetBlue!20}
Metric & GraphUnet & PhysicsNFP & \cellcolor[HTML]{FFFFE0}\textbf{Ours} \\
\hline

\cellcolor{green!3} $\mathrm{MSE}$ & \cellcolor{green!3} $5.572 \times 10^{-5}$ & \cellcolor{green!3} $1.150 \times 10^{-2}$ & \cellcolor[HTML]{FFFFE0}\bm{$4.955 \times 10^{-5}$}$_{\textcolor{ForestGreen}{\uparrow 11.07\%}}$ \\
\cellcolor{pink!10} $\mathrm{KE}$ & \cellcolor{pink!3} $1.207 \times 10^{-7}$ & \cellcolor{pink!3} $9.560 \times 10^{-3}$ & \cellcolor[HTML]{FFFFE0}\bm{$9.980 \times 10^{-8}$}$_{\textcolor{ForestGreen}{\uparrow 17.31\%}}$ \\
\cellcolor{blue!3} $\mathrm{Sinkhorn}$ & \cellcolor{blue!3} $1.949 \times 10^{-6}$ & \cellcolor{blue!3} $2.130 \times 10^{-2}$ & \cellcolor[HTML]{FFFFE0}\bm{$1.682 \times 10^{-6}$}$_{\textcolor{ForestGreen}{\uparrow 13.70\%}}$ \\

\hline

\end{tabular}}}%
\caption{Comparison of MSE, Kinetic Energy (KE) error, and Sinkhorn
divergence on \textbf{fluid-solid interaction over an irregular domain}. The relative performance gain against the strongest baseline is computed as $\frac{\text{Baseline} - \text{Ours}}{\text{Baseline}} \times 100\%$.}
\label{tab:fsi}
\end{wraptable}

\subsection{Fluid-Solid Interaction Simulation}
\label{sec:cyl}
The voxelization strategy adopted in \citet{rochman-sharabi2025a} is restricted to fixed, rectangular grids, which prohibits generalization to irregular domains that frequently arise in fluid–solid interaction problems. We visualize the data in Figure~\ref{fig:elliptical}. To evaluate our model’s ability to handle such scenarios, we conduct experiments on the \textit{Flow Around Cylinder} task within an irregular domain, comparing against the two strongest baselines, GraphUnet and PhysicsNFP. We refer the readers to Appendix~\ref{appendix:fsi_data} for dataset details. The results, reported in Table~\ref{tab:fsi}, demonstrate that our proposed model consistently outperforms both baselines. These findings highlight not only the effectiveness of our design but also its broader applicability to complex geometries where voxelization-based approaches face fundamental limitations.

\begin{wraptable}{r}{0.5\textwidth}
\centering
\scriptsize{%
\resizebox{\linewidth}{!}{%
    \setlength\tabcolsep{3pt}%
    \renewcommand\arraystretch{1.4}%
\begin{tabular}{l || c c|| c}
\hline\thickhline
\rowcolor{CadetBlue!20}
Metric & GraphUnet & PhysicsNFP & \cellcolor[HTML]{FFFFE0}\textbf{Ours} \\
\hline

\cellcolor{green!3} $\mathrm{MSE}$ & \cellcolor{green!3} $3.602 \times 10^{-5}$ & \cellcolor{green!3} $1.379 \times 10^{-4}$ & \cellcolor[HTML]{FFFFE0}\bm{$2.421 \times 10^{-5}$}$_{\textcolor{ForestGreen}{\uparrow 32.79\%}}$ \\
\cellcolor{pink!10} $\mathrm{KE}$ & \cellcolor{pink!3} $1.469 \times 10^{-5}$ & \cellcolor{pink!3} $7.203 \times 10^{-6}$ & \cellcolor[HTML]{FFFFE0}\bm{$4.229 \times 10^{-6}$}$_{\textcolor{ForestGreen}{\uparrow 41.29\%}}$ \\
\cellcolor{blue!3} $\mathrm{Sinkhorn}$ & \cellcolor{blue!3} $5.742 \times 10^{-5}$ & \cellcolor{blue!3} $1.646 \times 10^{-4}$ & \cellcolor[HTML]{FFFFE0}\bm{$3.107 \times 10^{-5}$}$_{\textcolor{ForestGreen}{\uparrow 45.89\%}}$ \\

\hline

\end{tabular}}}%
\caption{Comparison of MSE, Kinetic Energy (KE) error, and Sinkhorn
divergence on \textbf{deforming domain}. The relative performance gain against the strongest baseline is computed as $\frac{\text{Baseline} - \text{Ours}}{\text{Baseline}} \times 100\%$.}
\label{tab:elliptical}
\end{wraptable}

\subsection{Deformable Domain Simulation}
\label{sec:elliptical}
\begin{figure}[t]
    \centering
    \includegraphics[width=0.09\linewidth]{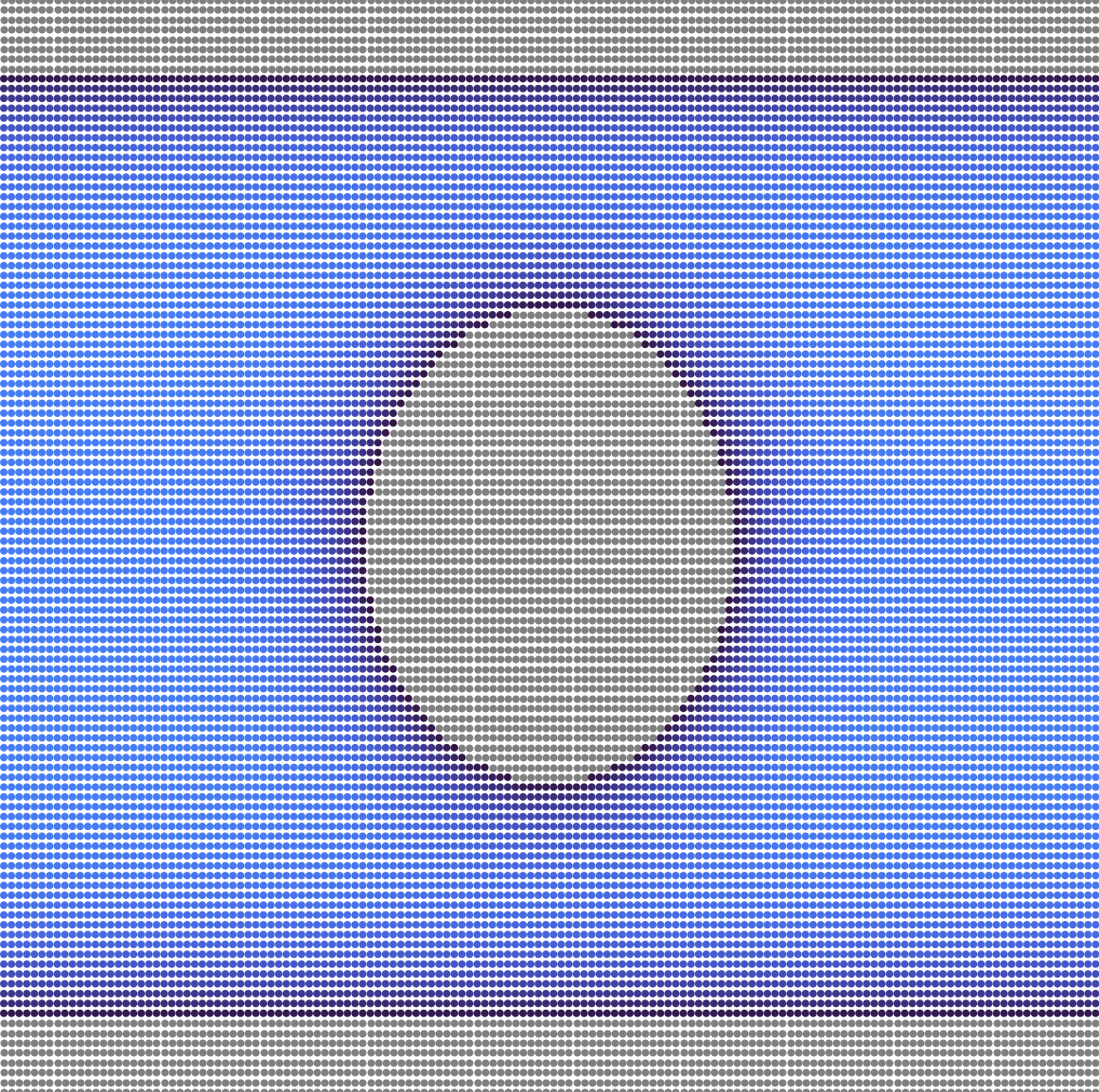}\hfill
    \includegraphics[width=0.09\linewidth]{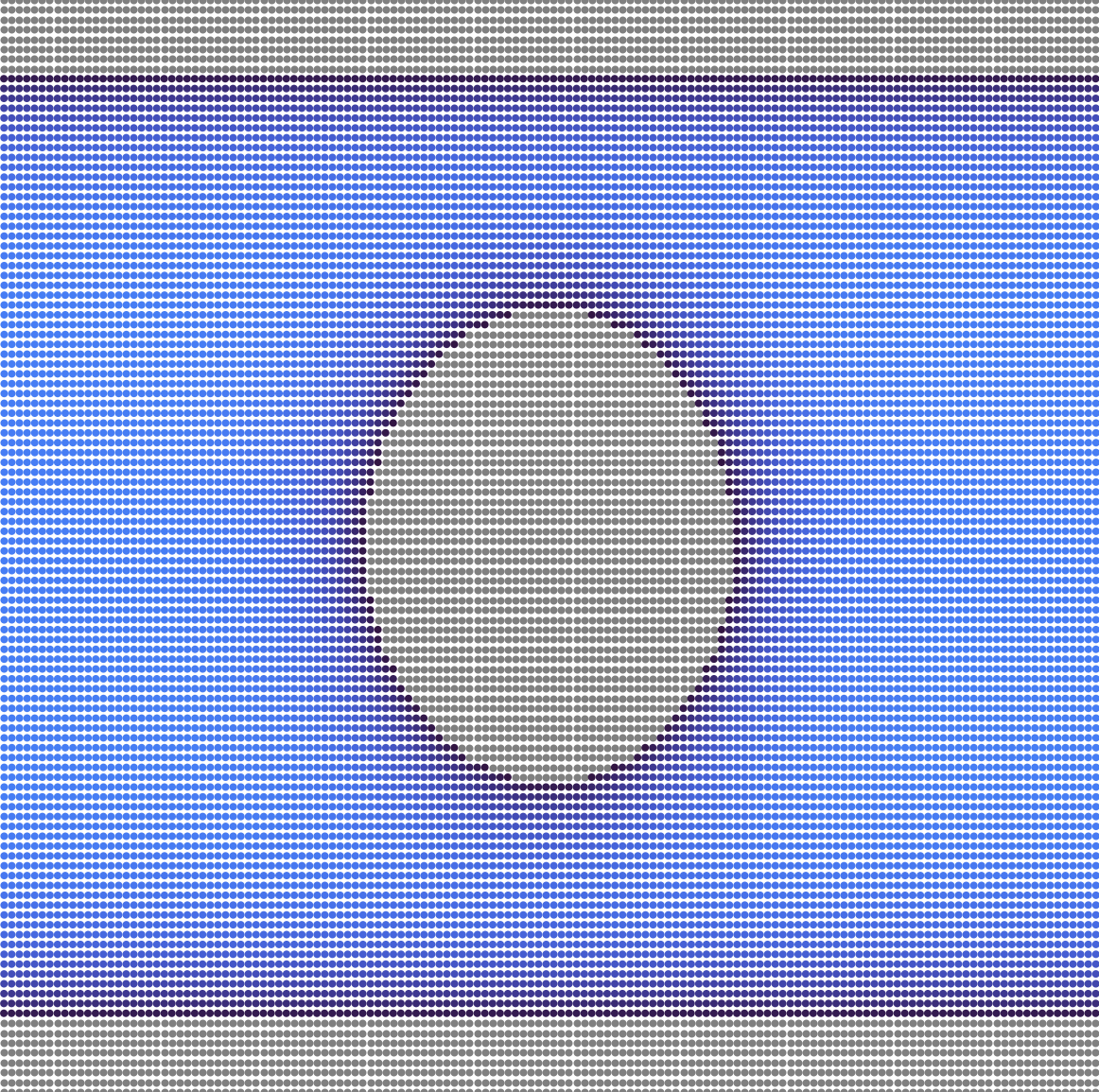}\hfill
    \includegraphics[width=0.09\linewidth]{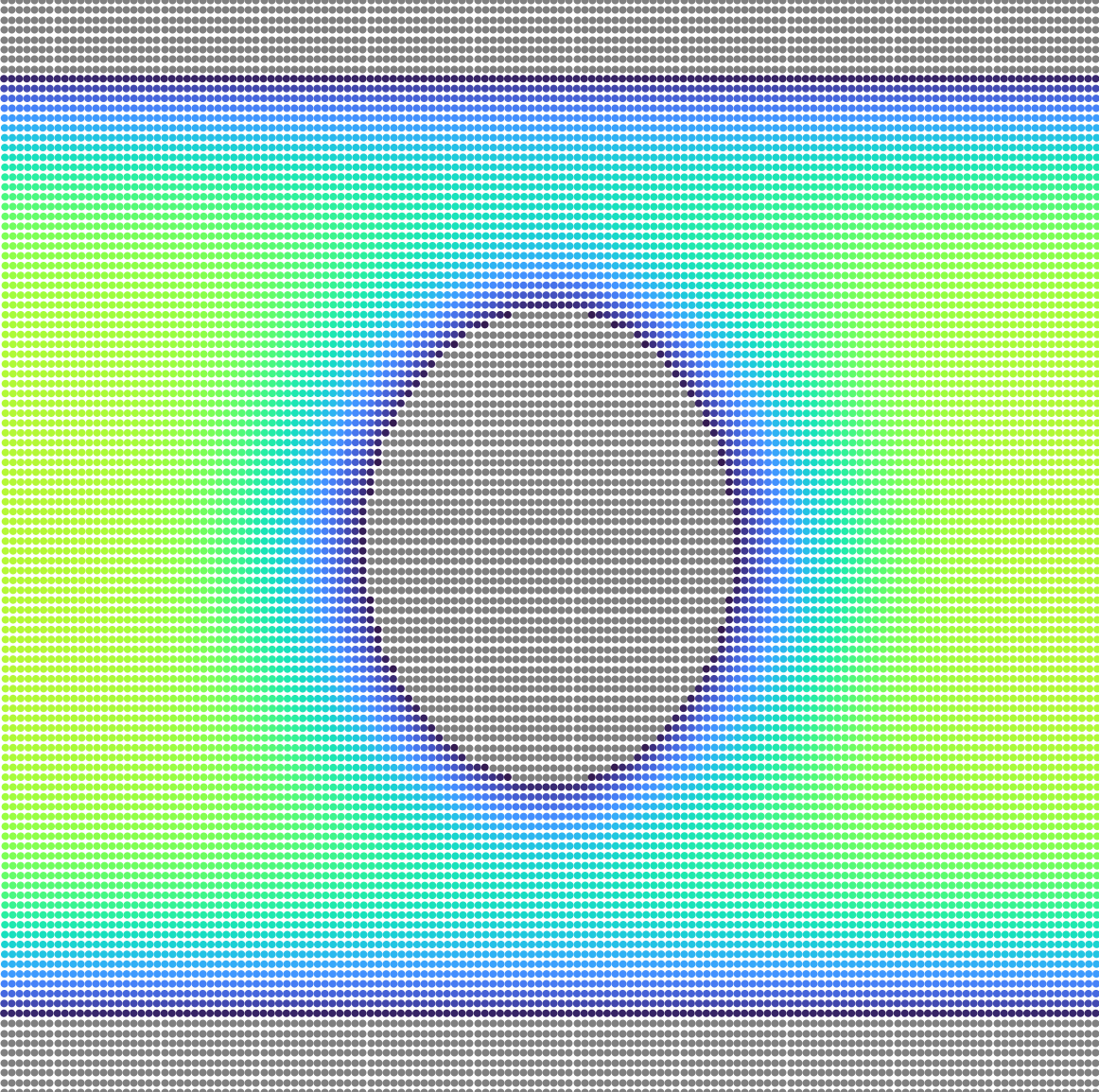}\hfill
    \includegraphics[width=0.09\linewidth]{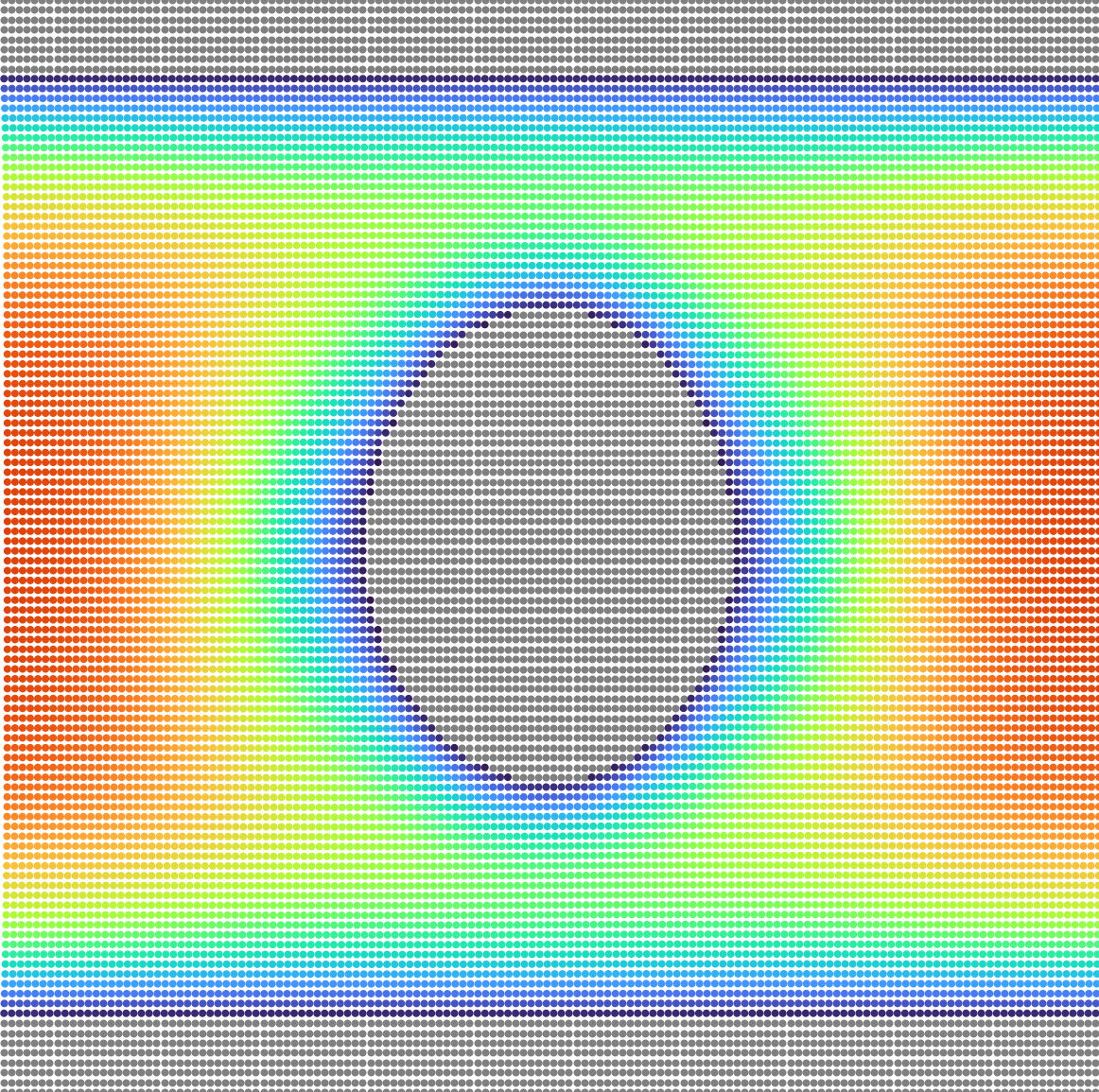}\hfill
    \includegraphics[width=0.09\linewidth]{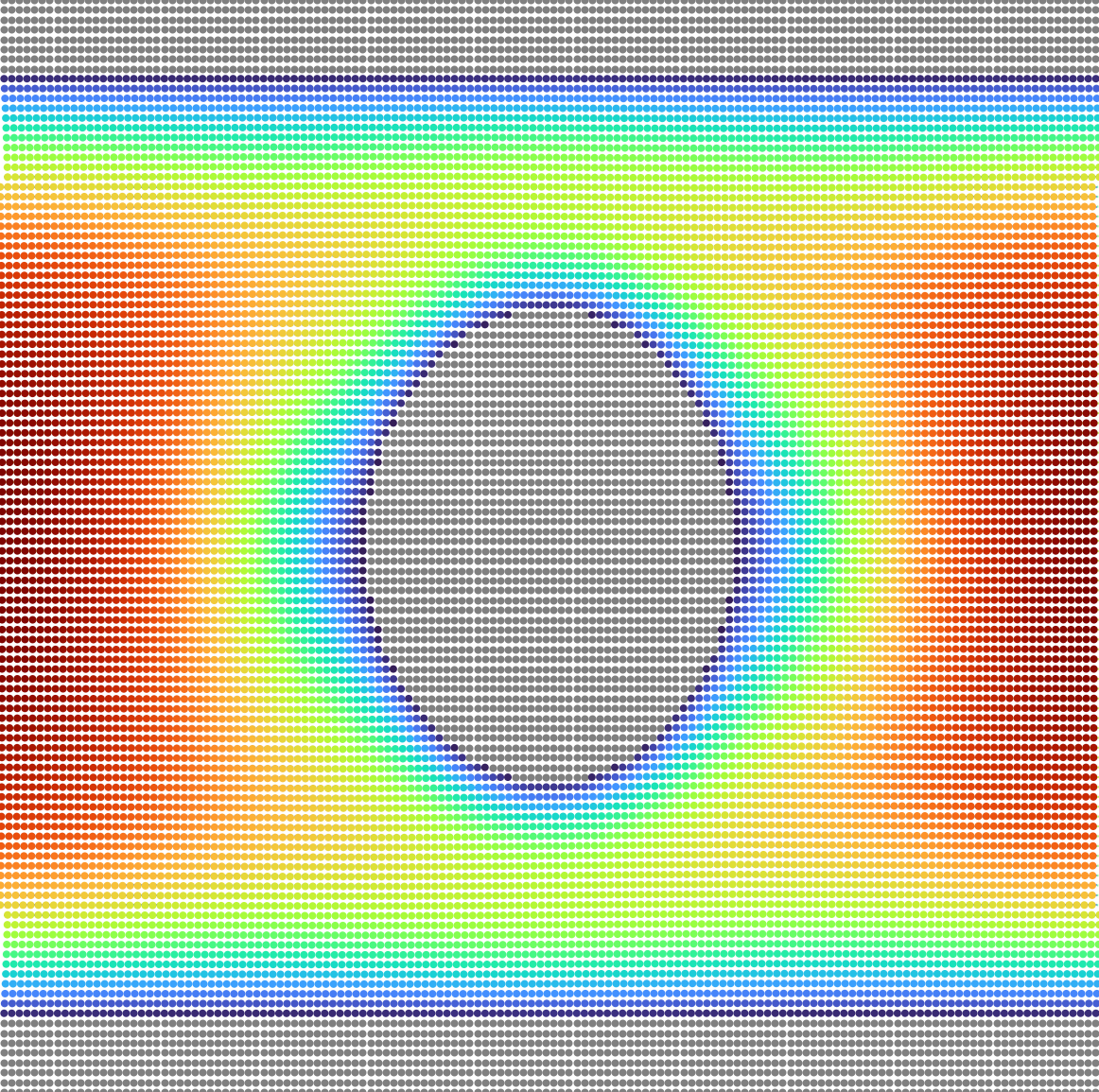}\hfill
    \includegraphics[width=0.09\linewidth]{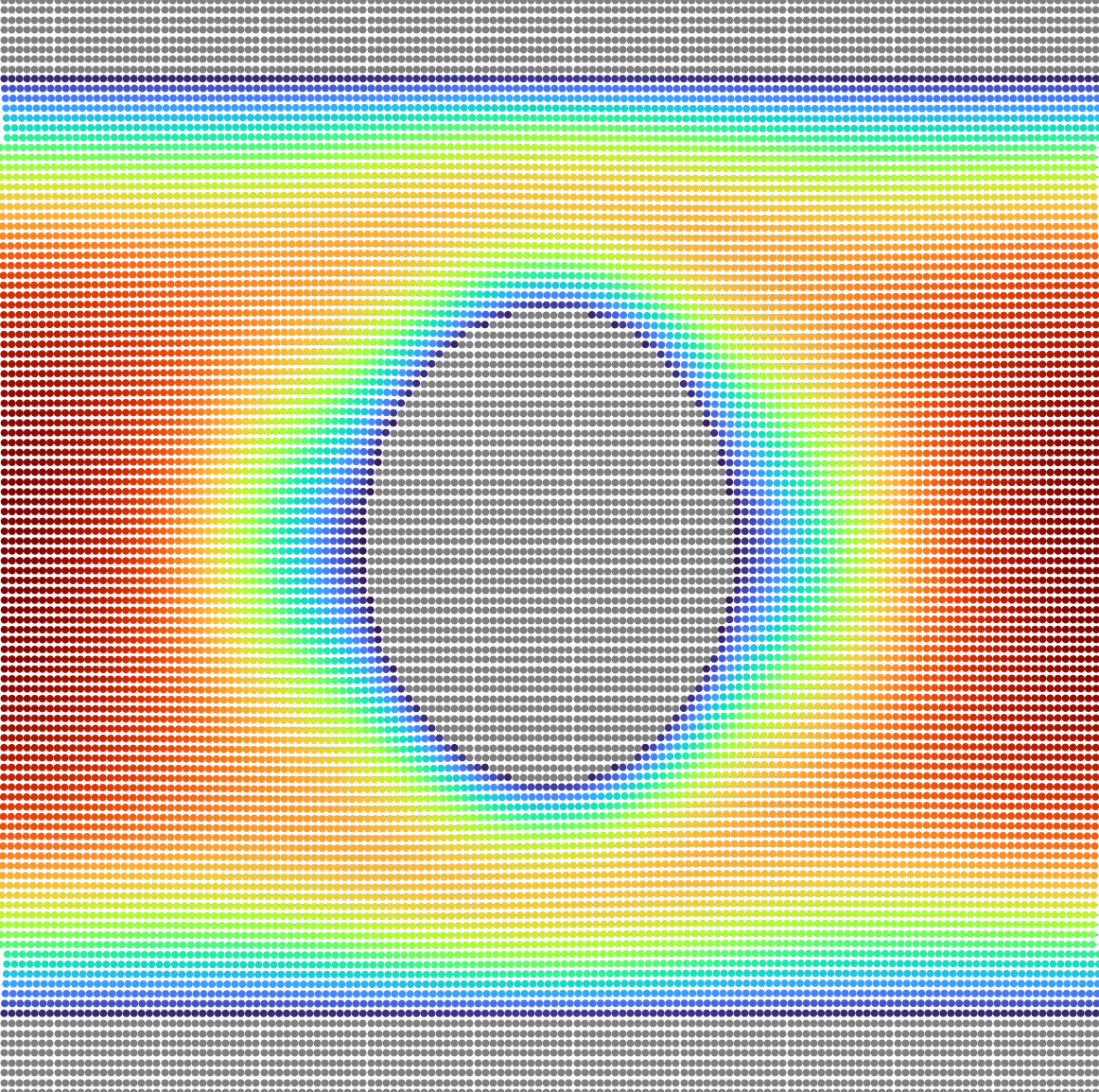}\hfill
    \includegraphics[width=0.09\linewidth]{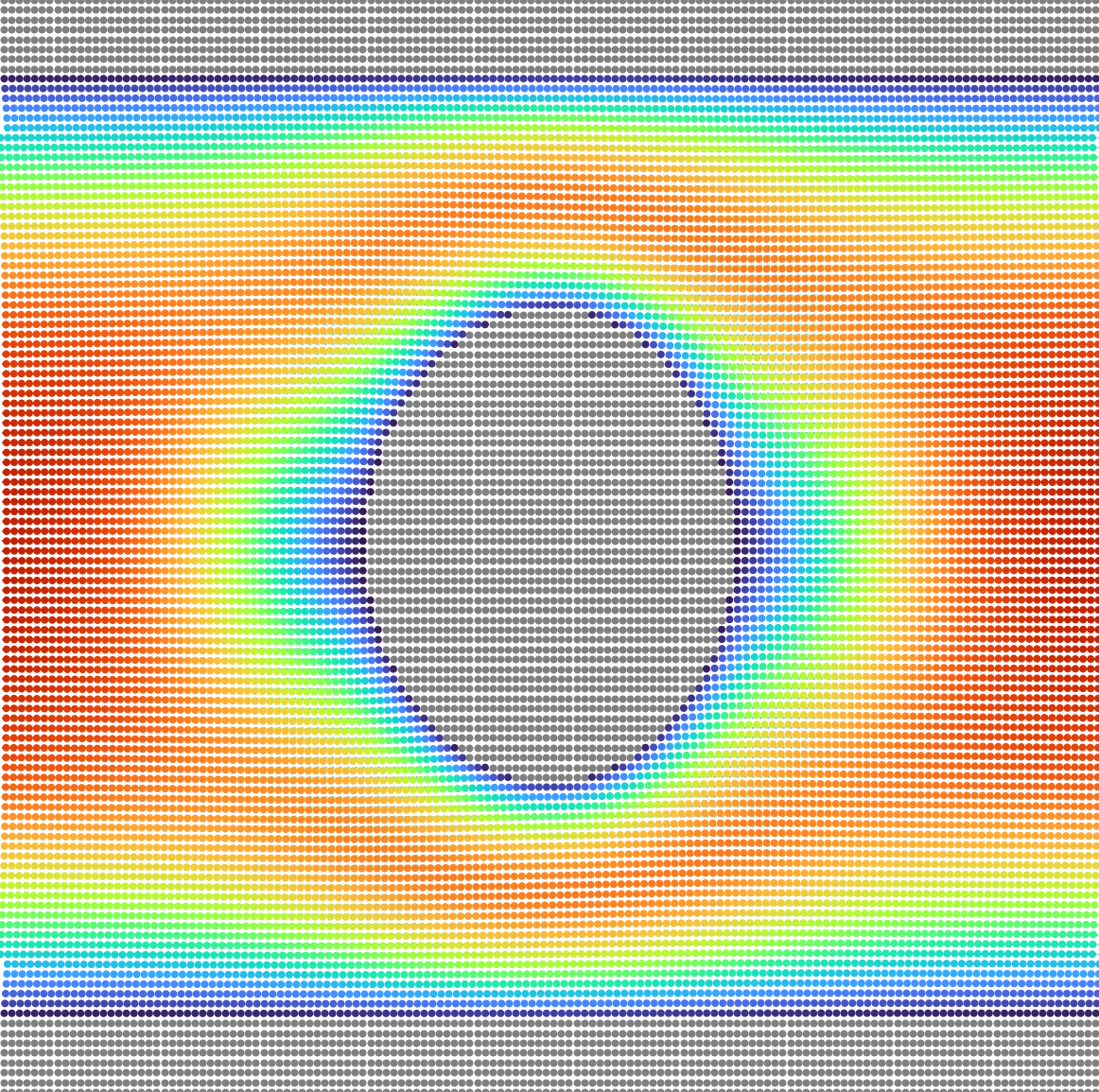}\hfill
    \includegraphics[width=0.09\linewidth]{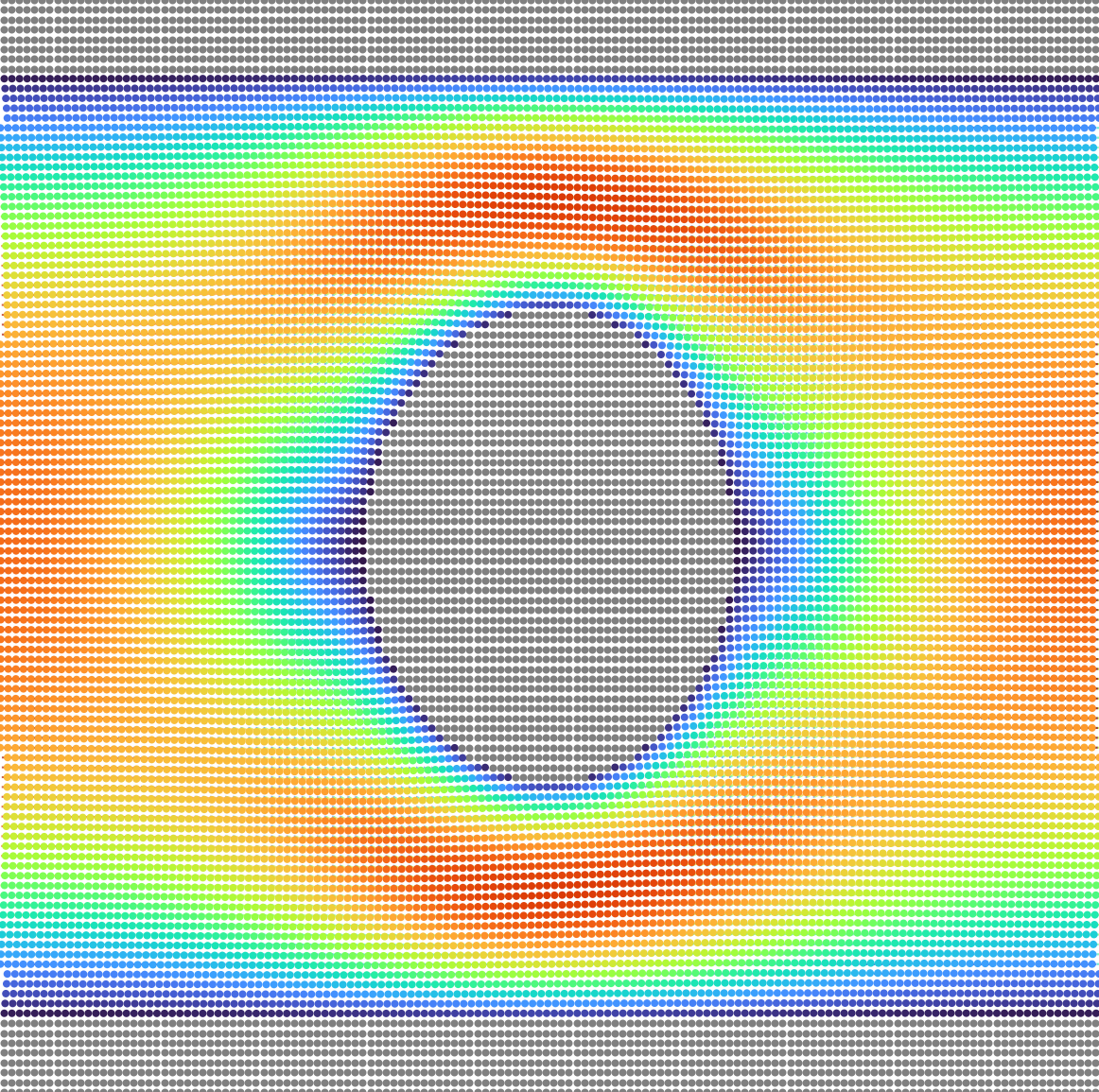}\hfill
    \includegraphics[width=0.09\linewidth]{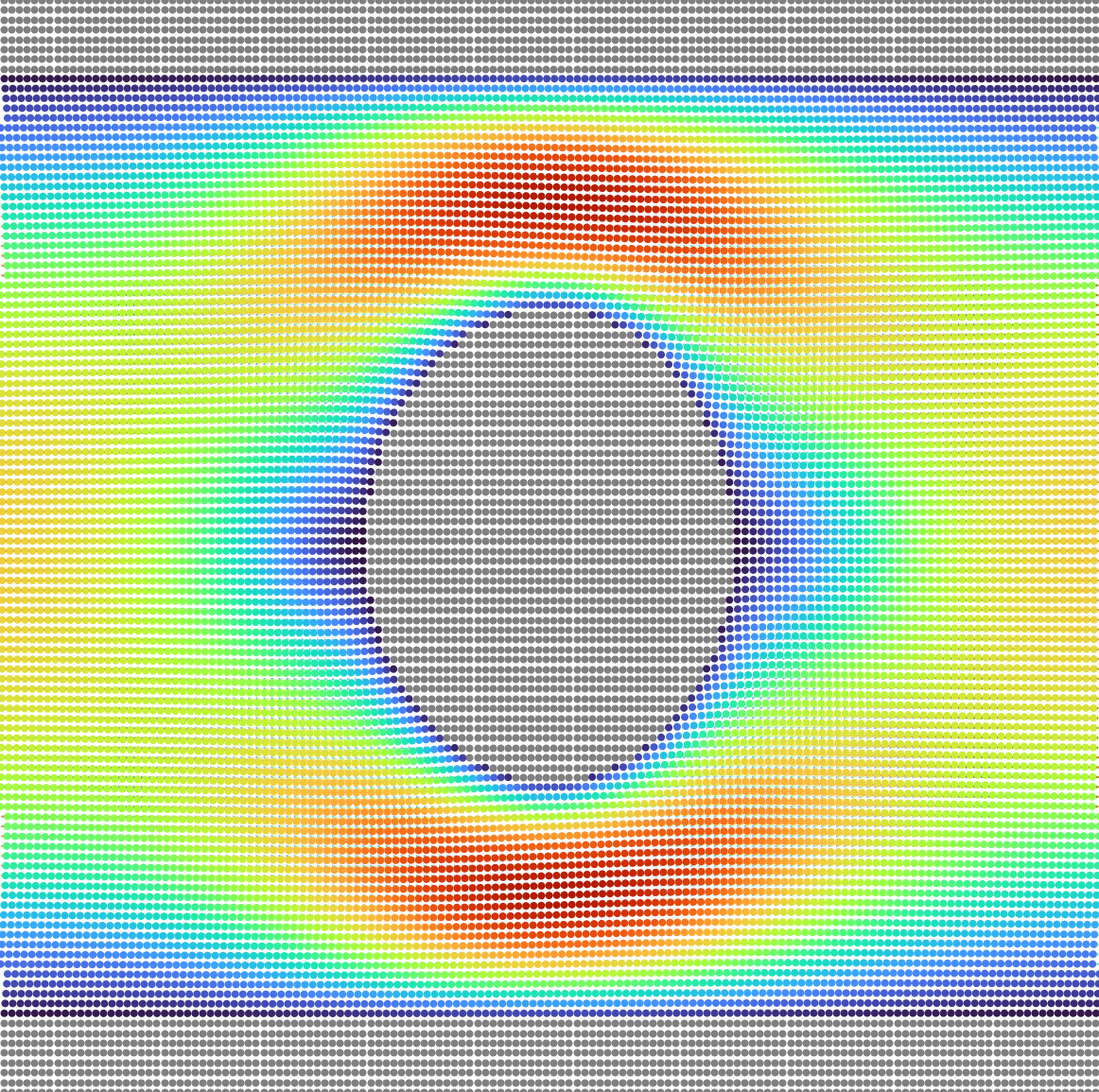}\hfill
    \includegraphics[width=0.09\linewidth]{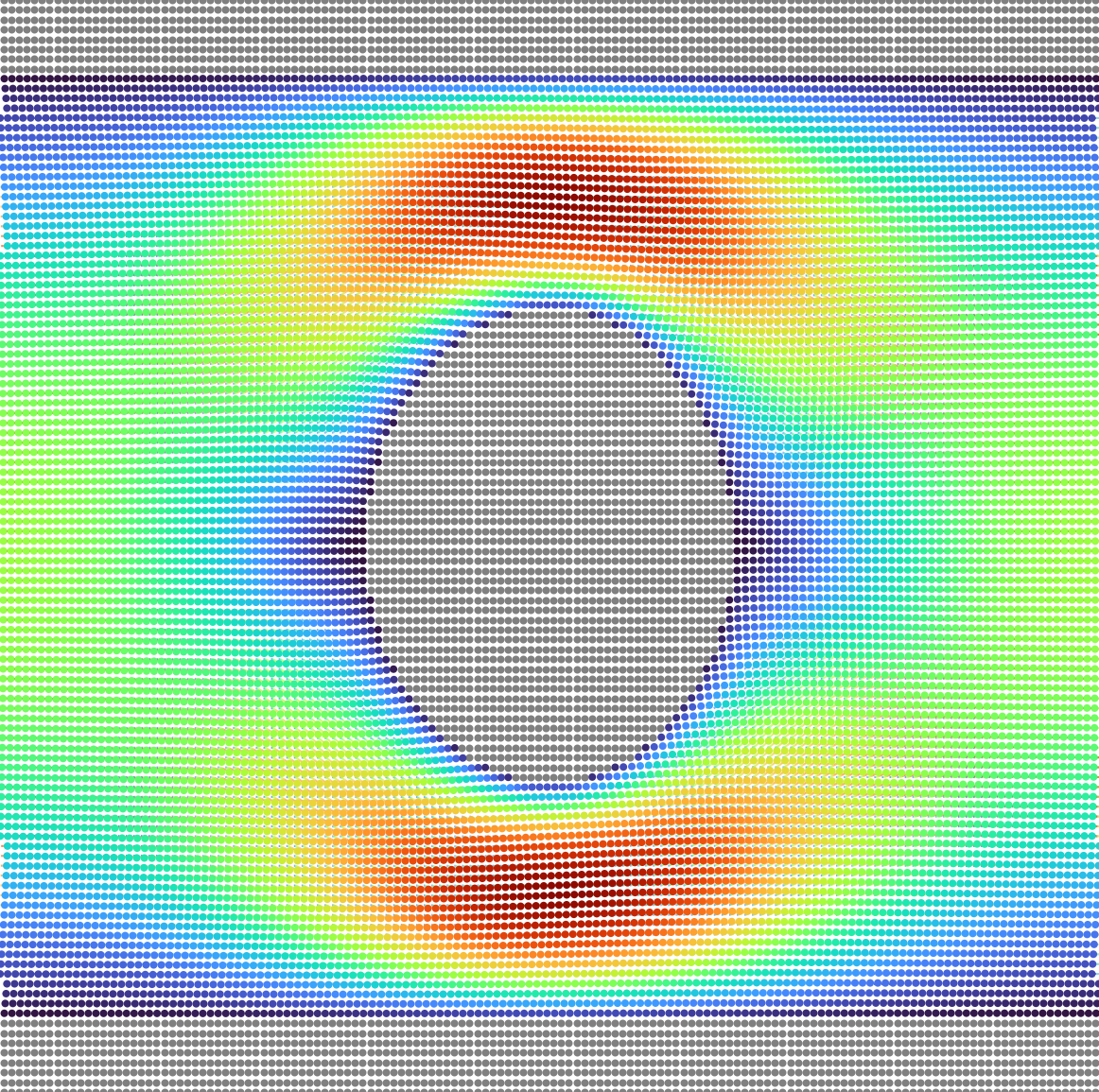}
    
    \vspace{0.5em}
    
    \includegraphics[width=0.09\linewidth]{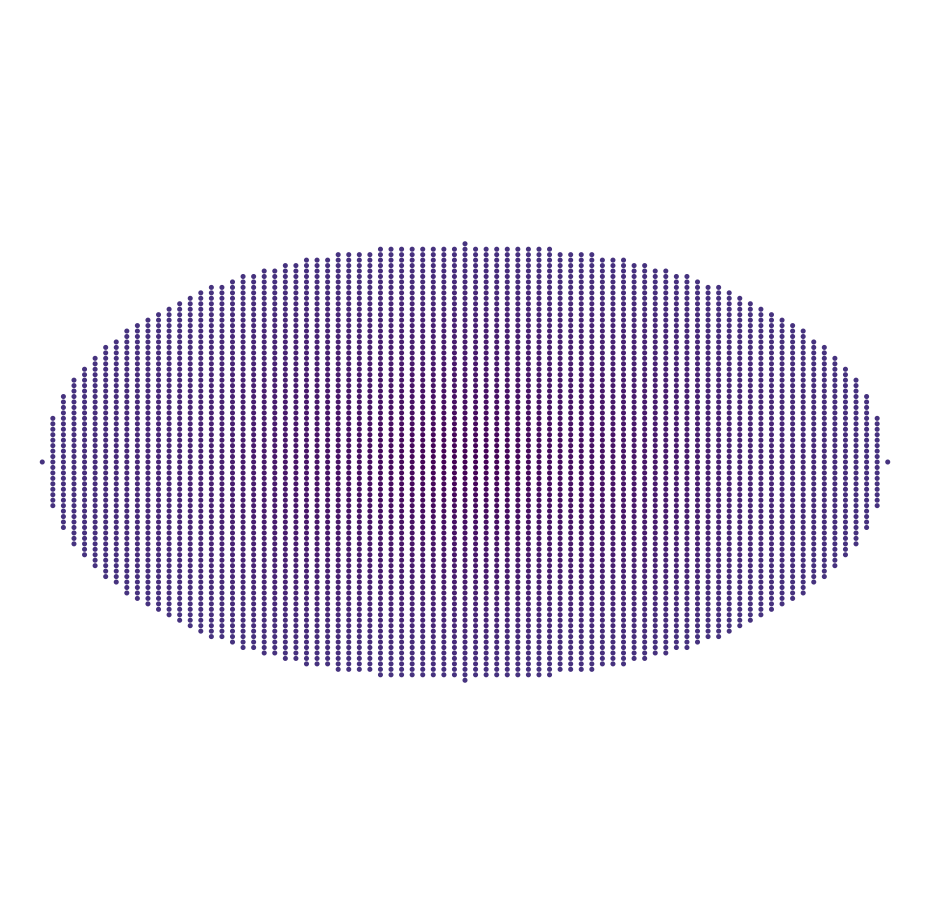}\hfill
    \includegraphics[width=0.09\linewidth]{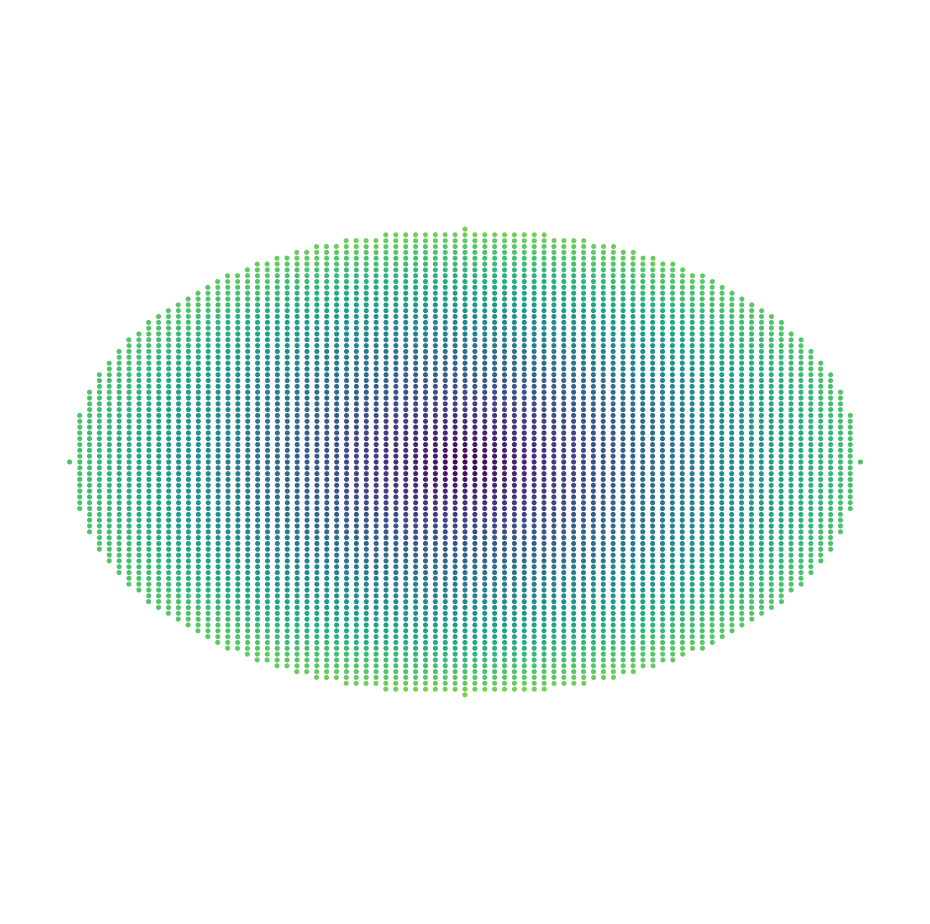}\hfill
    \includegraphics[width=0.09\linewidth]{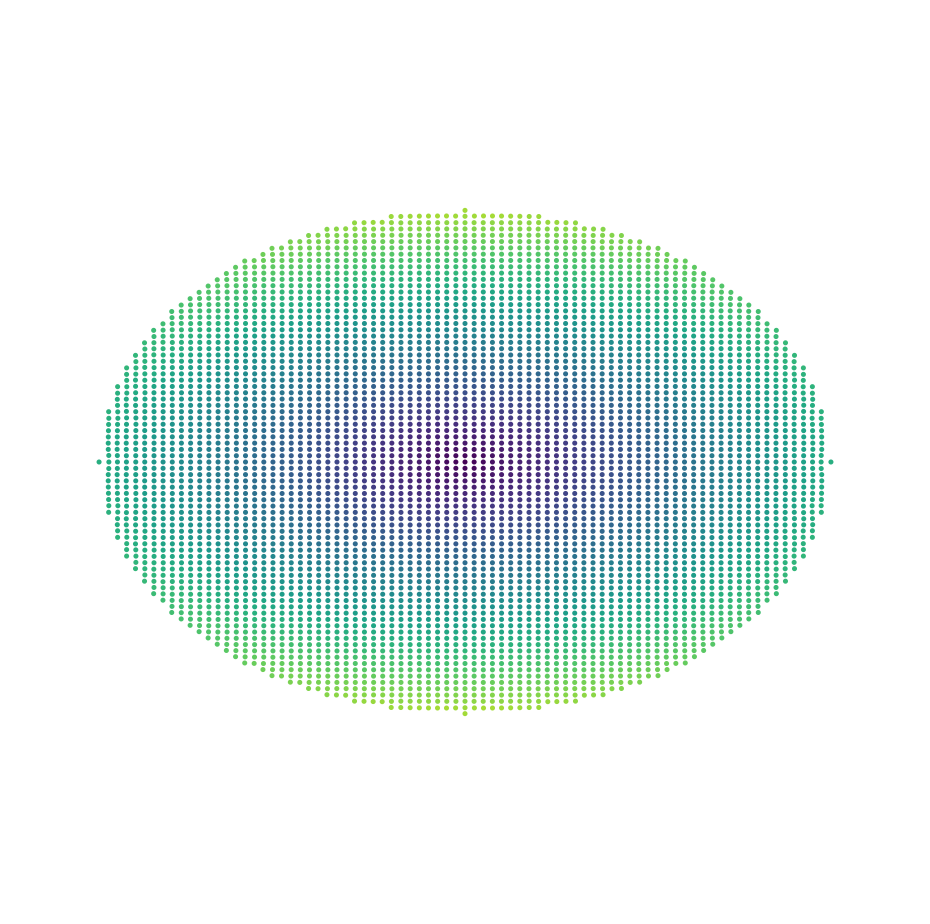}\hfill
    \includegraphics[width=0.09\linewidth]{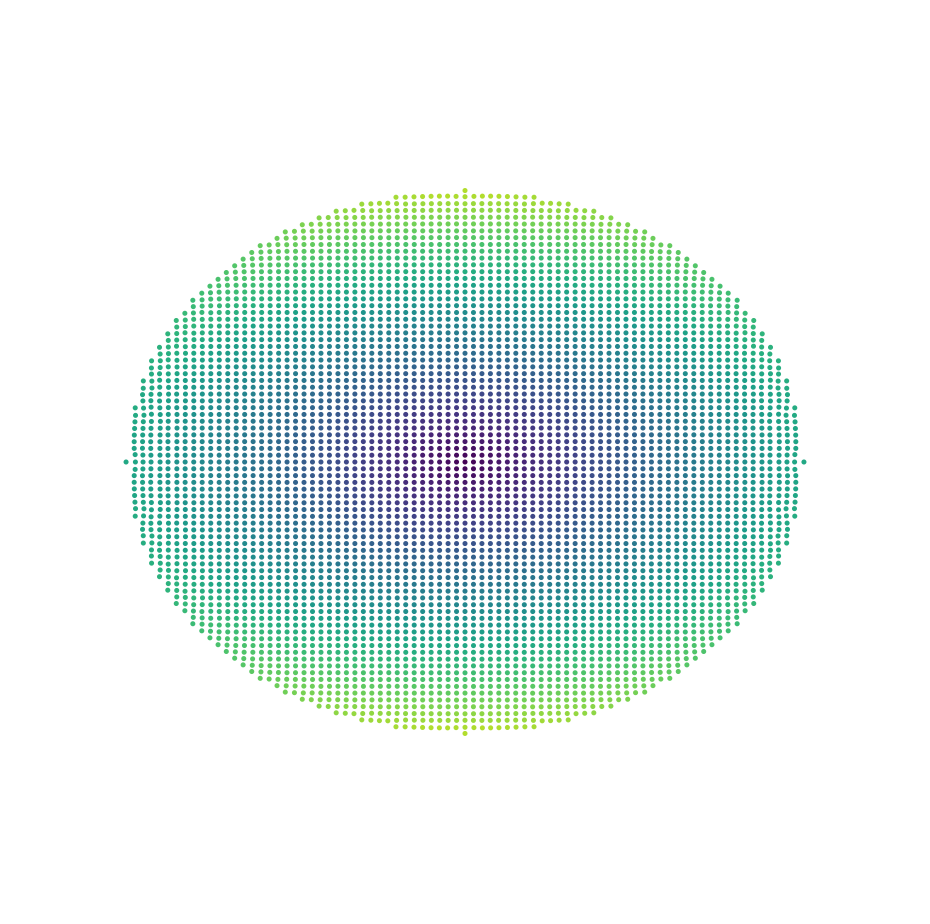}\hfill
    \includegraphics[width=0.09\linewidth]{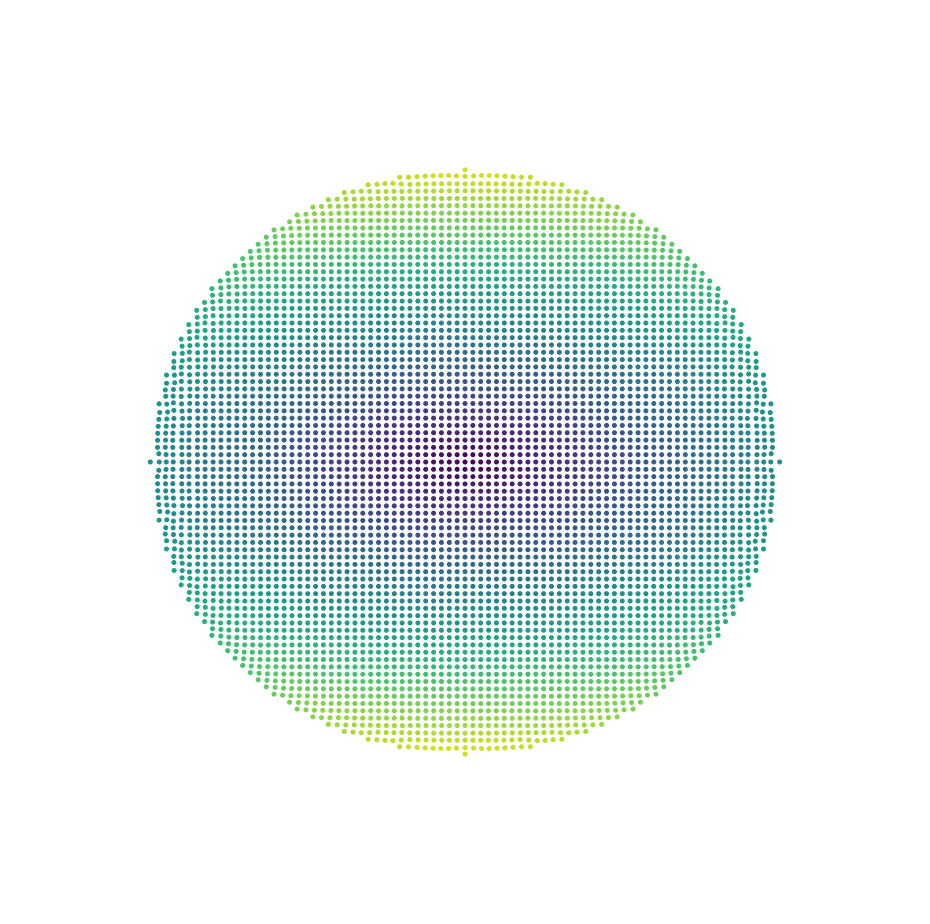}\hfill
    \includegraphics[width=0.09\linewidth]{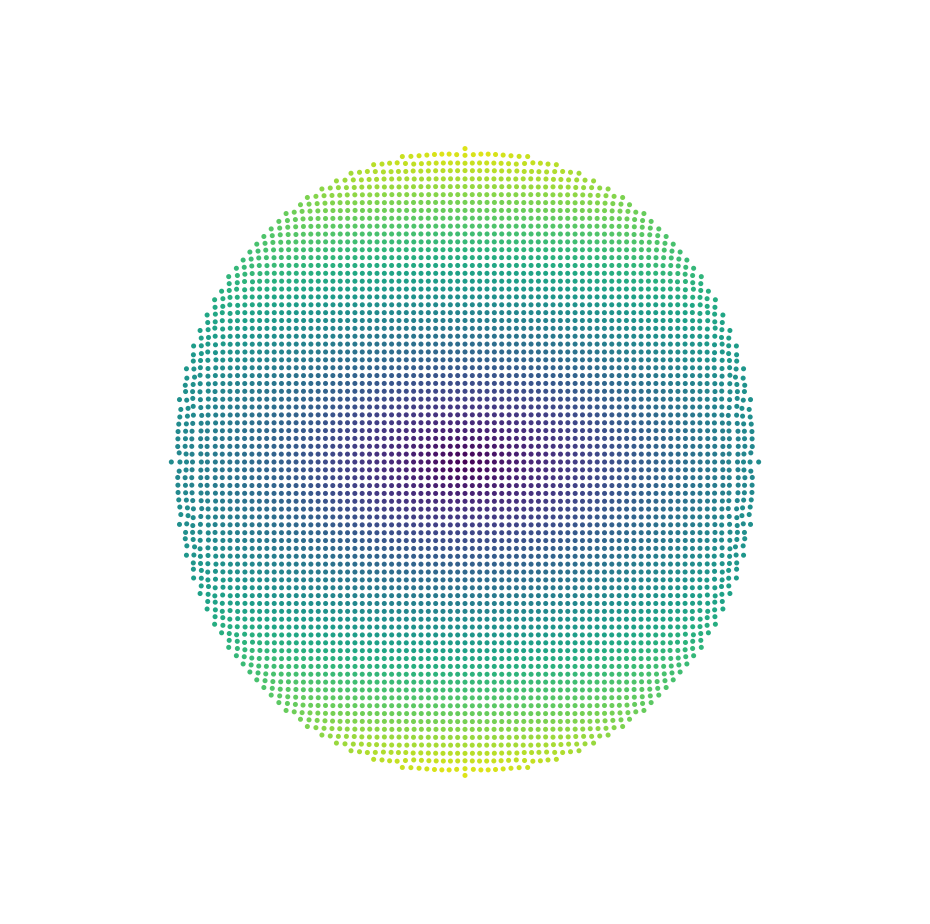}\hfill
    \includegraphics[width=0.09\linewidth]{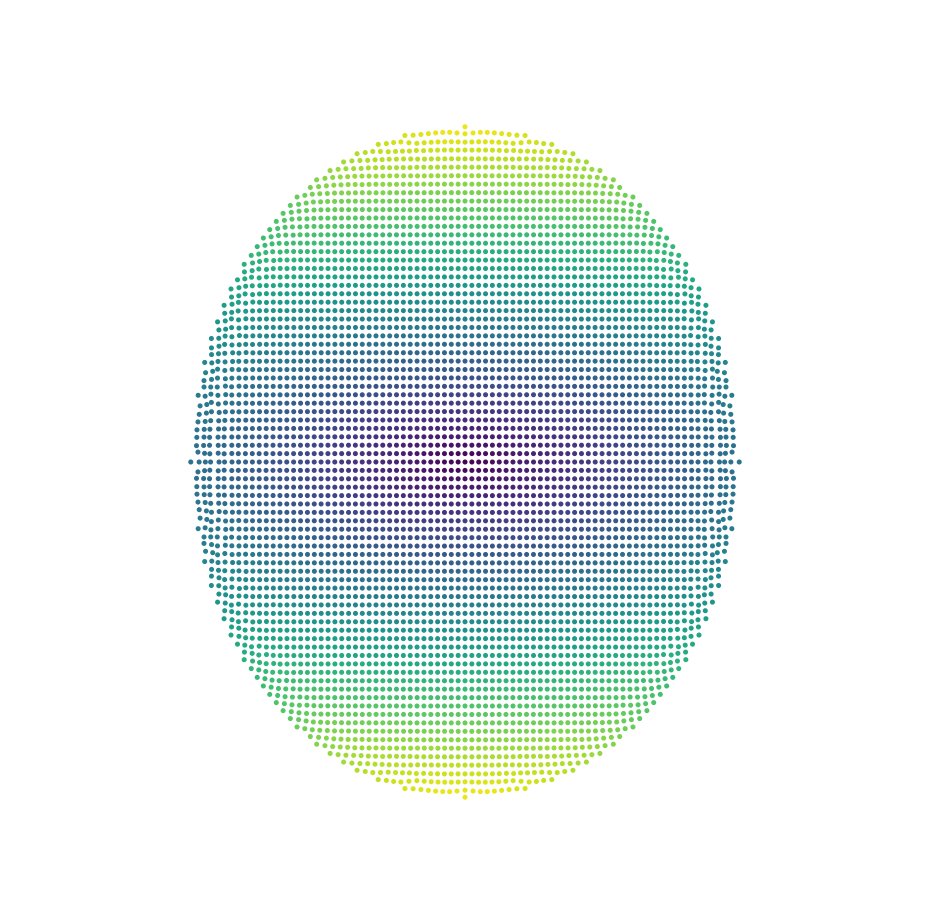}\hfill
    \includegraphics[width=0.09\linewidth]{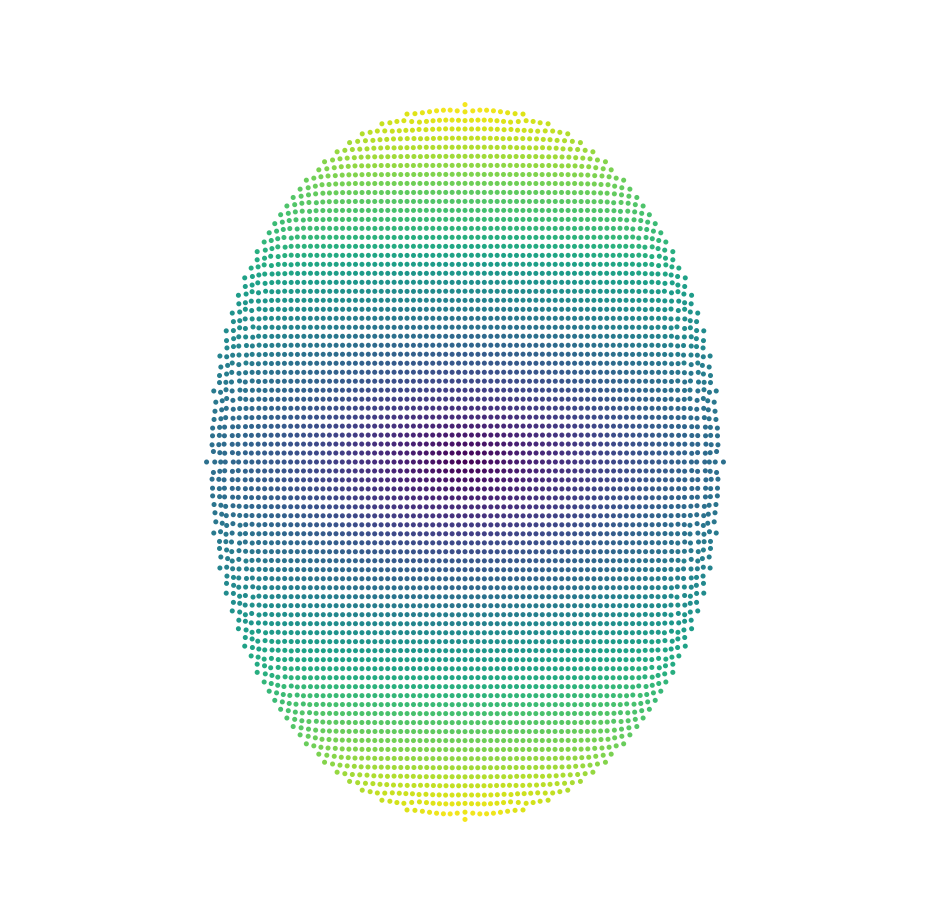}\hfill
    \includegraphics[width=0.09\linewidth]{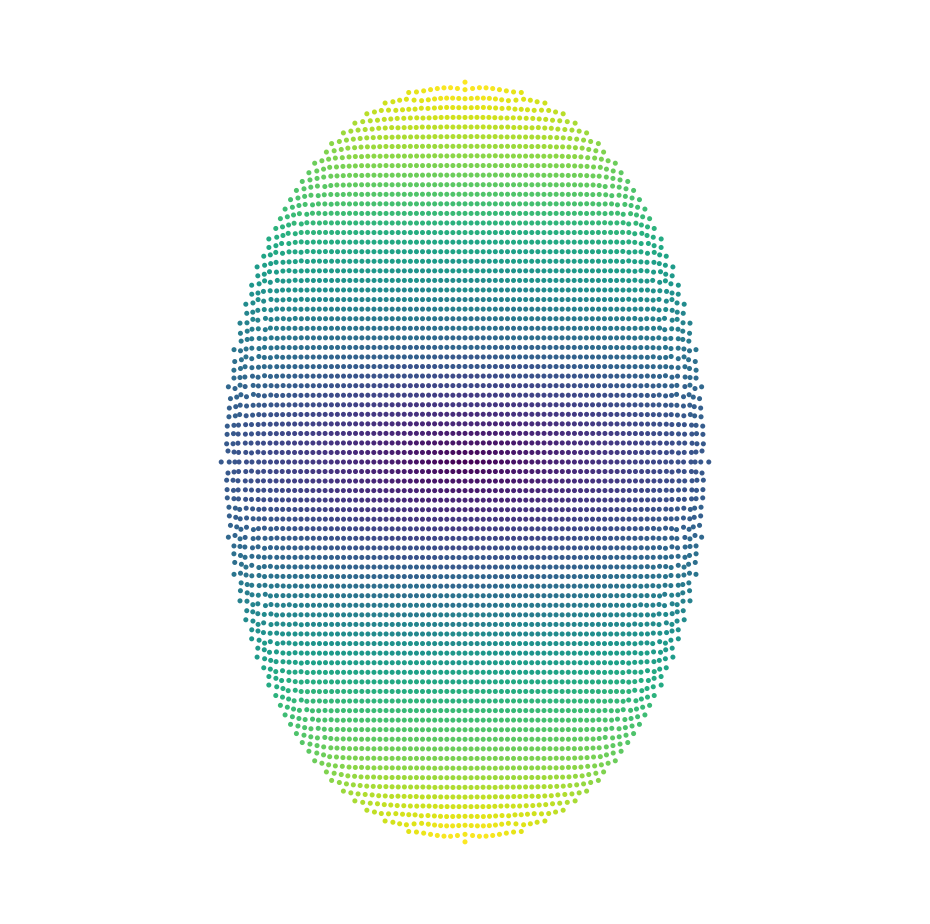}\hfill
    \includegraphics[width=0.09\linewidth]{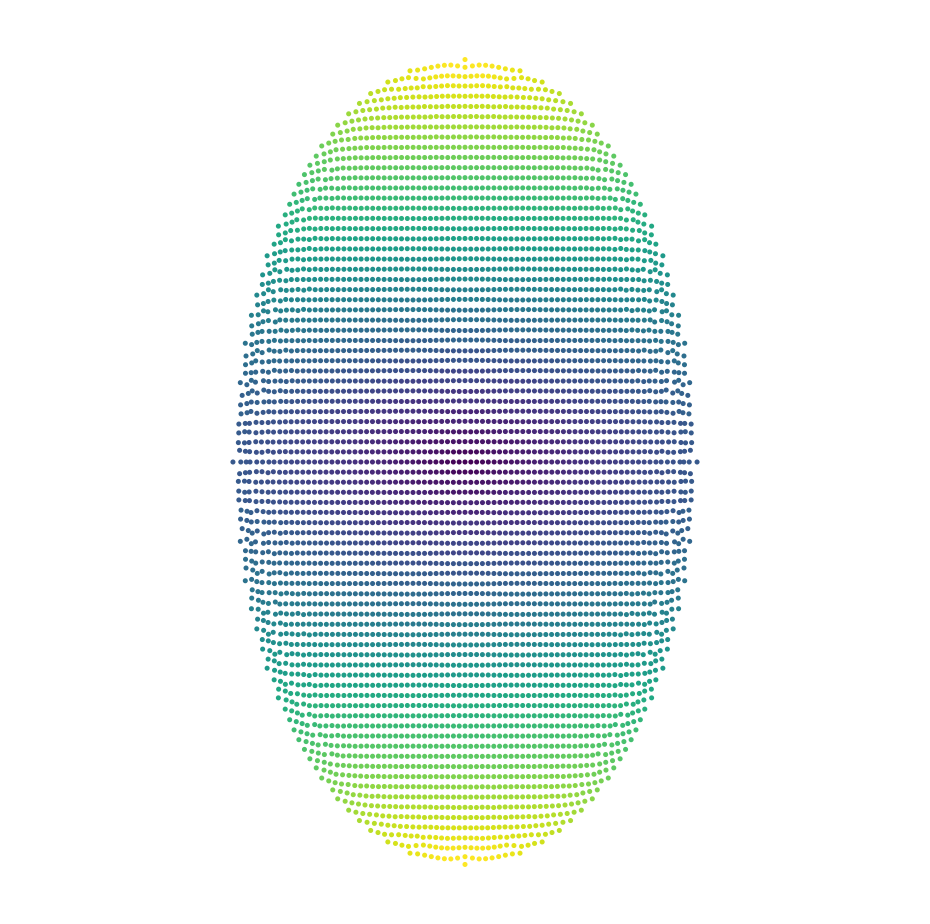}
    
    \caption{Visualization of datasets with irregular and deforming domains. \textbf{Top row}: Periodic flow around a fixed cylinder centered in the domain, creating an irregular flow field. \textbf{Bottom row}: A fluid droplet subjected to a velocity field, demonstrating continuous stretching and domain deformation.}
    \label{fig:elliptical}
\end{figure}

Lagrangian simulations excel at modeling deformations and free-surface boundaries. In addition to the dam break scenario evaluated in Section~\ref{sec:main_results}, we also test our model on a dynamically deforming domain. Specifically, we simulate a 2D circular drop of fluid that is subjected to a velocity field, stretching it into an ellipse~\citep{MONAGHAN1994399}. Because the fluid is incompressible, its total area remains strictly conserved during this deformation. We provide a visualization of this process in Figure~\ref{fig:elliptical}.

The experimental results are presented in Table~\ref{tab:elliptical}. As in previous experiments, we compare our approach against GraphUnet and PhysicsNFP. The results show that our proposed model achieves superior performance. This is expected, as our architectural design is inherently well-suited for continuously deforming domains.

\subsection{Extended Rollout}
We investigate our model’s performance on extended rollouts using the Water dataset~\citep{sanchezgonzalez2020learningsimulatecomplexphysics}. This dataset features a volume of water dropped into a box, involving complex splashing dynamics. We compare our model against GraphUnet, as PhysicsNFP is unable to handle the varying particle counts inherent in this dataset. We apply all auxiliary training strategies to both the baseline and our proposed models. These models predict higher-order temporal derivatives and utilize Euler integration to calculate subsequent positions. Additionally, we incorporate noise into the training data to enhance rollout stability. As shown in Table~\ref{tab:water}, our model maintains superior performance over extended horizons, surpassing the strongest baseline at every step. Notably, the kinetic energy error at the 200-step rollout is orders of magnitude smaller than that of the baseline. We attribute this to the cross-attention with Eulerian features, which helps the model correct particle drift.

\begin{table*}[h]
\centering
\scriptsize{%
\resizebox{\linewidth}{!}{%
    \setlength\tabcolsep{4pt}%
    \renewcommand\arraystretch{1.4}%
\begin{tabular}{l c || c c c c c c }
\hline\thickhline
\rowcolor{CadetBlue!20}
 & & \multicolumn{6}{c}{\textbf{Steps}} \\
\rowcolor{CadetBlue!20}
\multirow{-2}{*}{\textbf{Model}} & \multirow{-2}{*}{\textbf{Metric}} & \textbf{25} & \textbf{50} & \textbf{75} & \textbf{100} & \textbf{150} & \textbf{200} \\
\hline

\cellcolor{green!3} & \cellcolor{green!3} $\mathrm{MSE}$ & \cellcolor{green!3} $1.801 \times 10^{-4}$ & \cellcolor{green!3} $1.080 \times 10^{-3}$ & \cellcolor{green!3} $2.810 \times 10^{-3}$ & \cellcolor{green!3} $6.234 \times 10^{-3}$ & \cellcolor{green!3} $2.111 \times 10^{-2}$ & \cellcolor{green!3} $4.201 \times 10^{-2}$ \\
\cellcolor{pink!10} & \cellcolor{pink!10} $\mathrm{KE}$  & \cellcolor{pink!10} $3.252 \times 10^{-2}$ & \cellcolor{pink!10} $8.987 \times 10^{-1}$ & \cellcolor{pink!10} $6.726 \times 10^{0}$ & \cellcolor{pink!10} $3.114 \times 10^{1}$ & \cellcolor{pink!10} $3.652 \times 10^{2}$ & \cellcolor{pink!10} $1.545 \times 10^{3}$ \\
\cellcolor{blue!3}\multirow{-3}{*}{GraphUnet}  & \cellcolor{blue!3} $\mathrm{Sinkhorn}$ & \cellcolor{blue!3} $1.225 \times 10^{-4}$ & \cellcolor{blue!3} $4.905 \times 10^{-4}$ & \cellcolor{blue!3} $1.398 \times 10^{-3}$ & \cellcolor{blue!3} $3.966 \times 10^{-3}$ & \cellcolor{blue!3} $1.967 \times 10^{-2}$ & \cellcolor{blue!3} $3.235 \times 10^{-2}$ \\
\hline

\rowcolor[HTML]{FFFFE0}
  &   MSE &   \bm{$1.612 \times 10^{-5}$} &   \bm{$8.718 \times 10^{-5}$} &   \bm{$5.244 \times 10^{-5}$} &   \bm{$1.895 \times 10^{-3}$} &   \bm{$9.250 \times 10^{-3}$} &   \bm{$1.681 \times 10^{-2}$} \\
\rowcolor[HTML]{FFFFE0}
 & KE  & \bm{$1.015 \times 10^{-5}$} & \bm{$1.237 \times 10^{-4}$} & \bm{$3.230 \times 10^{-3}$} & \bm{$4.894 \times 10^{-2}$} & \bm{$1.212 \times 10^{0}$} & \bm{$3.735 \times 10^{0}$} \\
\rowcolor[HTML]{FFFFE0}
\multirow{-3}{*}{\textbf{Ours}} & Sinkhorn & \bm{$1.397 \times 10^{-5}$} & \bm{$1.314 \times 10^{-4}$} & \bm{$9.414 \times 10^{-4}$} & \bm{$3.680 \times 10^{-3}$} & \bm{$1.716 \times 10^{-2}$} & \bm{$3.042 \times 10^{-2}$} \\
\hline

\end{tabular}}}
\caption{Comparison of MSE, Kinetic Energy (KE) error, and Sinkhorn divergence on the \textbf{Water} dataset over 200 rollout steps. We compare against the strongest baseline, GraphUNet. PhysicsNFP is excluded, as it cannot handle a varying number of particles.}
\label{tab:water}
\end{table*}

\section{Additional Experiment}

\textbf{Ablation Study and Model Analysis: } We conduct extensive ablation studies to evaluate each component of our model, alongside an analysis of key parameters, in Appendix~\ref{appendix:ablation}. We demonstrate that every proposed component contributes positively to model performance across all metrics.

\textbf{Additional Baselines on Eulerian Mesh Simulator: } Because Eulerian mesh-based neural simulators operate on graph structures analogous to our Lagrangian framework, they are inherently compatible with our setting. Consequently, we compare our approach against a comprehensive suite of state-of-the-art Eulerian mesh models in Appendix~\ref{appendix:baselines}.

\textbf{Additional Experiments on Diverse Materials: } While the primary focus of this work is Lagrangian fluid simulation, other materials such as sand and deformable goop can also adopt a Lagrangian representation. In Appendix~\ref{appendix:diverse}, we demonstrate that our model generalizes well to these diverse material domains and achieves strong performance.

\textbf{Computational Complexity: } We report the runtime and memory consumption of the evaluated models in Appendix~\ref{appendix:complexity}. Despite processing both Lagrangian and Eulerian representations, our approach leverages parallelization to maintain highly competitive inference speeds.

\section{Conclusion}
In this work, we investigate the limitations of Lagrangian neural simulators and identify the sources of their performance gap relative to Eulerian approaches. To overcome these challenges, we introduced a \textbf{Hybrid Lagrangian–Eulerian} model that unifies the complementary strengths of both representations through particle downsampling, particle aggregation, and cross-attention. Our experiments demonstrate that this hybrid framework achieves higher accuracy, stronger robustness, and improved stability across diverse fluid regimes. These results highlight the promise of bridging particle- and grid-based formulations to advance the next generation of neural fluid simulators.

\bibliographystyle{neurips_2026}
\bibliography{neurips_2026}


\appendix

\section{Impact Statement}
Our neural simulator has the potential to transform computational fluid dynamics (CFD) and the broader sciences by enabling faster, more scalable, and more adaptable modeling tools. The development of approaches that combine data-driven learning with physically grounded representations opens new opportunities for fast simulations in areas such as engineering design, climate modeling, and multiphase flow analysis. By improving the efficiency and robustness of CFD, these methods can accelerate scientific discovery and expand access to large-scale simulations that were previously computationally prohibitive.

We do not identify any adverse societal or ethical impacts associated with this paper.

\section{Limitation and Future Work}
While this framework significantly enhances the rollout performance of Lagrangian neural simulators, a primary limitation remains the divergence of trajectories over extreme temporal horizons, such as those spanning 100 physical hours. This drift is not unique to our architecture. Rather, it stems from the irreducible local truncation errors inherent in both neural simulators and numerical solvers. As these microscopic errors compound over millions of integration steps, the global truncation error eventually accumulates, leading to a potential breakdown in physical fidelity.

This phenomenon represents a fundamental, systemic challenge in numerical analysis and the simulation of dynamical systems. Because error accumulation is an intrinsic property of discrete solvers, achieving stability over infinite horizons remains an open problem for the community. Future research could investigate specialized neural architectures that explicitly regularize long-term stability or incorporate error-correction manifolds. By moving beyond the imitation of classical solvers, there is a distinct opportunity to develop neural simulators that eventually surpass the stability and accuracy of traditional numerical methods. Resolving these foundational issues regarding long-term numerical stability will require a sustained, collective effort from the broader physics-ML community. Fully addressing these systemic error accumulation patterns is beyond the scope of the current work.

\section{Datasets}
\label{appendix:datasets}
We provide a brief description of the main benchmark dataset used in this work and refer the readers to \citet{toshev2024lagrangebench} for additional details.

\textbf{\textit{Taylor Green Vortex}}
The dataset represents a decaying flow system characterized by periodic boundary conditions and a specific initial velocity field that leads to kinetic energy decay through viscous interactions. Dataset generation employs a Lagrangian SPH scheme, solving the weakly-compressible Navier-Stokes equations coupled with a barotropic equation of state.  The solver setup involves randomly drawing particle positions, which are then relaxed under periodic boundary conditions via 1000 steps of SPH relaxation to create the initial state. The 2D TGV case is defined on a $1 \times 1$ spatial domain with 2,500 particles at a Reynolds number of 100, using a particle spacing ($\Delta x$) of $20 \times 10^{-3}$ and a time step ($\Delta t$) of $40 \times 10^{-3}$. Conversely, the 3D TGV case utilizes a $2\pi \times 2\pi \times 2\pi$ domain populated by 8,000 particles at a Reynolds number of 50, with a $\Delta x$ of $314.16 \times 10^{-3}$ and a $\Delta t$ of $500 \times 10^{-3}$. The 2D version follows an analytical solution of exponentially decaying velocity.

\textbf{\textit{Lid-Driven Cavity Flow}}
The dataset simulates a flow within a confined domain driven by a moving top wall. This dataset is generated using the Lagrangian SPH solver, which solves the weakly-compressible Navier-Stokes equations. The ground truth setup initializes the fluid velocity to zero and runs the simulation until the system reaches a statistically stationary equilibrium state, at which point data collection begins. While the solver uses a generalized wall boundary condition with multiple layers of dummy particles to enforce impermeability and no-slip conditions, the dataset retains only the innermost layer. In the 2D case, the simulation comprises 2,708 particles within a $1.12 \times 1.12$ spatial domain at a Reynolds number of 100, utilizing a particle spacing ($\Delta x$) of $20 \times 10^{-3}$ and a time step ($\Delta t$) of $40 \times 10^{-3}$55. The 3D case involves 8,160 particles in a $1.25 \times 1.25 \times 0.5$ domain, also at a Reynolds number of 100, with a $\Delta x$ of $41.667 \times 10^{-3}$ and a $\Delta t$ of $90 \times 10^{-3}$. Notably, the 3D version applies periodic boundary conditions in the z-direction to recover the dynamics of the 2D solution.

\textbf{\textit{Reverse Poiseuille Flow}}
The dataset simulates a fully periodic flow driven by a spatially varying external force field. This dataset is generated using the Lagrangian SPH solver, which solves the weakly-compressible Navier-Stokes equations. The ground truth setup initializes with zero fluid velocity and applies a force of magnitude 1 in the lower half of the domain and -1 in the upper half, running the simulation until it reaches a statistically stationary equilibrium state before data collection begins. In the 2D case, the system consists of 3,200 particles within a $1 \times 2$ spatial domain at a Reynolds number of 10, utilizing a particle spacing ($\Delta x$) of $25 \times 10^{-3}$ and a time step ($\Delta t$) of $40 \times 10^{-3}$. The 3D counterpart involves 8,000 particles in a $1 \times 2 \times 0.5$ domain, also at a Reynolds number of 10, with a $\Delta x$ of $50 \times 10^{-3}$ and a $\Delta t$ of $100 \times 10^{-3}$5. Although theoretically similar to laminar channel flow, this specific setup at $Re=10$ is not fully laminar and exhibits mixing, making the dynamics more diverse for learning tasks.

\textbf{\textit{Dam Break}}
The dataset is designed to benchmark the simulation of free-surface flows involving significant topological changes. Generated using the Lagrangian SPH solver to solve the weakly-compressible Navier-Stokes equations, this dataset distinctly calculates density via evolution equations rather than summation to accurately account for the lack of full kernel support at the free surface. The ground truth setup initializes the system by randomly drawing particle positions and letting them relax under non-periodic boundary conditions for 1000 steps. To manage the solid boundaries, the solver employs generalized wall boundary conditions with dummy particles to enforce no-slip and impermeability, though only the innermost particle layer is retained in the final dataset. The 2D simulation is defined on a spatial domain of $5.486 \times 2.12$ and contains 5,740 particles. While this problem is typically modeled as inviscid, this specific implementation includes a small amount of physical viscosity to minimize wall artifacts, resulting in a Reynolds number of 40,000, with a particle spacing ($\Delta x$) of $20 \times 10^{-3}$ and a time step ($\Delta t$) of $30 \times 10^{-3}$.

\subsection{Higher Resolution Simulation Data}
\label{appendix:resolution_data}
To generate higher-resolution simulation data for the Taylor-Green Vortex, we systematically reduce the initial particle spacing, $\Delta x$, starting from $20 \times 10^{-3}$ and progressively decreasing it to $15 \times 10^{-3}$, $10 \times 10^{-3}$, and $5 \times 10^{-3}$. These configurations yield simulations with increasing particle counts of 3,600, 4,489, 10,000, and 40,000, respectively. Reducing the inter-particle spacing enhances the spatial resolution of the Lagrangian discretization, which is critical for numerical accuracy. A denser particle distribution enables more precise approximation of spatial derivatives, resulting in more accurate gradient computations and improved fidelity when querying field variables such as velocity and pressure. Furthermore, this setup serves as a scalability experiment for the neural surrogate models. The significant increase in particle number challenges the model to generalize to larger, more computationally intensive graph structures while maintaining predictive accuracy.

\subsection{Fluid-Solid Interaction Data}
\label{appendix:fsi_data}
The simulation takes place in a rectangular fluid domain with physical dimensions of $18.0 \times 18.0$ units, extending from $x=0$ to $18.0$ and $y=-9.0$ to $9.0$. A cylinder with a diameter of $1.2$ units is positioned at coordinates $(6.0, 0.0)$ to obstruct the flow. The simulation uses a fixed time step ($dt$) of approximately $0.0027$ seconds, calculated initially based on the minimum of CFL and viscous stability criteria, without utilizing adaptive timestepping. The system saves the simulation state every 10 time steps, effectively capturing the dynamics of the flow at a Reynolds number of 200. The dataset is generated using \citet{ramachandran2021a}.

\subsection{Deformable Domain Data}
\label{appendix:deformable_data}
The data is generated by simulating the Navier-Stokes equations for an incompressible fluid, modeling the evolution of a circular patch of fluid deformed into an ellipse by an initial linear velocity field. This process perfectly conserves the area of the fluid. The simulation utilizes approximately 5,026 particles at standard resolution and dynamically calculates an adaptive time step of roughly $\Delta t = 5.27 \times 10^{-6}$ seconds using a CFL constraint. We generate 10000 frames for training, 1000 for validation, and 1000 for testing. The dataset is generated using \citet{ramachandran2021a}.

\section{Model Training and Implementation Details}
\label{appendix:model_training_and_implementation_details}
All models are implemented using the PyTorch framework~\citep{paszke2019pytorchimperativestylehighperformance} and trained with the Adam optimizer~\citep{kingma2017adammethodstochasticoptimization}. All models are configured to directly predict particle positions from unperturbed inputs. We deliberately exclude these auxiliary training strategies found in prior literature~\citep{pfaff2021learning, sanchezgonzalez2020learningsimulatecomplexphysics} to isolate the specific contributions of our proposed modules.

\textbf{GNS} The model employs a message-passing GNN architecture~\citep{sanchezgonzalez2020learningsimulatecomplexphysics}, adhering to the hyperparameter configuration detailed in \citet{toshev2024lagrangebench}. Both the encoder and decoder are composed of 4-layer MLPs with ReLU activations, while the processor utilizes 10 message-passing layers with a latent dimension of 128. 

\textbf{SEGNN} We implement the geometric architecture proposed by \citet{brandstetter2021geometric}, adopting the hyperparameter guidelines from \citet{toshev2024lagrangebench}. The network comprises 10 message-passing layers with a latent dimension of 64, using 2-layer MLP blocks throughout. We restrict the maximum spherical harmonic degree ($L_{max}$) to 1 for both hidden representations and attributes. The input features are treated as isotropic.

\textbf{GraphUnet} We adopt the hierarchical architecture proposed by \citet{gao2019graph}, which features three downsampling and upsampling stages with a pooling ratio of $0.6$. The encoder and decoder modules are composed of 4-layer MLPs with ReLU activations, mapping to a latent dimension of $128$. At the coarsest resolution (the bottom level), the processor employs a stack of 4 message-passing GNN layers.

\textbf{GraphTransformer} We adopt the model architecture proposed in \citet{dwivedi2021generalizationtransformernetworksgraphs}. We use $6$ layers with a latent dimension of $128$. The encoder and decoder modules are composed of 4-layer MLPs with ReLU activations.

\textbf{AdvDIFFormer} "We adopt the model architecture proposed in \citet{wu2025supercharging}, configured with a hidden channel size of 128. We set the approximation order $K$ to 1 and utilize 2 attention heads.

\textbf{NeuralMPM} We adopt the architecture proposed by \citet{rochman-sharabi2025a}, which operates on a fixed $32 \times 32$ Eulerian grid. The central processor employs a U-Net structure with a base channel dimension of 64.

\textbf{PhysicsNFP} We adopt the topology-aware architecture proposed by \citet{jiang2025topologyaware}. The model comprises 10 layers with a hidden channel dimension of $128$.

\textbf{Ours} Our architecture incorporates $M=2$ downsampling and upsampling stages with a ratio of $0.8$. The model operates with a latent dimension of $128$ and utilizes a central processor comprising $L=4$ steps. Both the encoder and decoder modules are constructed as 4-layer MLPs with ReLU activations. The model is trained using a learning rate schedule that decays from $10^{-4}$ to $10^{-6}$.

\section{Model Analysis}
\label{appendix:ablation}
We conduct extensive analysis on the \textit{Lid-Driven Cavity Flow} dataset to isolate the contribution of each module.

\begin{figure}[h!]
    \centering
    \includegraphics[width=0.32\columnwidth]{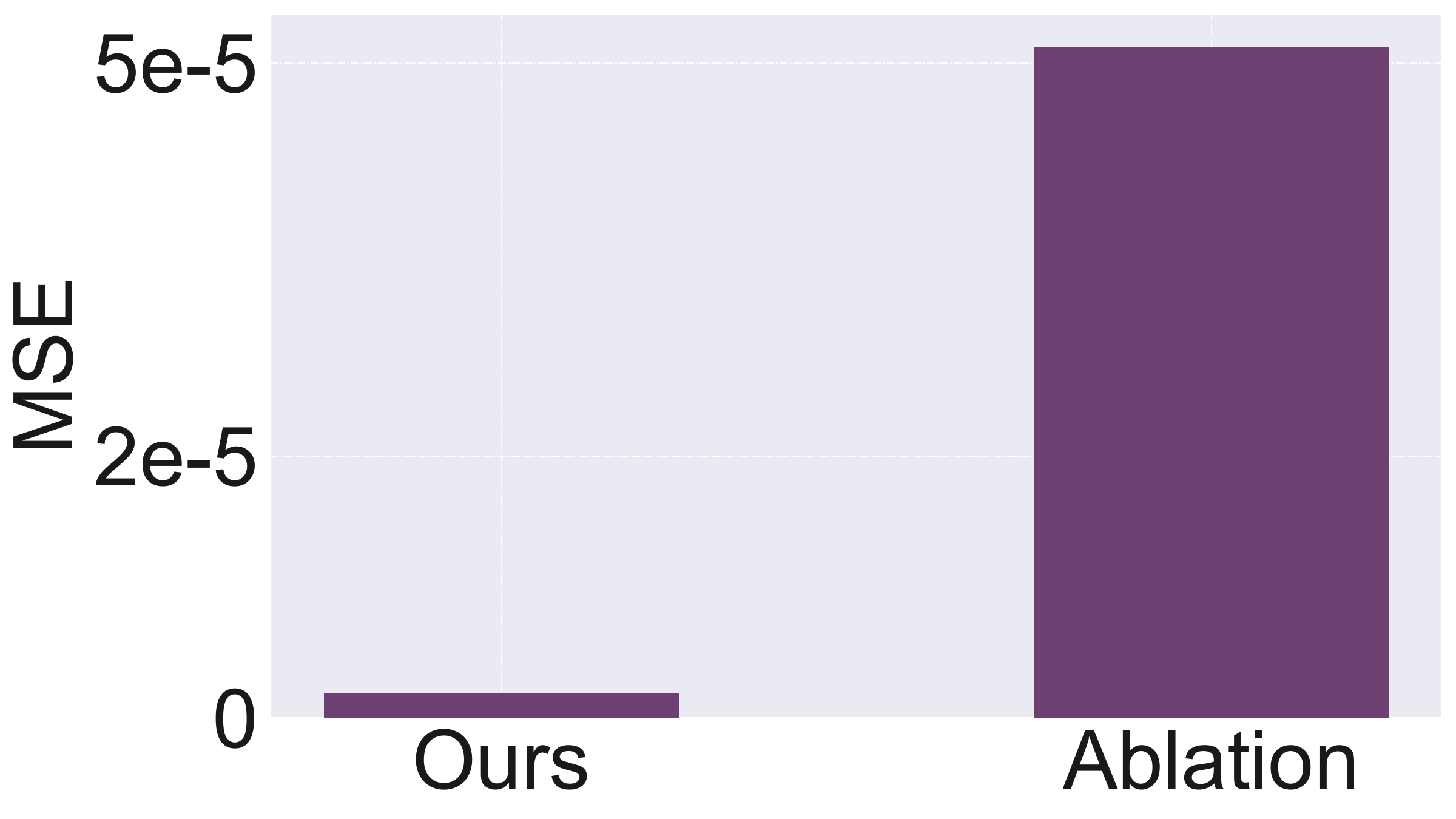}
    \includegraphics[width=0.32\columnwidth]{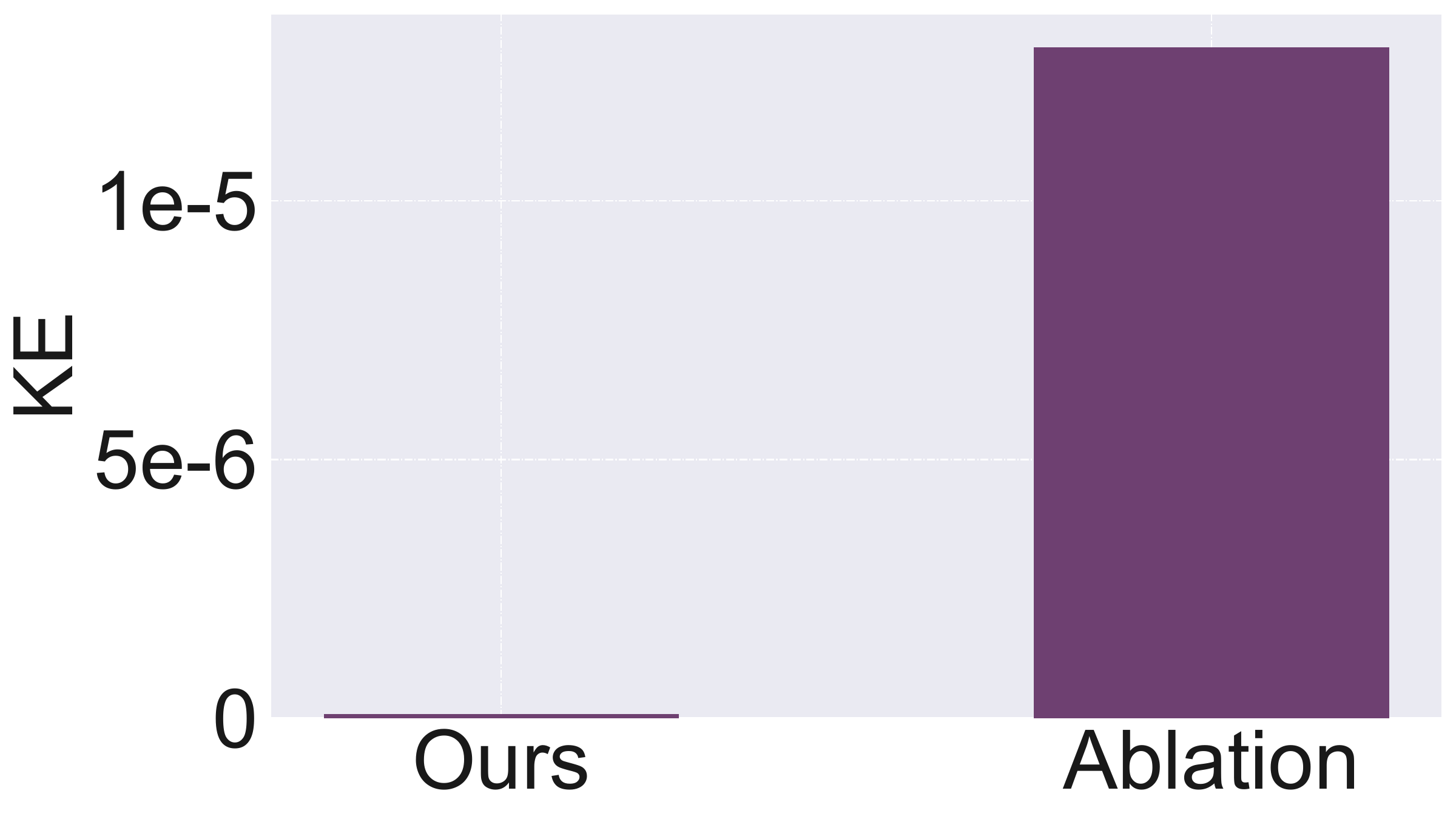}
    \includegraphics[width=0.32\columnwidth]{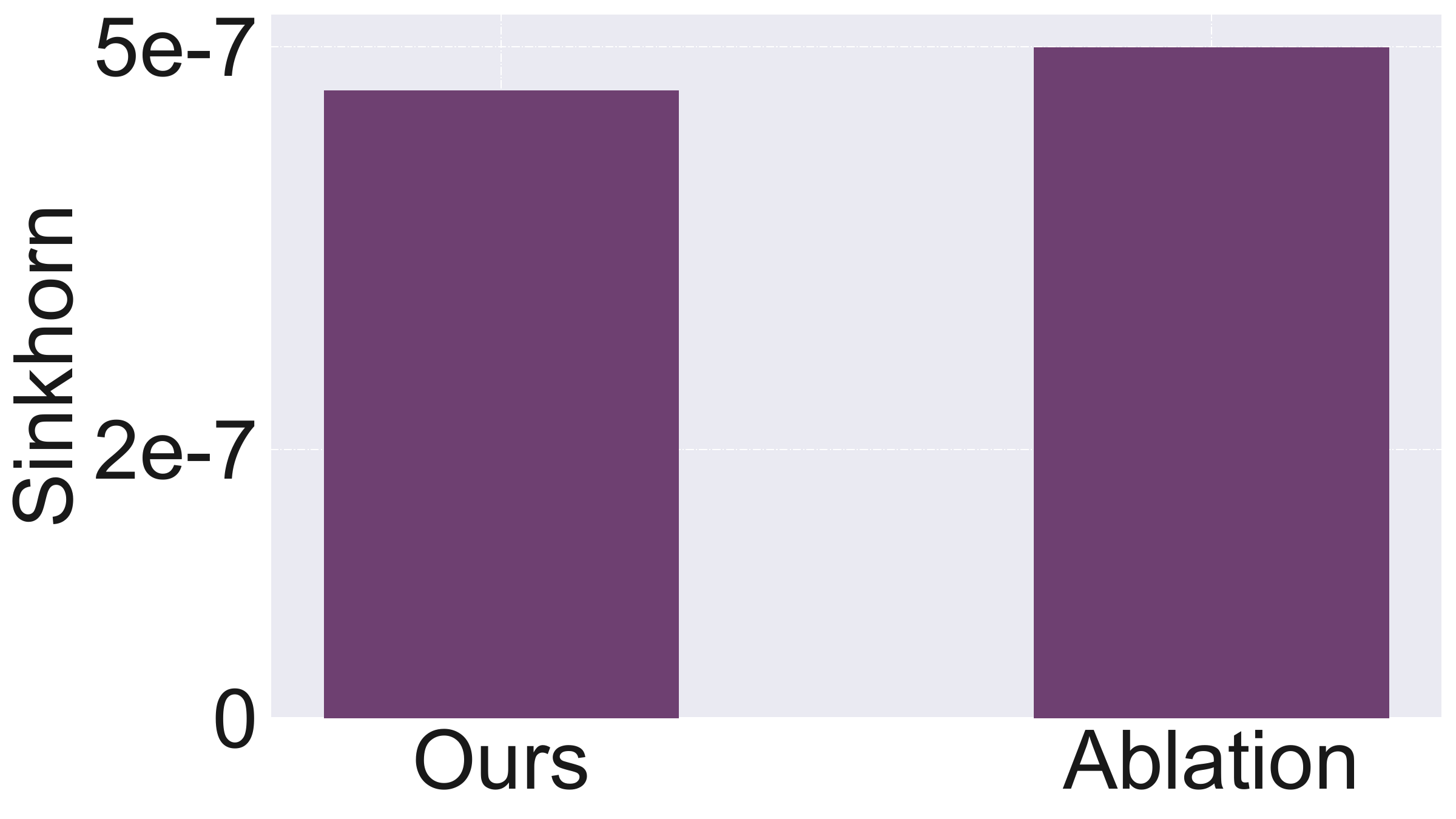}
    \caption{Five-step rollout results of the ablation study on the downsampler and upsampler, reported on the Lid-Driven Cavity flow dataset. Ours denotes the proposed model, while Ablation refers to the variant without the downsampler and upsampler.}
    \label{fig:ablation_downsample}
\end{figure}

\textbf{Ablation Study on Downsampler: } We perform an ablation study to evaluate the effect of the downsampler and upsampler. Results are reported in Figure~\ref{fig:ablation_downsample}. Across all metrics, the model equipped with these components consistently outperforms the variant without them over the five-step rollout. This observation aligns with our design rationale. In the Lid-Driven Cavity flow dataset, many particles in the lower-left region remain nearly stationary, while significant dynamics occur elsewhere. The downsampler and upsampler encourage the model to focus its capacity on regions with strong flow variation. As a result, the model with these modules achieves notably lower MSE and kinetic energy error. We attribute this to the difference in the processor, as the processor of the ablated model must explicitly model interactions among all particles, whereas our hierarchical framework delegates the coarse forward PDE dynamics to the processor and leaves the fine-scale refinements to the downsampler and upsampler.

\begin{figure}[h!]
    \centering
    \includegraphics[width=0.32\columnwidth]{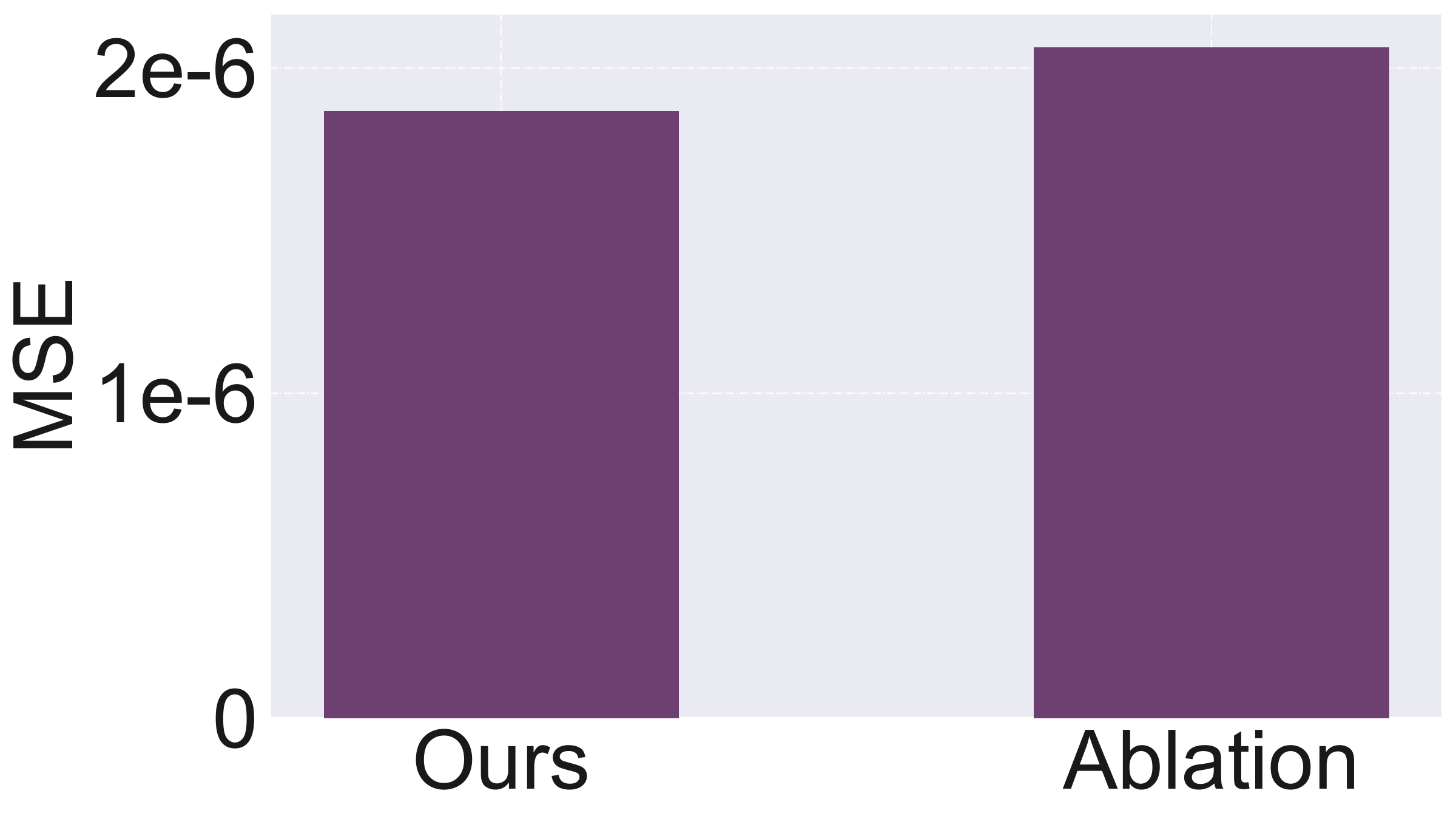}
    \includegraphics[width=0.32\columnwidth]{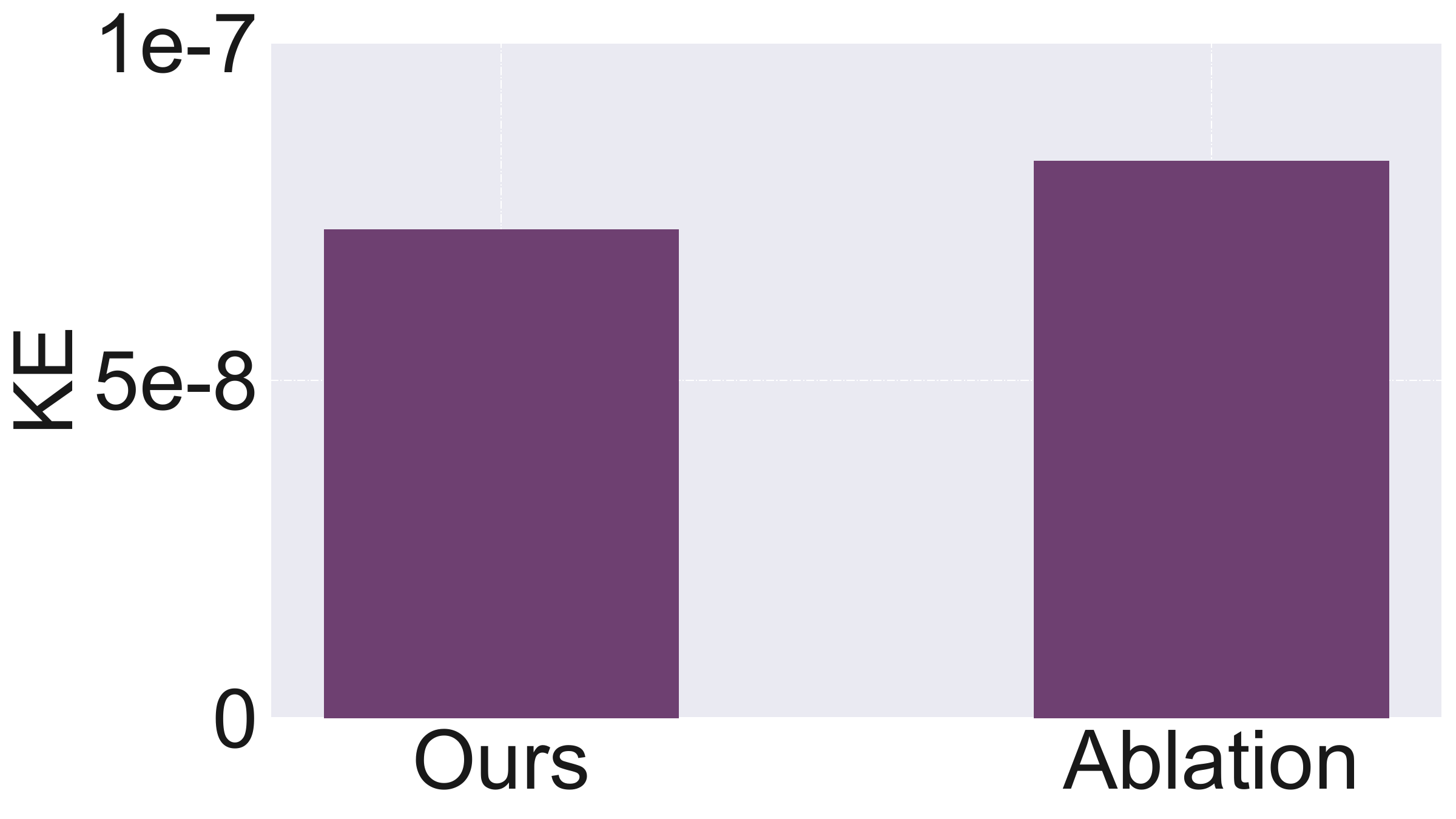}
    \includegraphics[width=0.32\columnwidth]{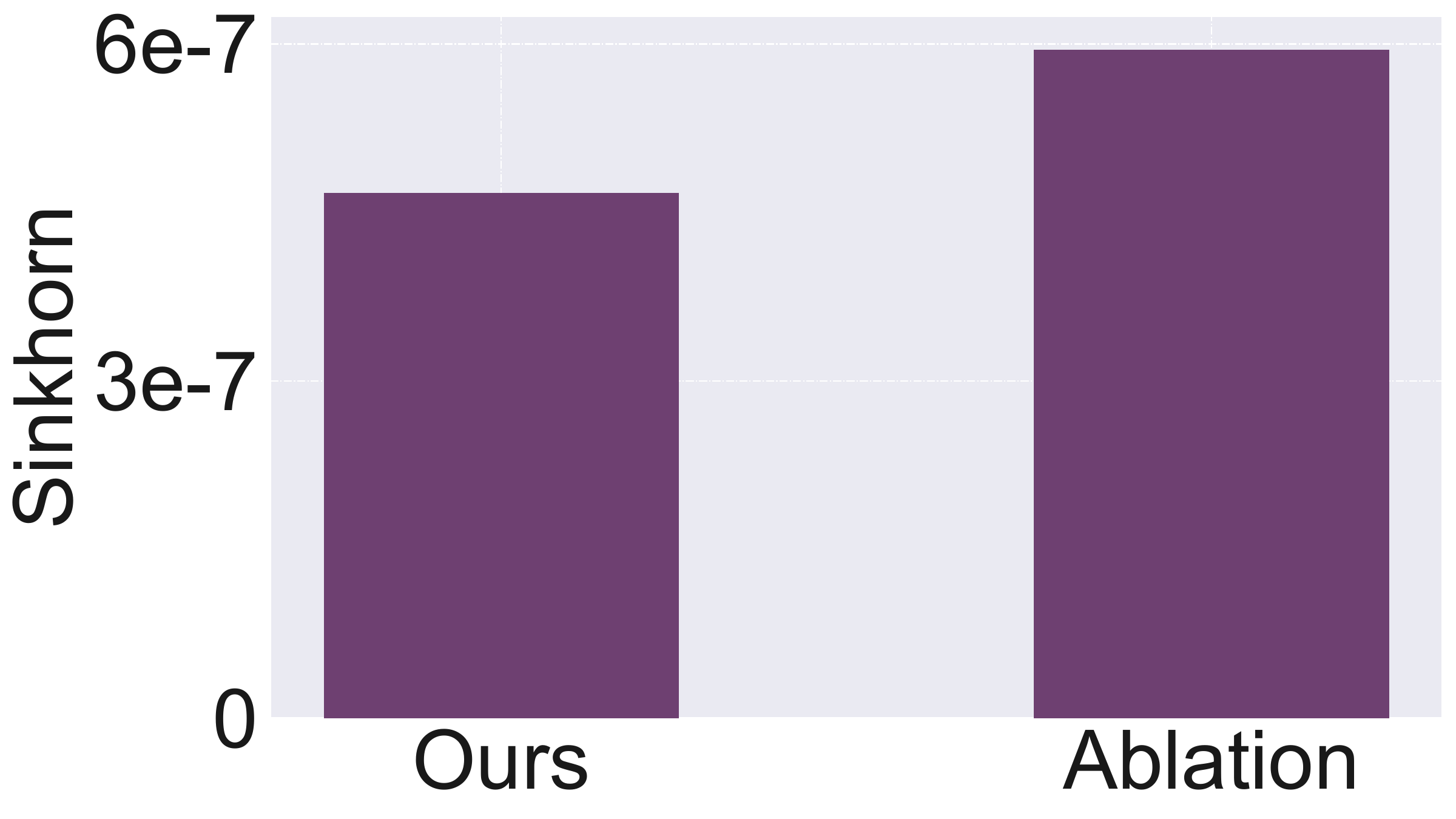}
    \caption{Five-step rollout results of the ablation study on the combined self-cross attention feature, reported on the Lid-Driven Cavity flow dataset. Ours denotes the proposed model, while Ablation refers to the variant without cross attention.}
    \label{fig:ablation_eu_plus}
\end{figure}

\textbf{Ablation Study on Combined Self–Cross Attention Feature: } We conduct an ablation study to examine the role of the combined self–cross attention feature. Specifically, we remove the cross-attention between Lagrangian and Eulerian features and retain only the Lagrangian self-attention, replacing Equation~\ref{eq:eu_plus} with Equation~\ref{eq:eu_plus_ablation}. Results are shown in Figure~\ref{fig:ablation_eu_plus}.
\begin{align}
    \bm{F} = \bm{A}, \quad \hat{\bm{q}}_{\mathrm{lag}} = \bm{W}\,\bm{F} \in \mathbb{R}^{\tilde{N}_{\mathrm{lag}} \times d_{\mathrm{latent}}}.
    \label{eq:eu_plus_ablation}
\end{align}

The model that integrates both Lagrangian self-attention and Lagrangian–Eulerian cross-attention consistently outperforms the ablated variant. This highlights the importance of cross-attention in enhancing overall performance by leveraging information from both representations.

\begin{figure}[h!]
    \centering
    \includegraphics[width=0.32\columnwidth]{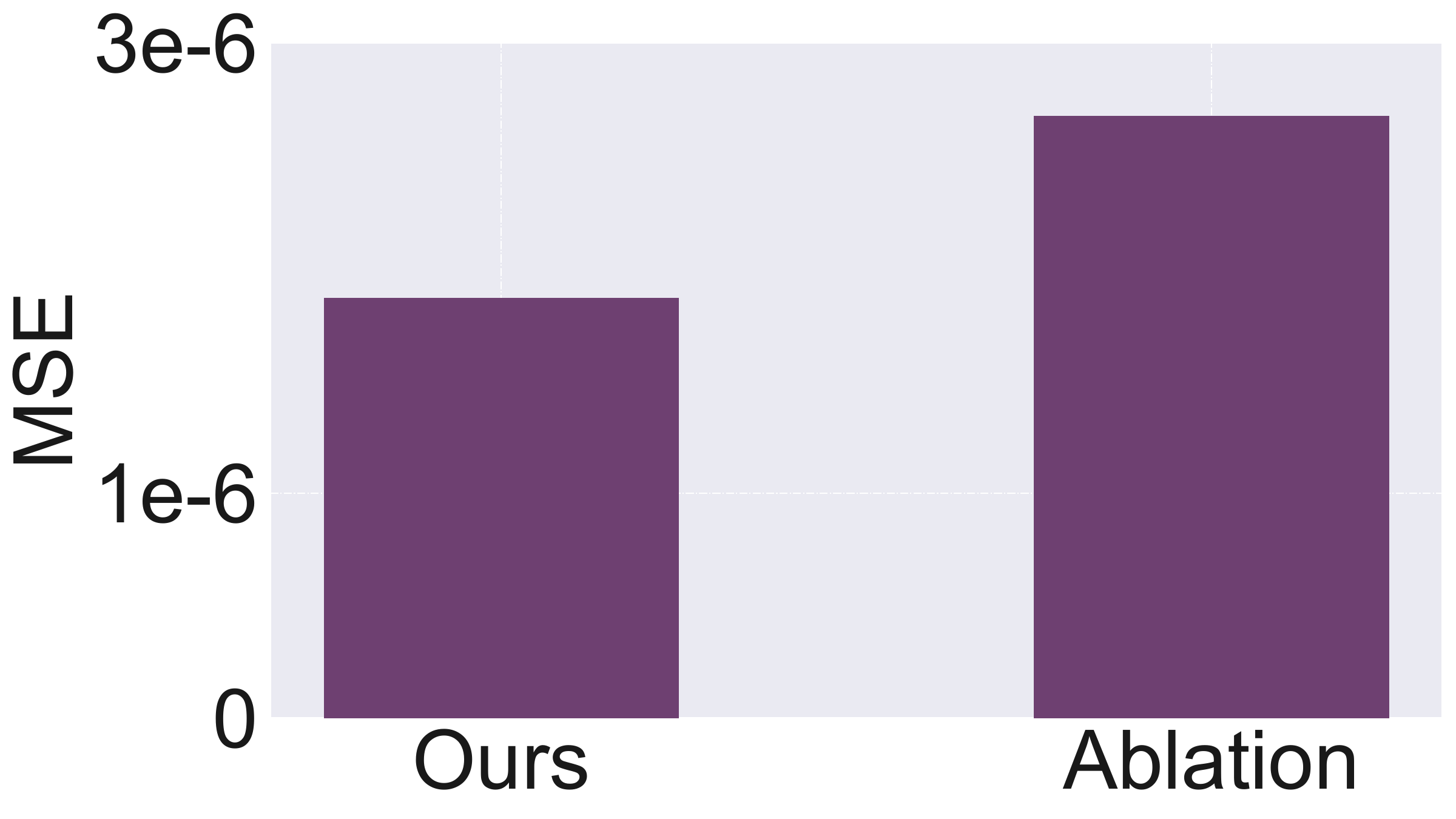}
    \includegraphics[width=0.32\columnwidth]{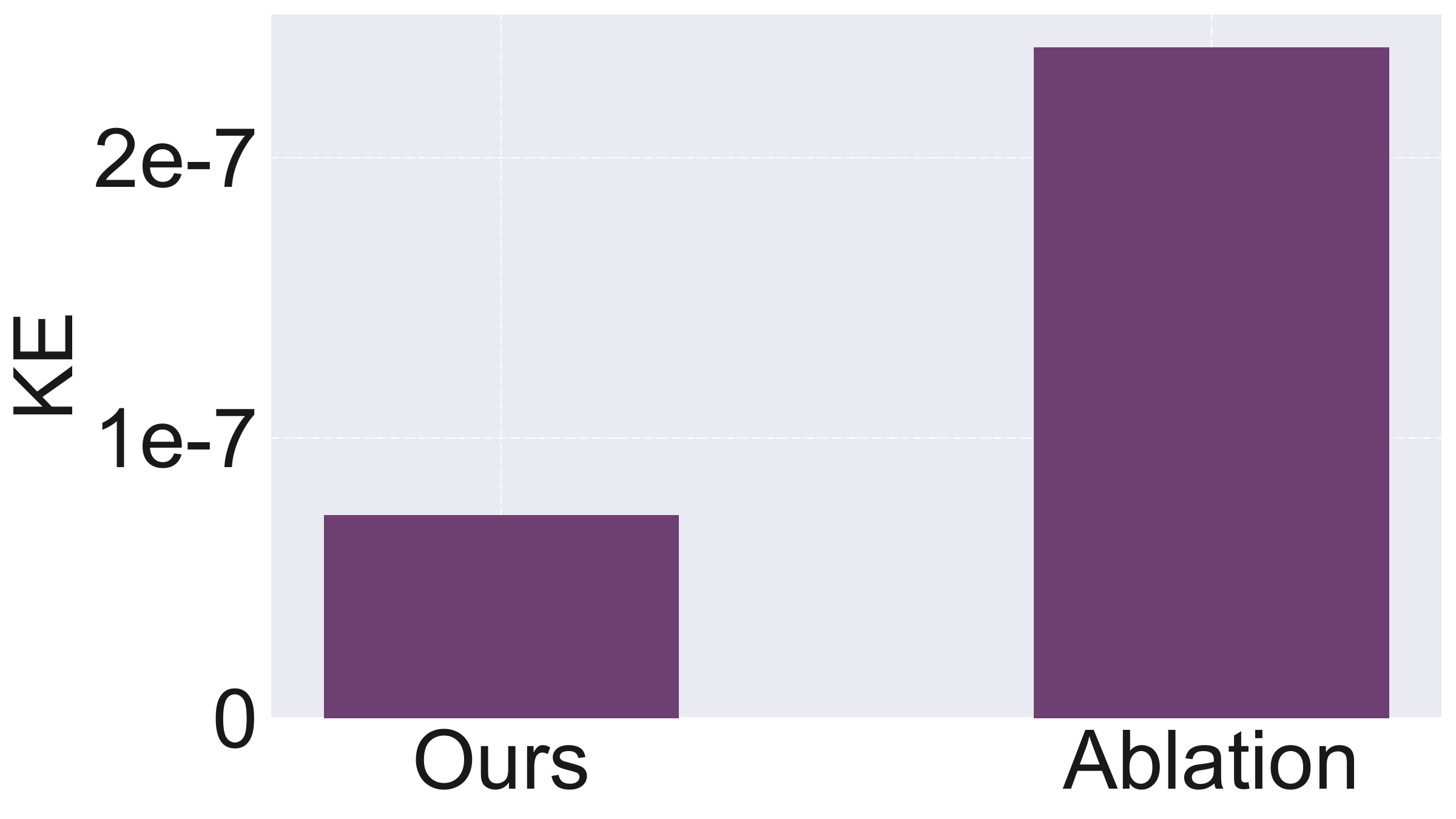}
    \includegraphics[width=0.32\columnwidth]{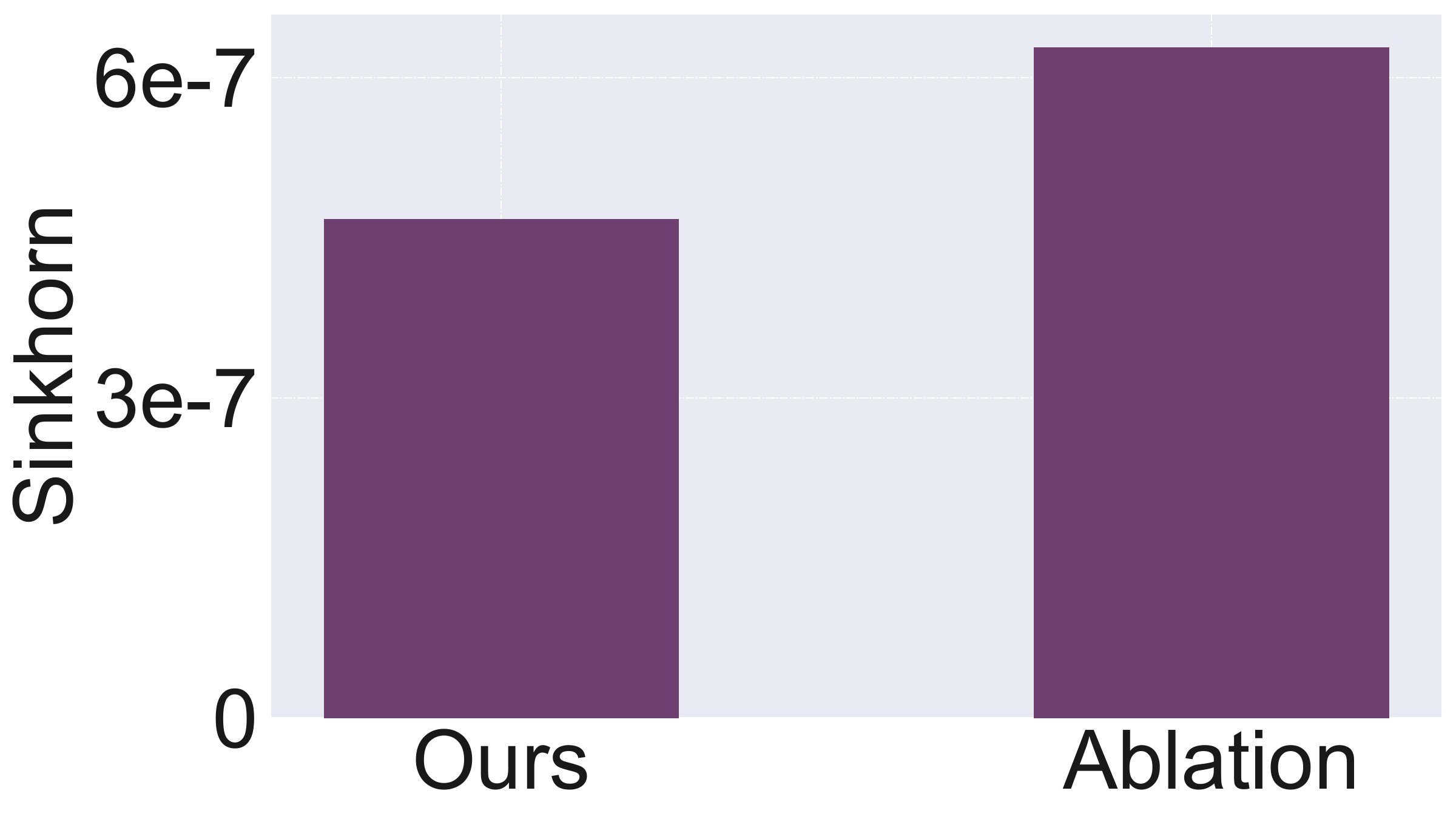}
    \caption{Five-step rollout results of the ablation study on the role of Eulerian feature in Lagrangian aggregation, reported on the Lid-Driven Cavity flow dataset. Ours denotes the proposed model, while Ablation refers to the variant that uses an MLP to determine aggregation points.}
    \label{fig:ablation_eu_pos}
\end{figure}

\textbf{Ablation Study on Aggregating Lagrangian Nodes to Eulerian Nodes: } We also examine the role of Eulerian features in defining aggregation points for Lagrangian features. In the full model, downsampled Eulerian nodes serve as anchors for aggregating Lagrangian representations. To assess their contribution, we design an ablation study where these nodes are replaced by a learnable alternative: an MLP that takes the positions of downsampled Lagrangian particles as input and outputs logits indicating the assignment of each particle to aggregation points. The number of aggregation points is kept equal to the number of downsampled Eulerian nodes, and the logits determine the contribution of each particle to its assigned aggregation node. Results are reported in Figure~\ref{fig:ablation_eu_pos}.

The full model that aggregates Lagrangian features onto downsampled Eulerian nodes consistently achieves stronger performance across all metrics. 
The full model that aggregates Lagrangian features onto downsampled Eulerian nodes consistently achieves stronger performance across all metrics. We attribute these gains to two factors. First, the downsampled Eulerian nodes naturally share a spatial distribution similar to that of the downsampled Lagrangian particles. Aggregating Lagrangian features onto this Eulerian structure produces representations that remain dynamically consistent with the Eulerian features, which is particularly beneficial during cross-attention, where the Eulerian branch provides a global perspective while the Lagrangian branch supplies fine-scale details. Second, the MLP-based aggregation points discard explicit spatial structure, weakening the self-attention’s ability to resolve PDE dynamics at a coarse level and diminishing the effectiveness of cross-attention in aligning complementary representations.

\begin{figure}[h!]
    \centering
    \includegraphics[width=0.32\columnwidth]{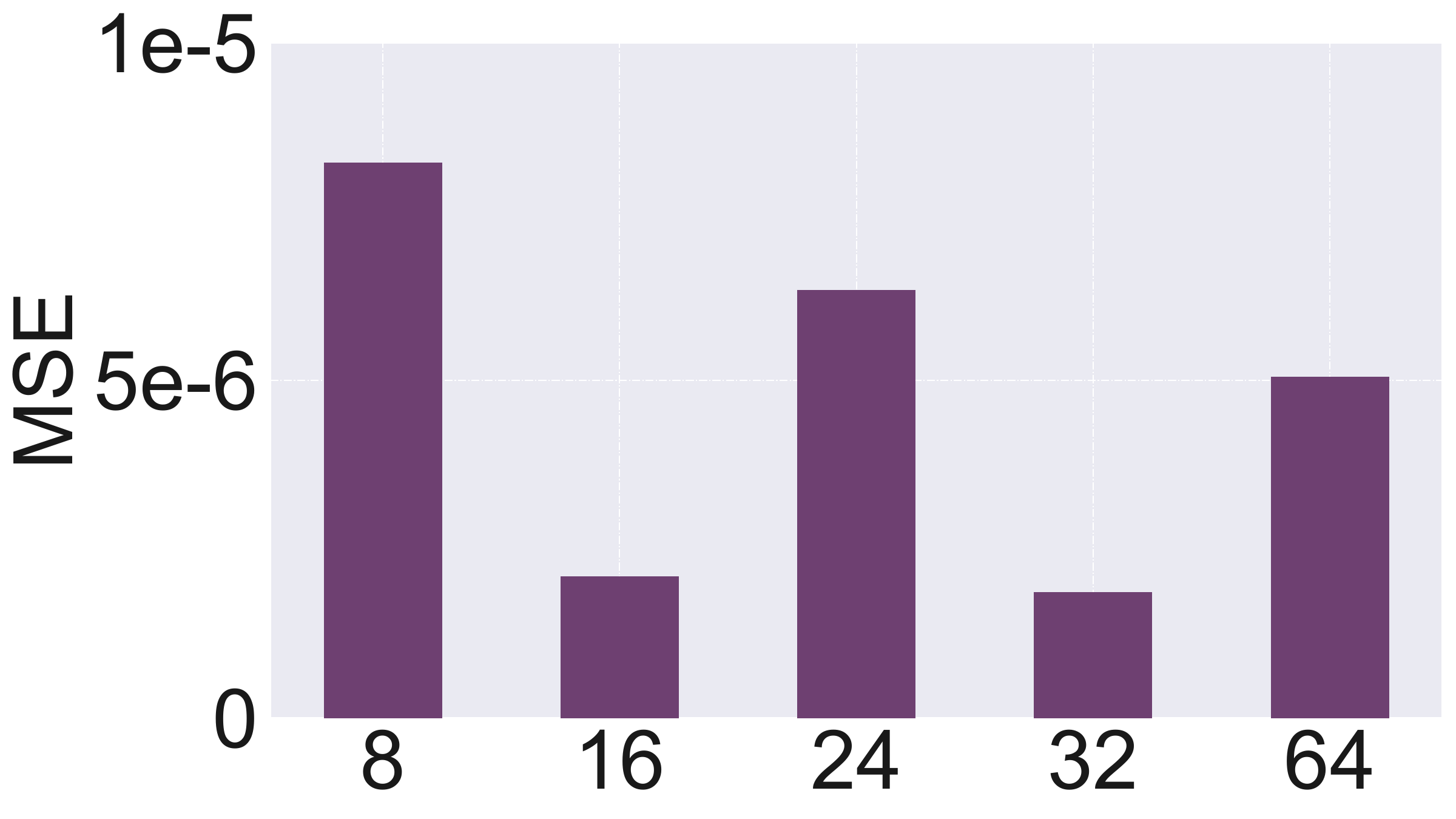}
    \includegraphics[width=0.32\columnwidth]{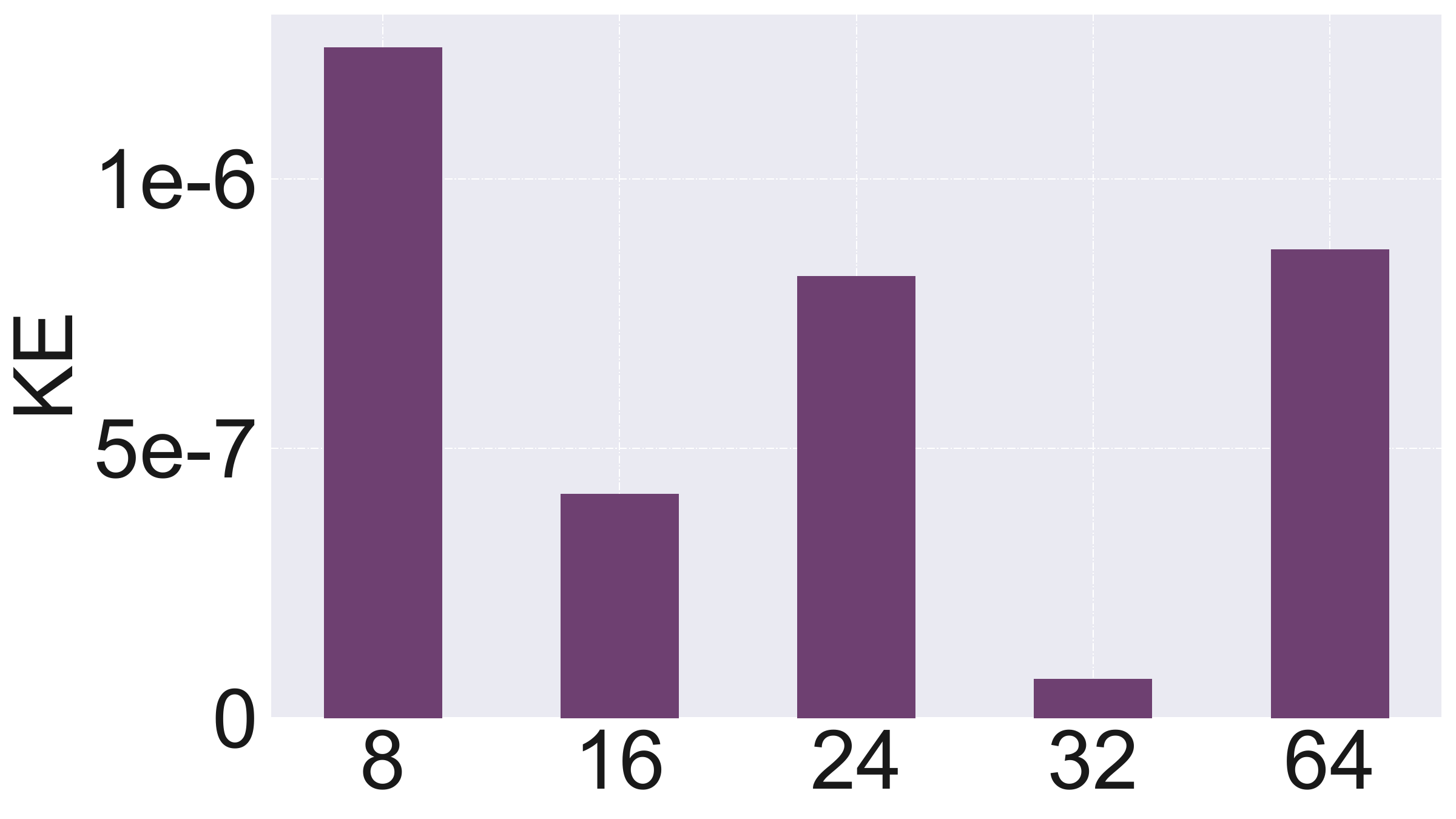}
    \includegraphics[width=0.32\columnwidth]{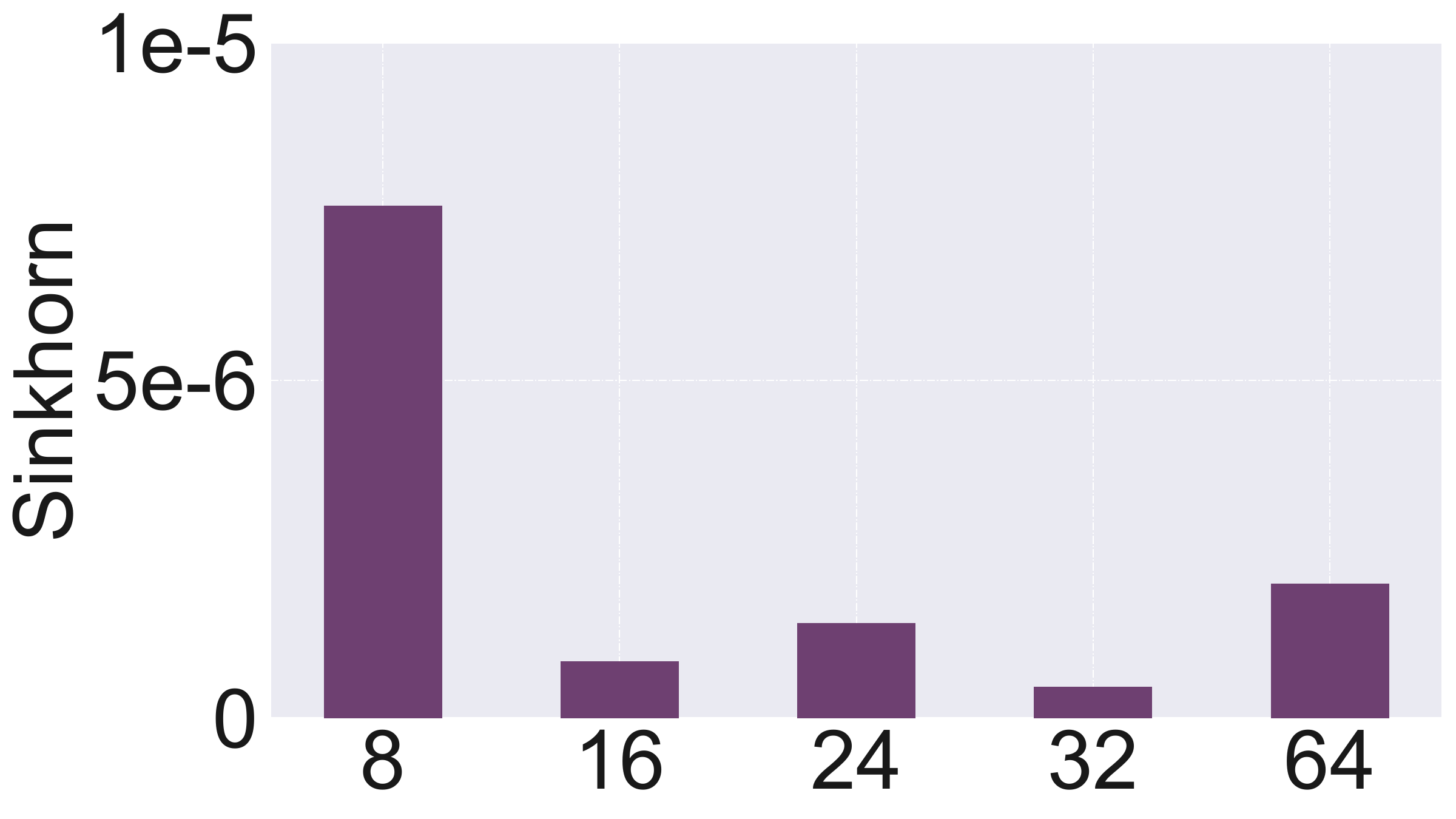}
    \caption{Five-step rollout results of varying the discretization resolution of Eulerian nodes, reported on the Lid-Driven Cavity flow dataset.}
    \label{fig:ablation_eu_dis}
\end{figure}

\textbf{Analysis on Eulerian Discretization: } The Eulerian features are represented on a discretized grid with $N_{\mathrm{eu}}$ nodes. We investigate how the discretization resolution affects model performance using the Lid-Driven Cavity flow dataset, defined on a bounded $1.12 \times 1.12$ domain. Each direction is discretized with $s$ nodes, and the results are shown in Figure~\ref{fig:ablation_eu_dis}. Among the tested resolutions, $s=32$ yields the best performance. Extremely coarse grids such as $s=8$, or overly fine grids such as $s=64$, perform the worst.  With too few Eulerian nodes, the representation lacks sufficient information about the global fluid field. In addition, aggregating Lagrangian features onto such a small set of nodes leads to significant information loss, as particles from diverse locations and flow behaviors are forced to map to the same averaged values.

\begin{figure}[h!]
    \centering
    \includegraphics[width=0.32\columnwidth]{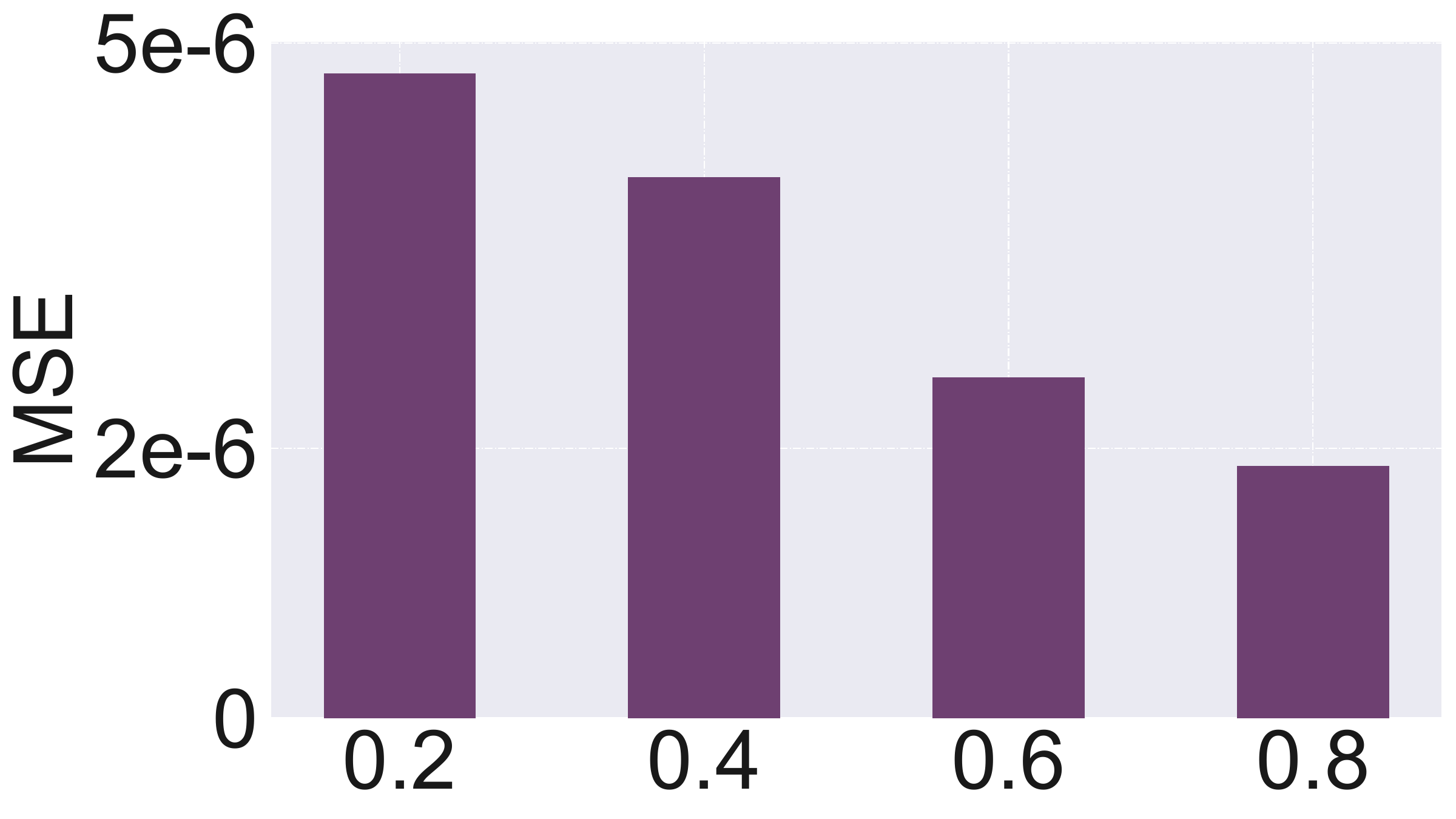}
    \includegraphics[width=0.32\columnwidth]{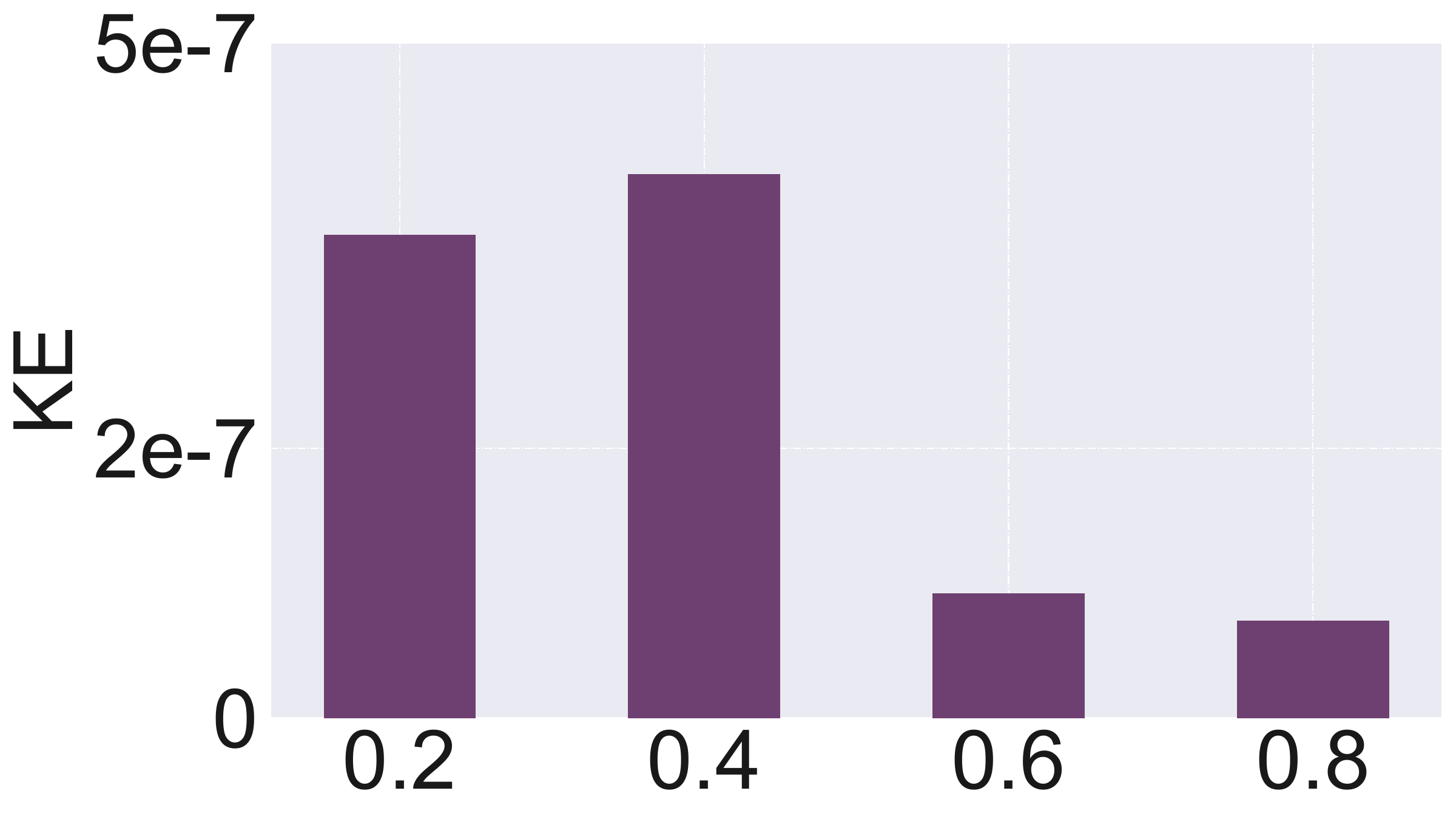}
    \includegraphics[width=0.32\columnwidth]{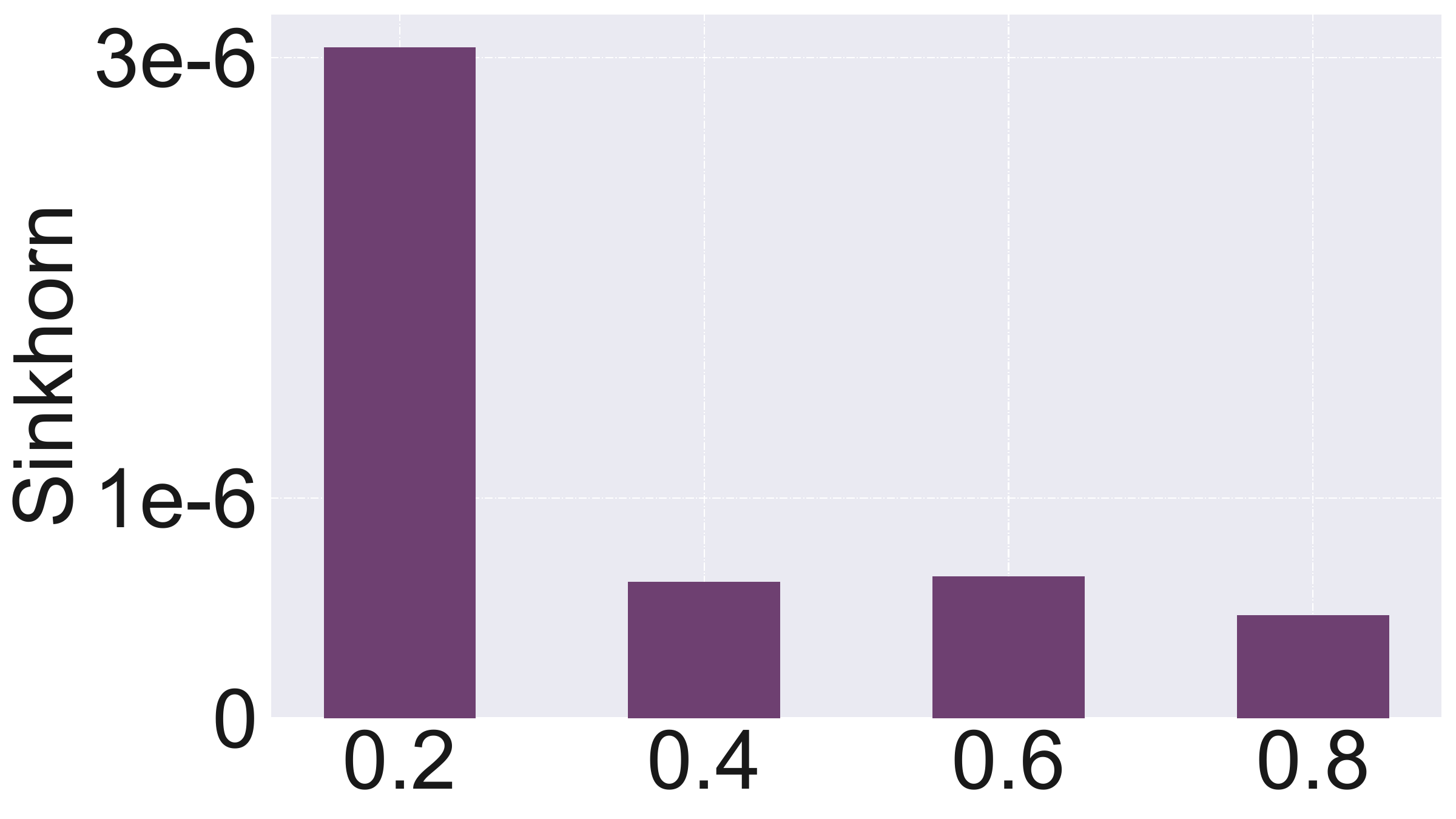}
    \caption{Five-step rollout results of varying the downsampling ratio, reported on the Lid-Driven Cavity flow dataset.}
    \label{fig:ablation_ratio}
\end{figure}

\textbf{Analysis on Downsampling Ratio: } We study the impact of the downsampling ratio, $\lambda$, on model performance, with results shown in Figure~\ref{fig:ablation_ratio}. The best accuracy is obtained at $\lambda=0.8$, while more aggressive choices such as $\lambda=0.2$ or $\lambda=0.4$ lead to a sharp decline in performance. This behavior is expected, as the purpose of downsampling is to merge particles that move coherently or remain nearly static, since these contribute little additional information when considered separately. However, if the ratio is too aggressive, the model also discards particles carrying essential information about the fluid dynamics, which in turn degrades the solution.

\begin{figure}[h!]
    \centering
    \includegraphics[width=0.32\columnwidth]{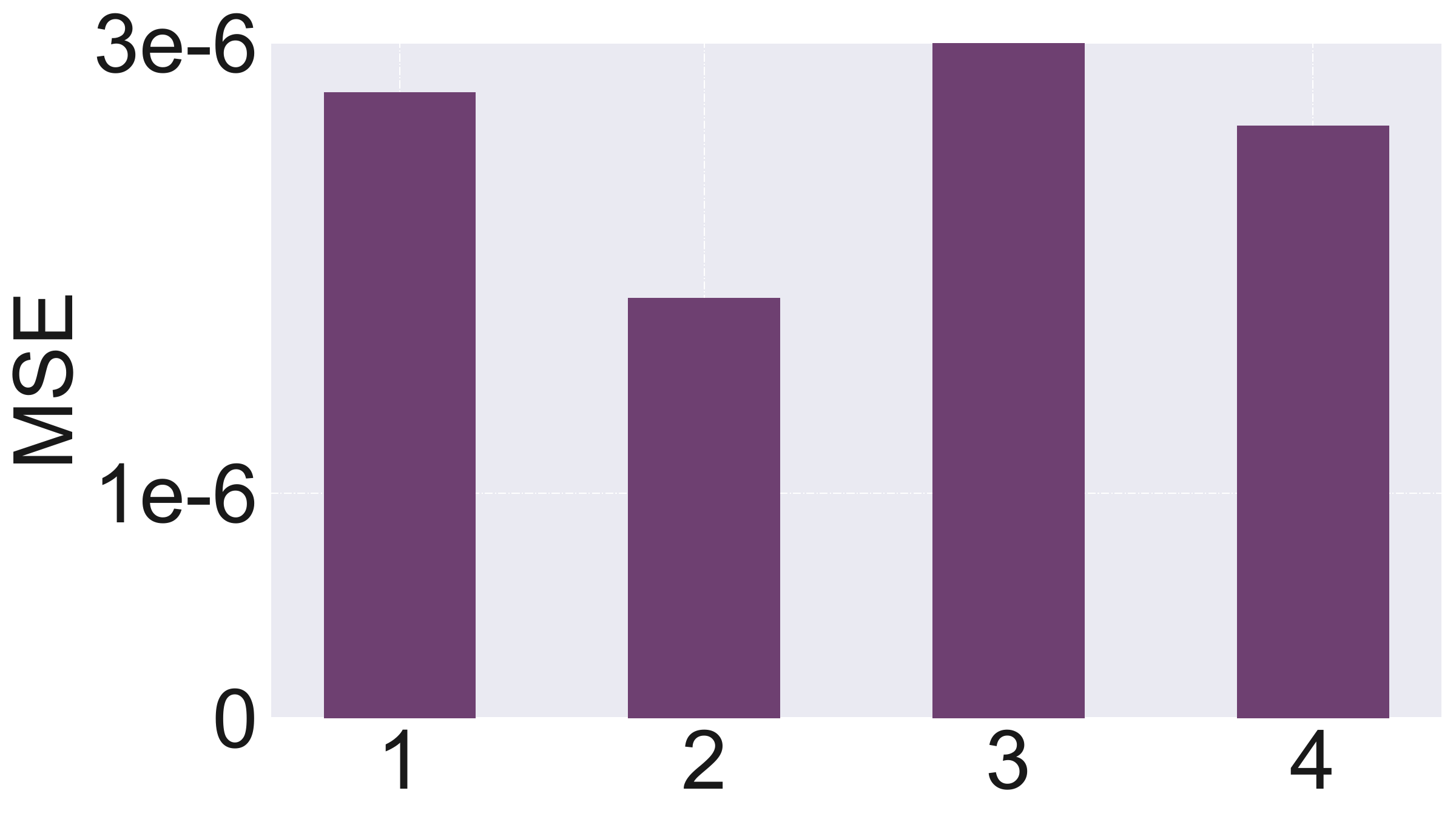}
    \includegraphics[width=0.32\columnwidth]{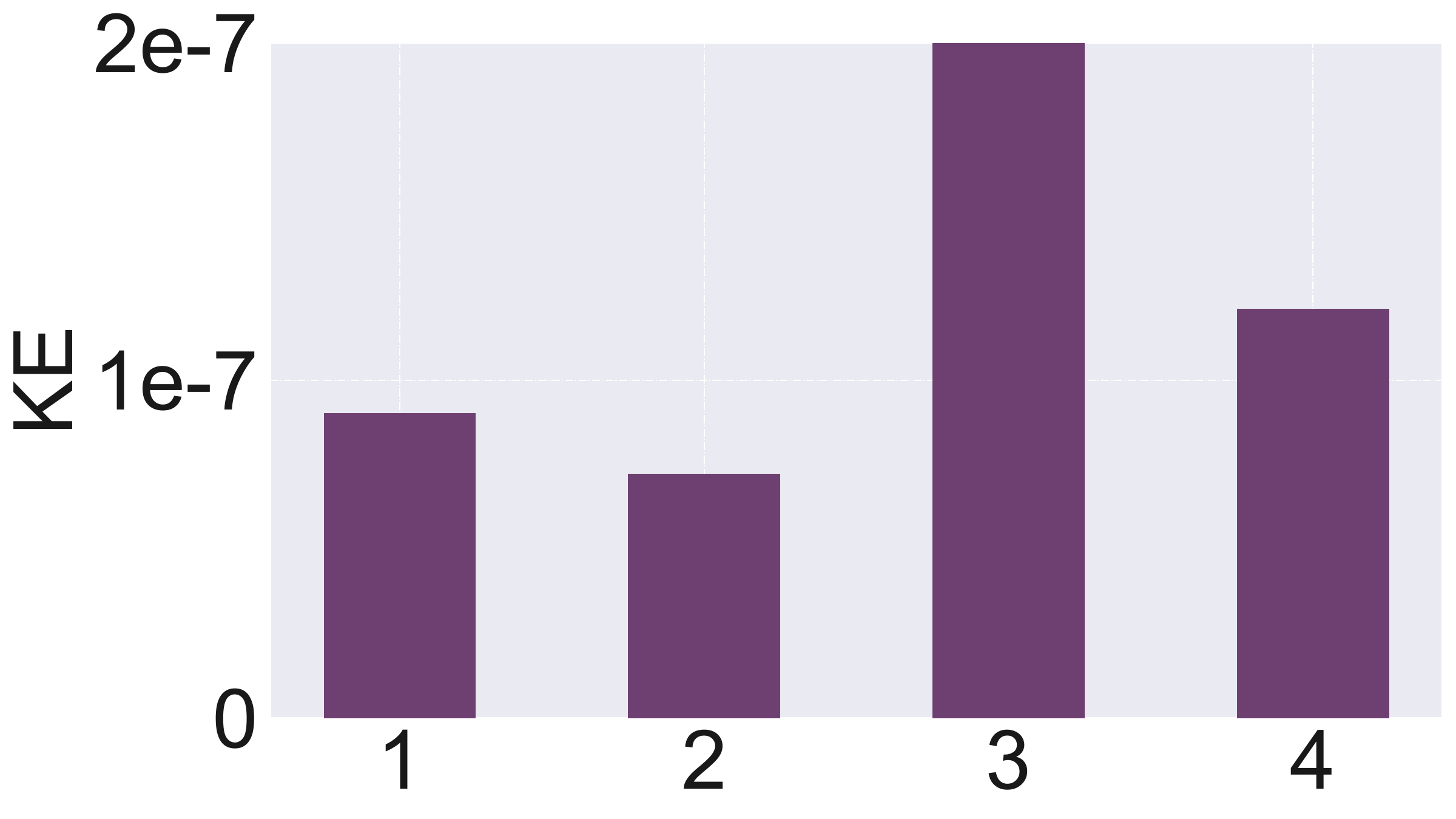}
    \includegraphics[width=0.32\columnwidth]{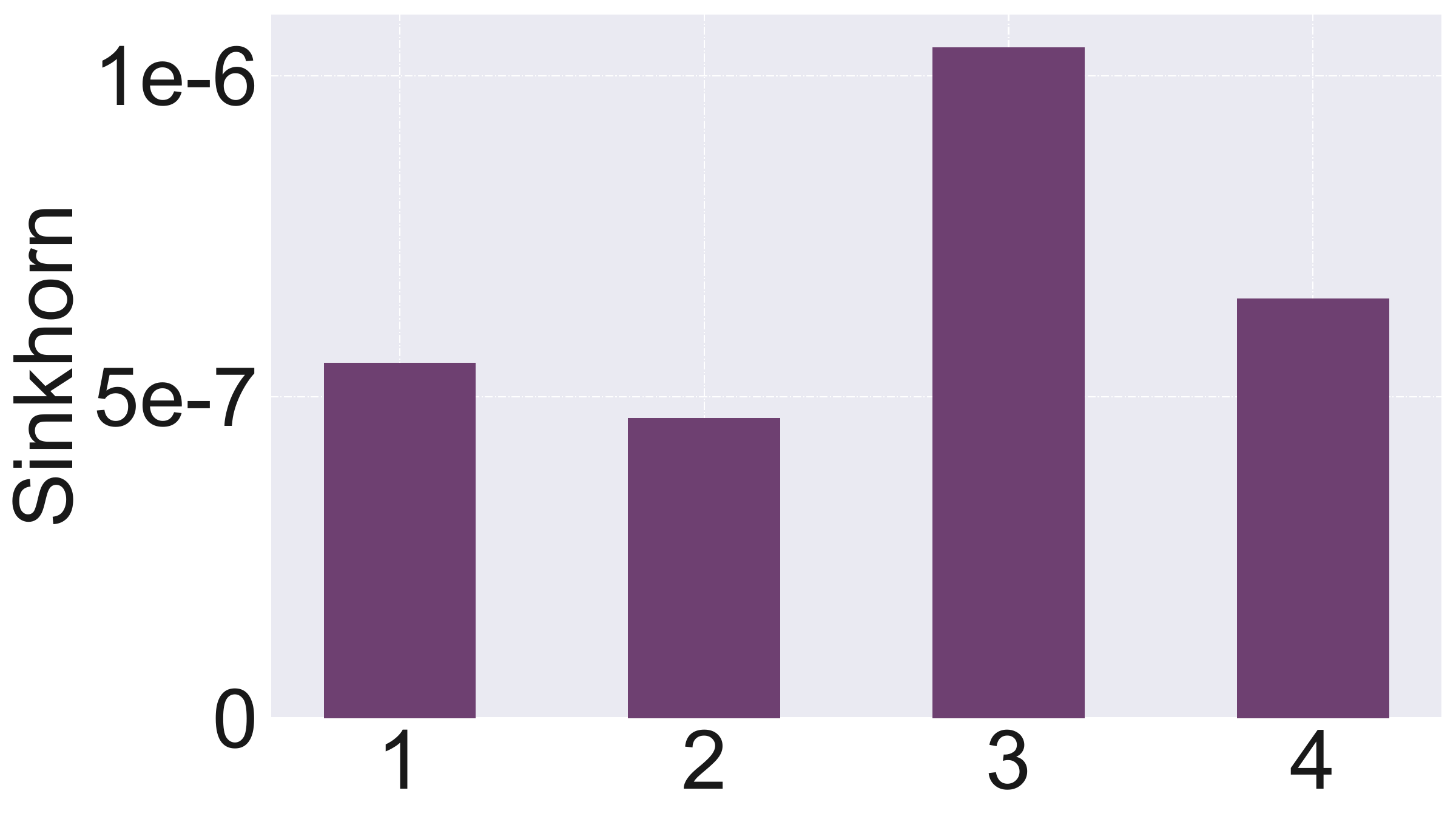}
    \caption{Five-step rollout results of varying the number of downsampling layers, reported on the Lid-Driven Cavity flow dataset.}
    \label{fig:ablation_layer}
\end{figure}

\textbf{Analysis on Number of Downsampling Layers: } We analyze how the number of downsampling layers, $M$, influences model performance, with results shown in Figure~\ref{fig:ablation_layer}. The best results are obtained with $M=2$, followed by $M=1$, while deeper configurations with $M=3$ or $M=4$ lead to a significant drop in performance. These findings are consistent with the results from the study on the downsampling ratio. Increasing the number of downsampling layers progressively reduces the particle set, which can cause the loss of particles carrying critical information about the fluid dynamics. Interestingly, two layers outperform a single layer. We attribute this to the hierarchical structure. Applying downsampling twice enables the model to better identify and preserve important particles, while additional rounds of message passing improve the transfer of information from distant particles to the aggregation nodes.

\section{Additional Baselines on Eulerian Mesh Simulator}
\label{appendix:baselines}
Since Eulerian mesh-based neural simulators operate on graph structures analogous to our Lagrangian framework, they are inherently compatible with our setting. Consequently, we evaluate our approach against a broad suite of state-of-the-art Eulerian mesh models, including the Graph Kernel Operator (GKO)~\citep{li2020neuraloperatorgraphkernel}, General Neural Operator Transformer (GNOT)~\cite{hao2023gnot}, Directional Transport-Aware Graph Neural Network (DTA-GNN)~\citep{li2025flow}, Graph Interaction Operator for Reduced-Order Modeling (GIOROM)~\citep{viswanath2024reduced}, Geometry Informed Neural Operator (GINO)~\citep{li2023geometryinformed}, Transolver~\citep{wu2024Transolver}, and Transolver++~\citep{luo2025transolver}. The experiments are conducted on the \textit{Lid-Driven Cavity Flow} dataset to assess performance.

\begin{table*}[h]
\centering
\scriptsize{%
\resizebox{\linewidth}{!}{%
    \setlength\tabcolsep{2.5pt}%
    \renewcommand\arraystretch{1.4}%
\begin{tabular}{l || c c c c c c c || c}
\hline\thickhline
\rowcolor{CadetBlue!20}
Metric & GKO & GNOT & DTA-GNN & GIOROM & GINO & Transolver & Transolver++ & \cellcolor[HTML]{FFFFE0}\textbf{Ours} \\
\hline

\cellcolor{green!3} $\mathrm{MSE}$ & \cellcolor{green!3} $5.205 \times 10^{-4}$ & \cellcolor{green!3} $1.011 \times 10^{-5}$ & \cellcolor{green!3} $3.505 \times 10^{-6}$ & \cellcolor{green!3} $8.335 \times 10^{-6}$ & \cellcolor{green!3} $3.601 \times 10^{-4}$ & \cellcolor{green!3} $1.308 \times 10^{-5}$ & \cellcolor{green!3} $2.665 \times 10^{-5}$ & \cellcolor[HTML]{FFFFE0}\bm{$1.868 \times 10^{-6}$} \\

\cellcolor{pink!10} $\mathrm{KE}$ & 
\cellcolor{pink!3} $1.161 \times 10^{-4}$ & \cellcolor{pink!3} $9.113 \times 10^{-7}$ & \cellcolor{pink!3} $5.127 \times 10^{-7}$ & \cellcolor{pink!3} $1.765 \times 10^{-6}$ & \cellcolor{pink!3} $8.656 \times 10^{-5}$ & \cellcolor{pink!3} $1.376 \times 10^{-6}$ & \cellcolor{pink!3} $1.417 \times 10^{-5}$ & \cellcolor[HTML]{FFFFE0}\bm{$7.240 \times 10^{-8}$} \\

\cellcolor{blue!3} $\mathrm{Sinkhorn}$ & \cellcolor{blue!3} $6.609 \times 10^{-6}$ & \cellcolor{blue!3} $6.470 \times 10^{-7}$ & \cellcolor{blue!3} $1.560 \times 10^{-6}$ & \cellcolor{blue!3} $1.245 \times 10^{-6}$ & \cellcolor{blue!3} $7.321 \times 10^{-6}$ & \cellcolor{blue!3} $2.001 \times 10^{-6}$ & \cellcolor{blue!3} $2.693 \times 10^{-6}$ & \cellcolor[HTML]{FFFFE0}\bm{$4.675 \times 10^{-7}$} \\

\hline

\end{tabular}}}%
\caption{Comparison of MSE, Kinetic Energy (KE) error, and Sinkhorn divergence on Lid-Driven Cavity flow dataset. The relative performance gain against the strongest baseline is computed as $\frac{\text{Baseline} - \text{Ours}}{\text{Baseline}} \times 100\%$.}
\label{tab:mesh}
\end{table*}

While several baselines incorporate aggregation or downsampling techniques, our method introduces fundamental improvements in spatial awareness and domain flexibility. For instance, GINO projects particles onto fixed grids to employ Fourier Neural Operators (FNO), a reliance that renders it inapplicable to irregular domains where our method excels. Furthermore, while Transolver and Transolver++ aggregate points into physical tokens based on latent states, our approach leverages physical coordinates to explicitly preserve the underlying spatial structure. Similarly, although GIOROM generates sparse representations of the input graph, it relies on static heuristics such as random sampling or farthest-point selection. In contrast, our downsampler dynamically identifies and selects salient features based on the data, ensuring the preservation of critical information and the production of more meaningful sparse representations.

The results are summarized in Table~\ref{tab:mesh}. We observe that our proposed model achieves significantly superior performance compared to the Eulerian mesh baseline simulators. This highlights the effectiveness of our design choices, particularly regarding sparse representation construction and particle aggregation techniques.

\section{Additional Experiments on Diverse Materials}
\label{appendix:diverse}
Although this work focuses primarily on Lagrangian fluid simulation, we also evaluate our model’s performance on diverse materials, including sand, goop, and multi-material interactions. Using datasets from \citet{sanchezgonzalez2020learningsimulatecomplexphysics}, we report performance across 200 rollout steps. The results, summarized in Table~\ref{tab:diverse}, demonstrate that our model generalizes effectively to these alternative materials. We exclude PhysicsNFP from this comparison as it is unable to accommodate varying particle counts.

\begin{table*}[h]
\centering
\scriptsize{%
\resizebox{\linewidth}{!}{%
    \setlength\tabcolsep{4pt}%
    \renewcommand\arraystretch{1.4}%
\begin{tabular}{c | l c || c c c c c c }
\hline\thickhline
\rowcolor{CadetBlue!20}
 & & & \multicolumn{6}{c}{\textbf{Steps}} \\
\rowcolor{CadetBlue!20}
\multirow{-2}{*}{\textbf{Data}} & \multirow{-2}{*}{\textbf{Model}} & \multirow{-2}{*}{\textbf{Metric}} & \textbf{25} & \textbf{50} & \textbf{75} & \textbf{100} & \textbf{150} & \textbf{200} \\
\hline

\cellcolor{white} & \cellcolor{green!3} & \cellcolor{green!3} $\mathrm{MSE}$ & \cellcolor{green!3} $1.668 \times 10^{-4}$ & \cellcolor{green!3} $1.616 \times 10^{-3}$ & \cellcolor{green!3} $5.883 \times 10^{-3}$ & \cellcolor{green!3} $2.173 \times 10^{-2}$ & \cellcolor{green!3} $2.421 \times 10^{-2}$ & \cellcolor{green!3} $3.240 \times 10^{-2}$ \\
\cellcolor{white} & \cellcolor{pink!10} & \cellcolor{pink!10} $\mathrm{KE}$  & \cellcolor{pink!10} $6.536 \times 10^{-3}$ & \cellcolor{pink!10} $2.414 \times 10^{-1}$ & \cellcolor{pink!10} $8.072 \times 10^{0}$ & \cellcolor{pink!10} $4.850 \times 10^{1}$ & \cellcolor{pink!10} $3.271 \times 10^{1}$ & \cellcolor{pink!10} $7.792 \times 10^{1}$ \\
\cellcolor{white} & \cellcolor{blue!3}\multirow{-3}{*}{GraphUnet}  & \cellcolor{blue!3} $\mathrm{Sinkhorn}$ & \cellcolor{blue!3} $3.095 \times 10^{-4}$ & \cellcolor{blue!3} $3.849 \times 10^{-3}$ & \cellcolor{blue!3} $9.788 \times 10^{-3}$ & \cellcolor{blue!3} $3.947 \times 10^{-2}$ & \cellcolor{blue!3} $3.078 \times 10^{-2}$ & \cellcolor{blue!3} $3.099 \times 10^{-2}$ \\
\cline{2-9}
\rowcolor[HTML]{FFFFE0}
\cellcolor{white} &   &   MSE &   \bm{$1.321 \times 10^{-4}$} &    \bm{$9.191 \times 10^{-4}$} &    \bm{$3.458 \times 10^{-3}$} &    \bm{$9.409 \times 10^{-3}$} &    \bm{$9.422 \times 10^{-3}$} &    \bm{$1.367 \times 10^{-2}$} \\
\rowcolor[HTML]{FFFFE0}
\cellcolor{white} & & KE & \bm{$3.784 \times 10^{-3}$} & \bm{$2.153 \times 10^{-1}$} & \bm{$4.183 \times 10^{0}$} & \bm{$3.370 \times 10^{1}$} & \bm{$2.775 \times 10^{1}$} & \bm{$4.773 \times 10^{1}$} \\
\rowcolor[HTML]{FFFFE0}
\cellcolor{white}\multirow{-6}{*}{\rotatebox{90}{\textbf{Goop}}} & \multirow{-3}{*}{\textbf{Ours}} & Sinkhorn & \bm{$2.599 \times 10^{-4}$} & \bm{$1.847 \times 10^{-3}$} & \bm{$6.752 \times 10^{-3}$} & \bm{$1.838 \times 10^{-2}$} & \bm{$9.632 \times 10^{-3}$} & \bm{$1.938\times 10^{-2}$} \\
\hline

\cellcolor{white} & \cellcolor{green!3} & \cellcolor{green!3} $\mathrm{MSE}$ & \cellcolor{green!3} $5.578 \times 10^{-5}$ & \cellcolor{green!3} $3.152 \times 10^{-3}$ & \cellcolor{green!3} $1.378 \times 10^{-2}$ & \cellcolor{green!3} $2.729 \times 10^{-2}$ & \cellcolor{green!3} $4.462 \times 10^{-2}$ & \cellcolor{green!3} $5.060 \times 10^{-2}$ \\
\cellcolor{white} & \cellcolor{pink!10} & \cellcolor{pink!10} $\mathrm{KE}$ & \cellcolor{pink!10} $7.982 \times 10^{-4}$ & \cellcolor{pink!10} $1.145 \times 10^{0}$ & \cellcolor{pink!10} $2.373 \times 10^{1}$ & \cellcolor{pink!10} $9.424 \times 10^{1}$ & \cellcolor{pink!10} $2.675 \times 10^{2}$ & \cellcolor{pink!10} $3.590 \times 10^{2}$ \\
\cellcolor{white} & \cellcolor{blue!3}\multirow{-3}{*}{GraphUnet}  & \cellcolor{blue!3} $\mathrm{Sinkhorn}$ & \cellcolor{blue!3} $9.766 \times 10^{-5}$ & \cellcolor{blue!3} $5.228 \times 10^{-3}$ & \cellcolor{blue!3} $2.567 \times 10^{-2}$ & \cellcolor{blue!3} $5.222 \times 10^{-2}$ & \cellcolor{blue!3} $8.543 \times 10^{-2}$ & \cellcolor{blue!3} $9.783 \times 10^{-2}$ \\
\cline{2-9}
\rowcolor[HTML]{FFFFE0}
\cellcolor{white} & & MSE & \bm{$4.545 \times 10^{-5}$} & \bm{$2.913 \times 10^{-3}$} & \bm{$1.346 \times 10^{-2}$} &   \bm{$2.429 \times 10^{-2}$} & \bm{$2.348 \times 10^{-2}$} & \bm{$2.05 \times 10^{-2}$} \\
\rowcolor[HTML]{FFFFE0}
\cellcolor{white} & & KE & \bm{$6.196 \times 10^{-4}$} & \bm{$9.243 \times 10^{-1}$} & \bm{$2.153 \times 10^{1}$} & \bm{$7.226 \times 10^{1}$} & \bm{$7.446 \times 10^{1}$} & \bm{$5.995 \times 10^{1}$} \\
\rowcolor[HTML]{FFFFE0}
\cellcolor{white}\multirow{-6}{*}{\rotatebox{90}{\textbf{Sand}}} & \multirow{-3}{*}{\textbf{Ours}} & Sinkhorn & \bm{$8.544 \times 10^{-5}$} & \bm{$4.312 \times 10^{-3}$} & \bm{$2.536 \times 10^{-2}$} & \bm{$4.572 \times 10^{-2}$} & \bm{$4.102 \times 10^{-2}$} & \bm{$3.477 \times 10^{-2}$} \\
\hline

\cellcolor{white} & \cellcolor{green!3} & \cellcolor{green!3} $\mathrm{MSE}$ & \cellcolor{green!3} $1.909 \times 10^{-4}$ & \cellcolor{green!3} $9.311 \times 10^{-4}$ & \cellcolor{green!3} $2.244 \times 10^{-3}$ & \cellcolor{green!3} $3.854 \times 10^{-3}$ & \cellcolor{green!3} $8.534 \times 10^{-3}$ & \cellcolor{green!3} $1.770 \times 10^{-2}$ \\
\cellcolor{white} & \cellcolor{pink!10} & \cellcolor{pink!10} $\mathrm{KE}$ & \cellcolor{pink!10} $2.270 \times 10^{-2}$ & \cellcolor{pink!10} $7.012 \times 10^{-1}$ & \cellcolor{pink!10} $4.394 \times 10^{0}$ & \cellcolor{pink!10} $1.283 \times 10^{1}$ & \cellcolor{pink!10} $3.196 \times 10^{1}$ & \cellcolor{pink!10} $1.369 \times 10^{2}$ \\
\cellcolor{white} & \cellcolor{blue!3}\multirow{-3}{*}{GraphUnet}  & \cellcolor{blue!3} $\mathrm{Sinkhorn}$ & \cellcolor{blue!3} - & \cellcolor{blue!3} - & \cellcolor{blue!3} - & \cellcolor{blue!3} - & \cellcolor{blue!3} - & \cellcolor{blue!3} - \\
\cline{2-9}
\rowcolor[HTML]{FFFFE0}
\cellcolor{white} &   &   MSE &   \bm{$1.773 \times 10^{-4}$} &   \bm{$7.528 \times 10^{-4}$} &   \bm{$1.571 \times 10^{-3}$} &   \bm{$2.490 \times 10^{-3}$} &   \bm{$7.056 \times 10^{-3}$} &   \bm{$1.259 \times 10^{-2}$} \\
\rowcolor[HTML]{FFFFE0}
\cellcolor{white} & & KE & \bm{$1.923 \times 10^{-2}$} & \bm{$4.449 \times 10^{-1}$} & \bm{$1.981 \times 10^{0}$} & \bm{$4.214 \times 10^{0}$} & \bm{$2.013 \times 10^{1}$} & \bm{$6.836 \times 10^{1}$} \\
\rowcolor[HTML]{FFFFE0}
\cellcolor{white}\multirow{-6}{*}{\rotatebox{90}{\textbf{Multi Material}}} & \multirow{-3}{*}{\textbf{Ours}} & Sinkhorn & - & - & - & - & - & - \\
\hline

\end{tabular}}}
\caption{Comparison of MSE, Kinetic Energy (KE) error, and Sinkhorn divergence across datasets containing diverse materials. Sinkhorn divergence is omitted for the multi-material dataset because it does not account for material type.}
\label{tab:diverse}
\end{table*}

\begin{wrapfigure}{r}{0.5\columnwidth}
    \centering
    \begin{minipage}{0.24\columnwidth}
        \centering
        \includegraphics[width=\linewidth]{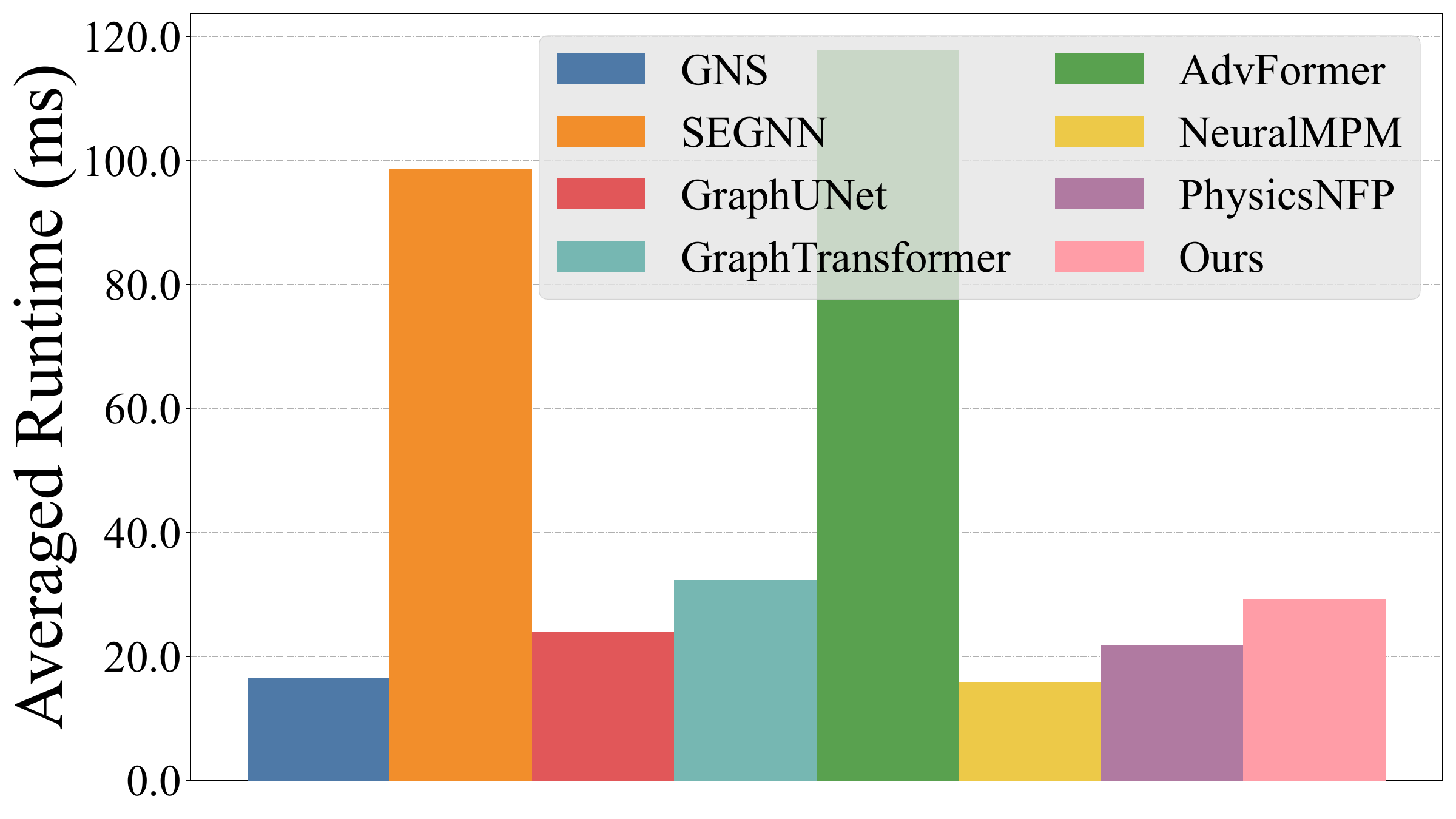}
    \end{minipage}\hfill
    \begin{minipage}{0.24\columnwidth}
        \centering
        \includegraphics[width=\linewidth]{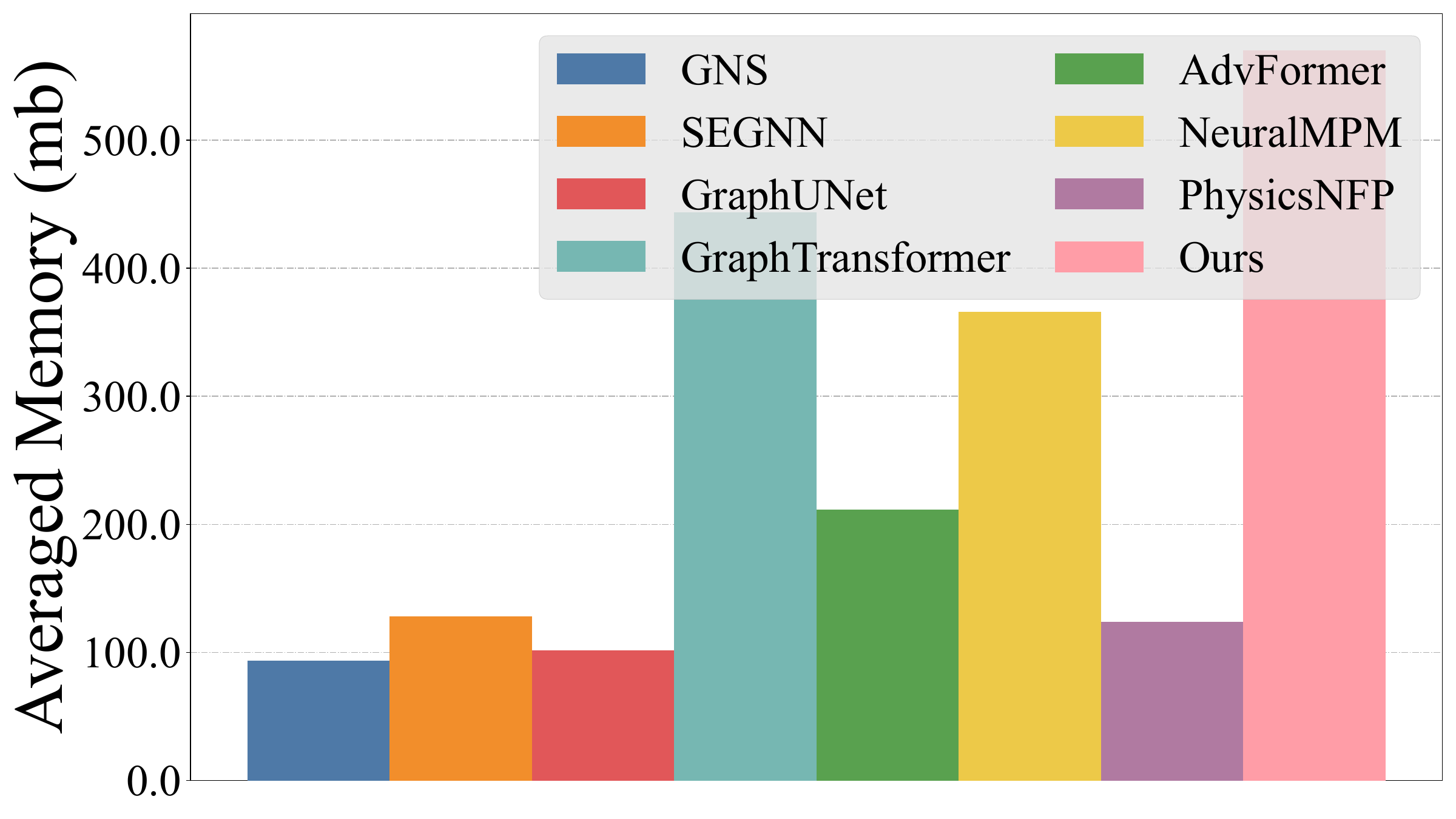}
    \end{minipage}
    \caption{Analysis of the runtime and memory usage of models.}
    \label{fig:runtime_memory}
\end{wrapfigure}

\section{Computational Complexity}
\label{appendix:complexity}
We evaluate the runtime and memory consumption of different models. Although our approach processes both Lagrangian and Eulerian representations, it is naturally amenable to parallelization at inference time. In particular, the encoding and downsampling stages for the two representations can be executed concurrently, which reduces computational time. As illustrated in Figure~\ref{fig:runtime_memory}, this design results in only a marginal increase in inference time compared to the baselines. We acknowledge that our model requires additional memory during inference, which could pose a limitation in certain scenarios. Nonetheless, we argue that the overhead remains modest, well within the capacity of widely available commercial GPUs.

\section{Hardware Specification}
We implement models in PyTorch~\citep{paszke2019pytorchimperativestylehighperformance}. All experiments can be run on servers/workstations with the following configuration:

\begin{itemize}
    \item 80 CPUs, 503G Mem, 8 x NVIDIA V100 GPUs.
    \item 48 CPUs, 220G Mem, 8 x NVIDIA TITAN XP GPUs.
    \item 96 CPUs, 1.0T Mem, 8 x NVIDIA A100 GPUs.
    \item 64 CPUs, 1.0T Mem, 8 x NVIDIA RTX A6000 GPUs.
    \item 224 CPUs, 1.5T Mem, 8 x NVIDIA L40S GPUs.
    \item 128 CPUs, 480G Mem, 8 × NVIDIA RTX 4090 GPUs.
\end{itemize}



\end{document}